\documentclass[10pt-default,11pt]{article}
\usepackage{graphicx}
\usepackage{amsmath}
\usepackage{amsfonts}
\usepackage{amssymb}
\usepackage{geometry}

\input{tcilatex}
\begin{document}

\bigskip

{\huge Optical knots and contact geometry II. }

{\huge From Ranada dyons to transverse and }

{\huge cosmetic knots}

\bigskip

\bigskip

\textbf{Arkady L. Kholodenko\ }

\bigskip

{\small 375 H.L.Hunter Laboratories, Clemson University, Clemson, SC
29634-0973,USA}

\bigskip

\textbf{Abstract\bigskip }

Some time ago Ranada (1989) obtained new nontrivial solutions of the
Maxwellian gauge fields without sources. These were reinterpreted \ in \
Kholodenko (2015a) (part I) as particle-like (monopoles, dyons, etc.). They
were obtained by the method of Abelian reduction of the non-Abelian
Yang-Mills functional. The developed method uses \ instanton-type
calculations normally employed for the non-Abelian gauge fields. By invoking
the electric-magnetic duality it then becomes possible to replace all known
charges/masses by the particle -like solutions of the source-free Abelian
gauge fields. To employ these results in high energy physics, it is
essential to to extend Ranada's results by carefully analysing and
classifying all \ dynamically generated knoted/linked structures in gauge
fields, including those discovered by Ranada. This task is completed in this
work. The study is facilitated by the recent progress made in solving the
Moffatt conjecture. Its essence is stated as follows: in steady
incompressible Euler-type fluids the streamlines could have knots/links of
all types. By employing \ the correspondence between the ideal hydrodynamics
and electrodynamics discussed in part I and by superimposing it with the
already mentioned method of Abelian reduction, it is demonstrated that in
the absence of boundaries only the iterated torus knots and links could be
dynamically generated. Obtained results allow to develop further
particle-knot/link correspondence studied in Kholodenko (2015b)\bigskip


Knot theory

Morse-Smale dynamical flows

Hydrodynamics\bigskip

\textbf{1}. \textbf{Introduction\bigskip }

The existence-type proofs [$1,2$] of Moffatt conjecture [$3$] are opening
Pandora's box of all kinds of puzzles. Indeed, from the seminal work by
Witten [$4$] (see also Atiyah [$5$]) it is known that the observables for
both the Abelian and non Abelian source-free gauge fields are knotted Wilson
loops. It is believed that only the non Abelian \ Chern-Simons (C-S)
topological field theory is capable of detecting nontrivial knots/links. \
By "nontrivial knots" we mean knots other than unknots, Hopf links and
torus-type knots/links. Being topological in nature the C-S \ functional is
not capable of taking into account boundary conditions. This is true for all
known to us path integral treatments of the Abelian \ and non-Abelian C-S \
field theories. In the meantime the boundary conditions do play an important
role in \ the work by Enciso and Peralta -Salas [$1$] on solving the Moffatt
conjecture. This conjecture can be formuated as follows: in steady
incompressible Euler-type fluids the streamlines could have knots/links of
all types. The correspondence between hydrodynamics and Maxwellian
electrodynamics discussed in our book $[6]$ makes results of [$1$]
transferable to the Abelian/Maxwellian electrodynamics where, in view of
this correspondence it is possible, in principle, to generate knots/links of
all types. The fact that the Abelian gauge fields are capable of producing
the nontrivial knots/links blurs the barriers between the Maxwellian
electrodynamics, Yang-Mills fields and gravity. As result, recently, there
has been (apparently uncorrelated) visibly large activity in
electromagnetism [$7$], non Abelian gauge fields [$8$], and gravity [$9$]
producing torus-type knots/links by using more or less the same methods.
Although the cited papers are presented as the most recent and
representative ones, there are many other papers describing the same type of
knotty structures in these fields. Also, in the magnetohydrodynamics, in
condensed matter physics, etc. Unlike other treatments, \ here we are
interested in study and clssification of all possible knots/links which can
be dynamically generated.

From knot theory and, now proven, geometrization conjecture it follows that
complements of knots/links embedded in $S^{3}$ are spaces of positive,
negative and zero curvature. Thus far the ability to curve the ambient space
was always associated with physical masses. With exception of neutrinos, the
Higgs boson is believed to supply physical masses to the rest of particles.
Now we encounter a situation when the space is being curved by knots/links
produced by stable (on some time scales) configurations of gauge fields of
both Abelian and non Abelian nature. In part I [$10$] (Kholodenko 2015a)) \
and in our book [$6$] we argued that the electric and magnetic charges can
be recreated by the Hopf-like links of the respective gauge fields. Surely,
such charges must also be massive. If such massive particle-like formations
are created by the pure gauge fields then, apparently, all known elementary
masses and charges \ can be topologically described. Attempts to do so is
described \ in a number of publications, begining with paper by Misner and
Wheeler [$11$] and Atiyah at all [$12$], and ending with our latest paper [$%
13$] (Kholodenko 2015b). Just mentioned replacement has many advantages. In
particular, if one believes that the non-Abelian gauge fields are just
natural generalizations of more familiar Maxwellian fields, then one
encounters a problem of existence of non-Abelian charges-analogs of
(seemingly) familiar charges in Maxwell's electrodynamics. Surprisingly, to
introduce the macroscopic charges \ into non-Abelian fields is a challenging
task which, to our knowledge, is not completed. By treating gravity as gauge
theory, the analogous problem exists in gravity too. In gravity it is known
as the problem of description of dynamics of extended bodies. The
difficulties in description of extended objects in both the Y-M and gravity
fields \ are summarized on page 97 of our book [$6]$.

This work is made of seven sections and seven appendices. Almost book-style
manner of presentation in this paper is aimed at making it accessible for
readers with various backgrounds: from purely physical to purely
mathematical. As in part I [$10$], for a quick introduction to ideas and
methods of contact geometry/topology our readers may consult either [$6$],
aimed mainly at readers with physics background, or [$14$] aimed at
mathematicians.

In section 2 we provide the statement of the problem to be studied written
in the traditional style of boundary value type problem. In the same section
we reformulate our problem in the language of contact geometry. In section 3
we introduce the Reeb and the Liouville vector fields and compare them with
the Beltrami vector field playing central role in both Part I and in this
work. Section 4 is essential for the whole paper. In it we establish the
chain of correspondences: Beltrami vector fields $\iff $Reeb vector fields$%
\iff $Hamiltonian vector fields. These correspondences allow us to introduce
the nonsingular Morse-Smale (NMS) flows. \ In section 5 we connect these
flows with the Hamiltonian flows discussed (independently of NMS flows) by
Zung and Fomenko [$15$]. In section 6 we explicitly derive the iterated
torus knot structures predicted by Zung and Fomenko in their paper [$15$] of
1990. These structures are obtained via cascade of bifurcations of
Hamiltonian vector fields which we describe in some detail. In the absence
of boundaries, these are the only knotted linked structures which can be
dynamically generated. We also reinterpret the obtained iterated torus
knots/links in terms of the transversely simple knots/links known in contact
geometry/topology. As a by product, we introduce the Legendrian and
transverse knots \ and links. Since the Legendrian knots/links \ were
christened by Arnol'd as \textit{optical knots/links} we use this
terminology in the titles of both parts I and II of our work. In section 7
we relate results of Birman and Williams papers [$16,17$] with what was
obtained already in previous sections in order to obtain other knots and
links of arbitrary complexity. These are obtainable only in the presence of
boundaries. Along this way we developed new method of designing the Lorenz
template. This template was \ originally introduced in Birman and Williams
paper [$16$] in order to facilitate the description of closed orbits
occuring in dynamics of Lorenz equations. The simplicity of our derivation
of this template enabled us to reobtain the universal template of Ghrist [$%
18 $] by methods different from those by Ghrist. \ Possible applications of
the obtained results to gravity are discussed in section 7 in the context of
cosmetic (not cosmic!) knots. Appendices- from A to G -contain all kinds of
support information needed for uninterrupted reading of the main
text.\bigskip \medskip

\bigskip \textbf{2}\textit{. \ }\textbf{Force-free/Beltrami equation from
the point of view of contact geometry\bigskip }

From Theorem 4.1. [$10$] (part I) it follows that force-free/Beltrami \
vector fields are solutions of the steady Euler flows. At the same time,
Corollary 4.2. is telling us that such flows minimize the kinetic energy
functional. This is achieved due to the fact that Beltrami/force-free fields
have nonzero helicity. The helicity is playing the central role in Ranada's
papers [$19,20$]. By studying helicity Ranada discovered his torus-type
knots/links. The same type of knots were recently reported by Kedia et al \ [%
$21$]. Based on results of part I, it should be obvious that study of
helicity is synonymous with the study of knots and links (at least of
torus-type). Can the same be achieved by studying the Beltrami/force-free
equation? We would like to demonstrate that this is indeed possible.
Although the literature on solving the Beltrami equation is large, only
quite recently the conclusive results on existence of knots and links in
Belrami flows have been published. An example of systematic treatment of
Beltrami flows using conventional methods of \ partial differential
equations is given in the\ pedagogically written monograph by Majda and
Bertozzi [$22$]. Our readers should be aware of many other examples existing
in literature. All these efforts culminated in \ the Annals of Mathematics
paper by Enciso and Peralta-Salas [$1$]. In this paper the authors proved
that the equation for (\textit{strong}) Beltrami fields%
\begin{equation}
\mathbf{\nabla }\times \mathbf{v}=\kappa \mathbf{v}  \tag{2.1a}
\end{equation}%
\ (that is \ the Beltrami equation with constant $\kappa )$ supplemented
with the boundary condition%
\begin{equation}
\mathbf{v\mid }_{\Sigma }=\mathbf{w}  \tag{2.1b}
\end{equation}%
,where $\Sigma $ is embedded oriented analytic surface \ in $\mathbf{R}^{3}$
so that \ the vector $\mathbf{w}$ is tangent to $\Sigma ,$ can have
solutions \ describing knots/links of \textit{any} type (that is not just
torus knots/links). Subsequent studies by the same authors [$23$]
demonstrated that $\Sigma $ is actually having a toral shape/topology%
\footnote{%
Recall that any knot is an embedding of $S^{1}$ into $S^{3}$ (or \textbf{R}$%
^{3}$).}. These authors were able to prove what Moffatt proposed/conjectured
long before heuristically [$3$].The same conclusion was reached in 2000, \
by Etnyre and Ghrist [$2$] who were using methods of contact geometry and
topology. Both Enciso and Peralta-Salas and Etnyre and Ghrist presented a
sort of existence-type proof of the Moffatt conjecture.

In this paper we present yet another proof (\textsl{constructive}) of
Moffatt's conjecture\footnote{%
This should be considered as our original contribution into solution of the
Moffatt conjecture.}. It is based on methods of contact geometry and
topology. Our results can be considered as some elaboration on the results
by Etnyre and Ghrist [$2$]. Unlike the existence-type results of previous
authors, we were able to find explicitly some of the knots/links being
guided (to some extent) by the seminal works by Birman and Williams [$16,17$%
], Fomenko [$24$] and Ghys [$25$]. It is appropriate to mention at this
point that recently proposed experimental methods of generating knots and
links in fluids [$26$] are compatible with those discussed by Birman and
Williams [$16,17$], Ghys [25] and Enciso and Peralta -Salas [$1$]. Following
Etnyre and Ghrist [$27$] we begin our derivation by rewriting the Beltrami
eq.(2.1a) as%
\begin{equation}
\ast d\alpha =\kappa \alpha ,  \tag{2.2a}
\end{equation}%
where $\alpha $ is any contact 1-form and $\ast $ is the Hodge star
operator. Although details of derivation of eq.(2.2a) are given in Chr.5 of
\ [$6$], for physics educated readers basics are outlined in the Appendix A.
Since $\ast \ast =id$ , the same equation can be equivalently rewritten as 
\begin{equation}
d\alpha =\kappa \ast \alpha .  \tag{2.2b}
\end{equation}%
Should $\kappa =1,$ the above equation would coincide with the standard
Hodge relation between 1 and 2 forms. Following Etnyre and Ghrist [$27$] we
need the following \bigskip

\textbf{\ Definition} \textbf{2.1.} \textit{The Beltrami field is called} 
\textit{rotational} if $\kappa \neq 0.$ \bigskip

For this case, we can introduce the \textit{volume 3-form} $\mu $ as follows%
\begin{equation}
\mu =\alpha \wedge d\alpha =\kappa \alpha \wedge \ast \alpha =\kappa dV. 
\tag{2.3a}
\end{equation}%
The volume form $\kappa dV$ can be re normalized so that the factor
(function) $\kappa $ can be eliminated. This is so because the volume form $%
dV$contains the metric factor $\sqrt{g}$ which can be readjusted. This fact
can be formulated as\bigskip

\textbf{Theorem 2.2.} (Chern and Hamilton [$28$]) \textit{Every contact form 
}$\alpha $\textit{\ on a 3-manifold has the adapted Riemannian metric }$g.$

\bigskip

The metric is adapted (that is normalized) if eq.(2.2b) \ can be replaced by 
\begin{equation}
d\alpha =\ast \alpha  \tag{2.2.c}
\end{equation}%
This result can be recognized as the standard result from the Hodge theory.
For such a case we obtain:

\bigskip 
\begin{equation}
\mu =\alpha \wedge \ast \alpha =dV.  \tag{2.3b}
\end{equation}%
Consider now the volume integral 
\begin{equation}
I=\dint\limits_{Y}\alpha \wedge \ast \alpha .  \tag{2.4}
\end{equation}%
On one hand, it can be looked upon as the action functional for the 3d
version of the Abelian/Maxwellian gauge field theory as discussed in
Sections 2 and 3 of part I, on another, the same functional can be used for
description of dynamics of 3+1 Einsteinian gravity \ [$29,30$]. In view of \
Theorem 2.2., eq.(2.2c) can be rephrased now as\bigskip

\textbf{Corollary 2.3. }\textit{Every 3-manifold admits a non-singular
Beltrami flow for some Riemannian structure on it. That is to say, study of
the Beltrami fields on 3-manifolds is equivalent to study of the Hodge
theory on 3-manifolds.\bigskip }

The non singularity of flows is assured by the requirement $\kappa \neq 0.$
The above statement does not include any mention about the existence of
knots/links in the Beltrami flows. Thus, the obtained results are helpful
but not constructive yet. To obtain constructive results we need to
introduce the Reeb vector fields associated with contact structures. For our
purposes it is sufficient to design the Reeb vector fields only for $S^{3}$%
.\bigskip

3. \ \ \ \textbf{Reeb vs Beltami vector fields on}\textit{\ }$S^{3}\bigskip $

\bigskip Following Geiges [$14$] we begin with the definition of the
Liouville vector field \textbf{X}. For this purpose we need to use the
symplectic 2-form introduced in (4.15a) of part I defined on \textbf{R}$%
^{4}. $ Up to a constant factor it is given by\footnote{%
With such normalization it coincides with 2-form given in Geiges [14], page
24.} 
\begin{equation}
\omega =dx_{1}\wedge dy_{1}+dx_{2}\wedge dy_{2}.  \tag{3.1}
\end{equation}%
If we use the definition of the Lie derivative $\mathcal{L}_{X}$ for the
vector field $\mathbf{X}$ 
\begin{equation}
\mathcal{L}_{X}=d\circ i_{X}+i_{X}\circ d,  \tag{3.2}
\end{equation}%
then, we arrvie at\ the following\bigskip

\textbf{Definition 3.1.} \ \textit{The vector field }$X$\textit{\ is called
Liouville if it obeys the equation} 
\begin{equation}
\mathcal{L}_{_{X}}\omega =\omega .  \tag{3.3}
\end{equation}%
The contact 1-form $\alpha $ can be defined now as 
\begin{equation}
\alpha =i_{X}\omega .  \tag{3.4}
\end{equation}%
This formula connects the symplectic and contact geometries in the most
efficient way. To find the Liouville vector field $\mathbf{X}$\textbf{\ }for 
$S^{3}$ we notice that eq.(3.3) may hold for any form and, therefore, such a
form could be, say, some function $\mathfrak{f}$. In this case eq.(3.3)
acquires the form \ [$31$] 
\begin{equation*}
\dsum\limits_{i}x_{i}\frac{\partial \mathfrak{f}}{\partial x_{i}}=\mathfrak{f%
}.
\end{equation*}%
Using this result, the Liouville vector field $\mathbf{X}$ on $S^{3}$,\
where $S^{3}$ is defined by the equation $%
r^{2}=x_{1}^{2}+y_{1}^{2}+x_{2}^{2}+y_{2}^{2},$ is given by 
\begin{equation}
\mathbf{X}=\frac{1}{2}(x_{1}\partial _{x_{1}}+y_{1}\partial
_{y_{1}}+x_{2}\partial _{x_{2}}+y_{2}\partial _{y_{2}}).  \tag{3.5}
\end{equation}%
To check correctness of this result, by combining eq.s(3.4)-(3.5) and using
properly normalized eq.(4.14) of part I, we obtain%
\begin{equation}
\mathcal{L}_{_{X}}\omega =d\circ i_{X}\omega =d\circ \alpha =d[\frac{1}{2}%
\dsum\nolimits_{i=1}^{2}(x_{i}dy_{i}-y_{i}dx_{i})]=\omega  \tag{3.6}
\end{equation}%
as required. Going back to eq.(3.4) we would like to demonstrate now that
the 1-form $\alpha $ can be also obtained differently. This is so because
the very same manifold has both the symplectic and the Riemannian structure.
In the last case the metric 2-form $g=g_{ij}dx^{i}\otimes dx^{j}$ should be
defined. Then, for the vector field $\bar{X}=X^{i}\frac{\partial }{\partial
x^{i}}$ we obtain (using definitions from Appendix A): $\bar{X}^{b}$ $=i_{%
\bar{X}}g=g(\bar{X},\cdot )\equiv g_{ij}X^{i}dx^{j}$. From the same appendix
we know that if the operator $\flat $ transforms vector fields into 1-forms,
then the inverse operator $\natural $ is transforming 1-forms into vector
fields, that is $[X^{\flat }]^{\natural }=g^{ij}X_{i}\frac{\partial }{%
\partial x^{j}}=X^{i}\frac{\partial }{\partial x^{i}}=\bar{X}.$ Suppose now
that $\alpha =\tilde{X}^{b}=i_{\tilde{X}}g$ for some vector field \~{X} such
that $\alpha =\tilde{X}^{b}=i_{\tilde{X}}g=i_{X}\omega .$ This can be
accomplished as follows. Suppose that \~{X} is the desired \ vector field
then, we can normalize it as 
\begin{equation}
i_{\tilde{X}}\alpha =i_{\tilde{X}}i_{\tilde{X}}g=1.  \tag{3.7}
\end{equation}%
Explicitly, this equation reads $i_{\tilde{X}}[g_{ij}\tilde{X}%
^{i}dx^{j}]=g_{ij}\tilde{X}^{i}\tilde{X}^{j}=1.$ This result is surely
making sense. Furthermore, the above condition can be safely replaced by $%
g_{ij}\tilde{X}^{i}\tilde{X}^{j}>0$. This is so, because in the case of $%
S^{3}$ the condition given by eq.(3.7) reads: $%
x_{1}^{2}+y_{1}^{2}+x_{2}^{2}+y_{2}^{2}=1.$ Therefore, it is clear that this
condition can be relaxed to $x_{1}^{2}+y_{1}^{2}+x_{2}^{2}+y_{2}^{2}=r^{2}$ $%
>0.$ The condition, eq.(3.7), is the 1st of two conditions defining \textit{%
the Reeb vector field}. \ The 2nd condition is given by 
\begin{equation}
i_{\tilde{X}}d\alpha =0.  \tag{3.8}
\end{equation}%
Suppose that, indeed, $\alpha =i_{\tilde{X}}g=i_{X}\omega ,$ where $X$ is
the Liouville and $\tilde{X}$ is the Reeb vector field. Then, we have to
require: $i_{\tilde{X}}d\alpha =i_{\tilde{X}}d\circ i_{\tilde{X}}g=i_{\tilde{%
X}}d\circ i_{X}\omega =i_{\tilde{X}}\omega =0\footnote{%
Here we used eq.s(3.6) and (3.8).}$. This requirement allows us to determine
the Reeb field. \ It also can be understood physically. For this purpose we
consider the volume 3-form $\ \mu =\alpha \wedge d\alpha $ and apply to it
the Lie derivative, i.e. 
\begin{eqnarray}
\mathcal{L}_{_{\tilde{X}}}\mu &=&(\mathcal{L}_{_{\tilde{X}}}\alpha )\wedge
d\alpha +\alpha \wedge (\mathcal{L}_{_{\tilde{X}}}d\alpha )  \notag \\
&=&(d\circ i_{\tilde{X}}+i_{\tilde{X}}\circ d)\alpha \wedge d\alpha +\alpha
\wedge (d\circ i_{\tilde{X}}+i_{\tilde{X}}\circ d)d\alpha =0.  \TCItag{3.9}
\end{eqnarray}%
This result is obtained after we used \ the \ two Reeb conditions. Clearly,
for the Reeb fields the equation 
\begin{equation}
\mathcal{L}_{_{\tilde{X}}}\mu =0  \tag{3.10}
\end{equation}%
is equivalent to the incompressibility condition for fluids, or to the
transversality condition div$\mathbf{A}=0$ for electromagnetic fields.
\bigskip From here we obtain the major

\textbf{Corollary 3.2}. \textit{From eq.(3.9) it follows that the condition
\ }$\mathcal{L}_{_{\tilde{X}}}\alpha =0$\textit{\ is implying that the Reeb
vector field \ flow preserves the form }$\alpha $\textit{\ and, with it, the
contact structure }$\xi =\mathit{ker}$\textit{\ }$\alpha .\bigskip $ $\
Alternatively,$\textit{\ the Reeb vector} \textit{field }$\tilde{X}$\textit{%
\ is determined by the condition} $\mathcal{L}_{_{\tilde{X}}}\alpha
=0.\bigskip $

Nevertheless, we would like to demonstrate now that the condition $i_{\tilde{%
X}}\omega =0$ is also sufficient for determination of the Reeb field. \ For
the tasks we are having in mind, it is sufficient to check this condition
for $S^{3}$ where the results are known \ [$6,14$]. Specifically, it is
known that for $S^{3}$ the Reeb vector field is given by 
\begin{equation}
\tilde{X}=2(x_{1}\frac{\partial }{\partial y_{1}}-y_{1}\frac{\partial }{%
\partial x_{1}}+x_{2}\frac{\partial }{\partial y_{2}}-y_{2}\frac{\partial }{%
\partial x_{2}}).  \tag{3.11}
\end{equation}%
By combining eq.s(3.1) and (3.11) we obtain:%
\begin{equation}
i_{\tilde{X}}\omega =-2(x_{1}dx_{1}+y_{1}dy_{1}+x_{2}dx_{2}+y_{2}dy_{2}). 
\tag{3.12}
\end{equation}%
However, in view of the fact that $%
x_{1}^{2}+y_{1}^{2}+x_{2}^{2}+y_{2}^{2}=r^{2},$ we obtain as well: $%
2rdr=2\left( x_{1}dx_{1}+y_{1}dy_{1}+x_{2}dx_{2}+y_{2}dy_{2}\right) =0,$ if
this result is to be restricted to $S^{3}.$ Thus, at least for the case of $%
S^{3},$ we just have obtained $i_{\tilde{X}}\omega =0$ as required.

Eq.(3.4), when combined with eq.(3.5), yields $\alpha =i_{X}\omega =\frac{1}{%
2}\{eq.(4.14),$part I\} in accord with eq.(3.6). Now we take again $\alpha =%
\tilde{X}^{b}=i_{\tilde{X}}g=i_{X}\omega $ and, since $\tilde{X}^{b}=g_{ij}%
\tilde{X}^{i}dx^{j},$ we obtain 
\begin{equation}
\alpha =\tilde{X}^{b}=g_{ij}\tilde{X}^{i}dx^{j}.  \tag{3.13}
\end{equation}%
By combining eq.s(3.11) and (3.13) we again recover $\alpha =i_{\tilde{X}}g=%
\frac{1}{2}\{eq.(4.14),$part I$\}.$ Therefore, we just demonstrated that,
indeed, $i_{\tilde{X}}g=i_{X}\omega ,$ where $\tilde{X}$ is the Reeb and $X$
is the Liouville vector fields$.$ By combining eq.(3.13) with eq.(A.5)\ and
(2.2a) we re obtain now the Beltrami equation 
\begin{equation}
\ast d\tilde{X}^{\flat }=\kappa \tilde{X}^{b}.  \tag{3.14}
\end{equation}%
Clearly, it is equivalent to either eq.(2.2a) or (2.2b). Furthermore, in
view of the Theorem 2.2., it is permissible to put $\kappa =1.$

Next, suppose that $\ast \alpha =i_{\tilde{X}}\mu $ then, for the r.h.s \ of
this equality we obtain: $i_{\tilde{X}}\mu =\left( i_{\tilde{X}}\alpha
\right) \wedge d\alpha -\alpha \wedge i_{\tilde{X}}d\alpha =d\alpha .$ This
result becomes possible in view of the 1st and 2nd Reeb conditions. Thus, we
just reobtained eq.(2.2c). The obtained results can be formulated as
\bigskip theorem.\footnote{%
Our derivation of this result differs from that in Etnyre and Ghrist [$27$].}
\ It is of major importance \ for this work\bigskip

\textbf{Theorem 3.3. }\textit{Any rotational Beltrami field on a Riemannian
3-manifold is Reeb-like and vice versa\bigskip }

\textbf{Corollary 3.4. }\textit{Every Reeb-like vector field generates a
non-singular steady solution to the Euler equations for a perfect
incompressible fluid with respect to some Riemannian structure.
Equivalently, every Reeb-like vector field \ which is solution of the
force-free equation generates non-singular solution of the source-free
Maxwell equations with respect to some Riemannian structure}\textbf{%
.\bigskip }

Appendix B provides an illustration of the Corollary 3.4. in terms of
conventional terminology used in physics literature.\bigskip

\textbf{4. \ \ \ \ Hamiltonian dynamics \ and Reeb vector fields\bigskip }

\bigskip Eq.(9.1a) of part I describes the conformation of the single vortex
tube. \ In view of the Beltrami condition, this equation can be equivalently
rewritten as 
\begin{equation}
\lbrack v_{x}\frac{\partial }{\partial x}+v_{y}\frac{\partial }{\partial y}%
+v_{z}\frac{\partial }{\partial z}]f=0.  \tag{4.1a}
\end{equation}%
If we add just one\ (compactification) point to \textbf{R}$^{3}$ we can use
the stereographic projection allowing us to replace \textbf{R}$^{3}$ by $%
S^{3}$ and to consider the conformation of the vortex tube in $S^{3}.$
Example 1.9. (page 123) from the book by Arnol'd and Khesin [$32$] is
telling us (without proof) that the components of the vector $\mathbf{v}$ on 
$S^{3}$ are $\mathbf{v}=[x_{1},-y_{1},x_{2},-y_{2}]\footnote{%
We have relabeled coordinates in Arnol'd -Kheshin book so that they match
those given in eq.(3.11).}.$ \ The same source (again without proof) is also
telling us that the vector \textbf{v} is the eigenvector of the force-free
equation curl$^{-1}\mathbf{v}=\lambda \mathbf{v}$ with the eigenvalue $%
\lambda =1/2.$

In view of these results and using eq.(3.11) for the Reeb vector field, we
replace eq.(4.1a) by 
\begin{equation}
(x_{1}\frac{\partial }{\partial y_{1}}-y_{1}\frac{\partial }{\partial x_{1}}%
+x_{2}\frac{\partial }{\partial y_{2}}-y_{2}\frac{\partial }{\partial x_{2}}%
)f=0  \tag{4.1b}
\end{equation}%
Now, in view of eq.(9.1b) \ of part I this equation can be equivalently
rewritten as 
\begin{equation}
\frac{dy_{1}}{x_{1}}=\frac{dx_{1}}{-y_{1}}=\frac{dy_{2}}{x_{2}}=\frac{dx_{2}%
}{-y_{2}}=dt  \tag{4.1c}
\end{equation}%
so that we recover \ the result of Arnol'd and Khesin for \textbf{v}. In
addition, we obtain: 
\begin{eqnarray}
\dot{x}_{1} &=&-y_{1};\dot{x}_{2}=-y_{2}  \TCItag{4.2} \\
\dot{y}_{1} &=&x_{1};\text{ \ \ }\dot{y}_{2}=x_{2}.  \notag
\end{eqnarray}%
These are the Hamiltonian-type equations describing dynamics of two
uncoupled harmonic oscillators. From mechanics it is known that all
integrable systems can be reduced by a sequence of canonical transformations
to the set of independent harmonic oscillators. The simplicity of the final
result is misleading though\ as can be seen from the encyclopedic book by
Fomenko and Bolsinov [$33$]. It is misleading because the dynamical system
described by eq.s (4.2) possesses several integrals of motion. In
particular, it has the energy $h=\frac{1}{2}%
(p_{1}^{2}+p_{2}^{2}+y_{1}^{2}+y_{2}^{2}),$where $p_{1}=x_{1},p_{2}=x_{2},$
as one of such integrals. The existence of $h$ indicates that the motion is
constrained to $S^{3}.$ Thus, the problem emerges of classification of all
exactly integrable systems whose dynamics is constrained to $S^{3}$.
Surprisingly, there are many dynamical systems fitting such a
classification. The full catalog is given \ in the book by Fomenko and
Bolsinov.\ Whatever these systems might be, once their description is
reduced to the set of eq.s(4.2) supplemented by, say, the constraint of
moving on $S^{3},$ their treatment follows the standard protocol. The
protocol can be implemented either by the methods of symplectic mechanics [$%
31,34$] or by the methods of sub-Riemannian geometry-a discipline which is
part of contact geometry \ [$6,35$]. The results of, say, symplectic
treatment indicate that the trajectories of the dynamical system described
by eq.s(4.2) are the linked (Hopf) rings. This result is consistent with
results of Ranada discissed in Part I. Furthermore, the same eq.s(4.1b) were
obtained by Kamchatnov [$36$]\footnote{%
Without any uses of contact geometry} whose analysis of these equations
demonstrates that, indeed, in accord with the result by Arnol'd and Khesin [$%
32$]\footnote{%
Which is given without derivation in this reference}, the largest eigenvalue 
$\lambda $ of \ the Beltrami equation curl$^{-1}\mathbf{v}=\lambda \mathbf{v}
$ is $\frac{1}{2}$. \ This result follows from eq.(16) of Kamchatnov's paper
where one should replace $x^{2}$ by 1 as required for description of\ a
sphere $S^{3}$ of unit radius. The Example 1.9. in the book by Arnol'd and
Khesin exhausts all possibilities available without further use of methods
of contact geometry and topology. \ These methods are needed, nevertheless,
if we are interested in obtaining solutions of Hamiltonian eq.s(4.2) more
complicated than Hopfian rings.

To begin \ our \ study of this topic, we would like to use the notion of
contactomorphism defined by eq.(4.9) of part I. Now we are interested in
applying it to the standard contact form on $S^{3}.$ To do so, we introduce
the complex numbers $z_{1}=r_{1}\exp \{i\phi _{1}\}$ and $z_{2}=r_{2}\exp
\{i\phi _{2}\}$ so that in terms of these variables the 3-sphere $S^{3}$ is
described by the equation $r_{1}^{2}+r_{2}^{2}=1.$ The 1-form, eq.(4.14) of
part I, can be rewritten in terms of just introduced variables as\footnote{%
Here the factor 1/2 is written in accord with eq.(3.6).} 
\begin{equation}
\alpha =\frac{1}{2}(r_{1}^{2}d\phi _{1}+r_{2}^{2}d\phi _{2}).  \tag{4.3}
\end{equation}%
By combining eq.s (3.11) and (3.6) we see that the factor $1/2$ is needed if
we want to preserve the Reeb condition, eq.(3.7). \ Accordingly, in terms of
\ just introduced new variables \ the Reeb vector $\tilde{X}$ acquires the
following form: \ $\tilde{X}=2(\frac{\partial }{\partial \phi _{1}}+\frac{%
\partial }{\partial \phi _{2}}).$ By design, it satisfies the Reeb condition 
$\alpha (\tilde{X})=1$. Consider now \ yet another Reeb vector $\check{X}=2(%
\frac{1}{r_{1}}\frac{\partial }{\partial \phi _{1}}+\frac{1}{r_{2}}\frac{%
\partial }{\partial \phi _{2}})$ and consider a contactomorphism 
\begin{equation}
\frac{1}{2}\varepsilon (r_{1}^{2}d\phi _{1}+r_{2}^{2}d\phi _{2})=\frac{1}{2}(%
\tilde{r}_{1}^{2}d\phi _{1}+\tilde{r}_{2}^{2}d\phi _{2})=\tilde{\alpha}. 
\tag{4.4}
\end{equation}%
For such defined $\tilde{\alpha}$ we obtain $\alpha (\check{X})=1,$ provided
that we can find such $\varepsilon >0$ that $\varepsilon (r_{1}+r_{1})=1.$
But this is always possible!

By analogy with eq.(4.1b) using $\ \tilde{X}=2(\frac{\partial }{\partial
\phi _{1}}+\frac{\partial }{\partial \phi _{2}})$ we obtain, 
\begin{equation}
(\frac{\partial }{\partial \phi _{1}}+\frac{\partial }{\partial \phi _{2}}%
)f=0\text{ or, equivalently, }\dot{\phi}_{1}=1,\text{ }\dot{\phi}_{2}=1. 
\tag{4.5a}
\end{equation}%
This solution describes the Hopf link [$34,36$]. At the same time, by using $%
\check{X}$ we obtain: 
\begin{equation}
(\frac{1}{r_{1}}\frac{\partial }{\partial \phi _{1}}+\frac{1}{r_{2}}\frac{%
\partial }{\partial \phi _{2}})f=0\text{ or, equivalently, }\dot{\phi}_{1}=%
\frac{1}{r_{1}},\text{ }\dot{\phi}_{2}=\frac{1}{r_{2}}.  \tag{4.5b}
\end{equation}%
If both $r_{1}$ and $r_{2}$ are rational numbers, eq.s(4.5b) describe torus
knots. Both cases were discussed in detail by Birman and Williams [$16$]. \
Understanding/appreciating of this paper is substantially facilitated by
supplemental reading of books by Ghrist et al [$37$] and by Gilmore and
Lefranc [$38$] . Reading of the review article by Franks and Sullivan [$39$]
is also helpful. The \ above arguments, as plausible as they are, cannot be
considered as final. This is so because of the following. According to the
Theorem 3.3. the Beltrami fields can be replaced by the Reeb fields and, in
view of eq.s(4.1) and (4.2), the Beltrami vector fields are equivalent to
the Hamiltonian vector fields. The question arises: Can we relate the
Beltrami fields to Hamiltonian fields without using specific examples given
by eq.s(4.1) and (4.2)? This indeed happens to be the case [$14$]. \ For
physics educated readers needed mathematical information about symplectic
and contact manifolds is given in Appendix C\footnote{%
\ Readers interested in more details are encouraged to read [$6$].} Using
this appendix we obtain: 
\begin{equation}
-dH=i_{\mathbf{v}_{H}}\omega \equiv \omega (\mathbf{v}_{H},\cdot ). 
\tag{4.6}
\end{equation}%
Suppose now that the Reeb vector field $\tilde{X}$ is just a
reparametrization of the Hamiltonian vector field \textbf{v}$_{H}$\ This
makes sense if we believe that examples given in eq.s(4.1) and (4.2) are
generic. If this would be indeed the case, we \ would obtain: $i_{\mathbf{v}%
_{H}}\omega =i_{\tilde{X}}d\alpha =i_{\tilde{X}}\omega =0.$ Here we used
results which follow after eq.(3.8). By looking at eq.(4.6) and by using
results of Appendix C we conclude that just obtained results are equivalent
to the requirement $dH=0$. But, since $dH=\dsum\nolimits_{i}(\dfrac{\partial
H}{\partial q_{i}}dq_{i}+\dfrac{\partial H}{\partial p_{i}}dp_{i})$ we
obtain,%
\begin{equation}
\frac{dp_{i}}{-\dfrac{\partial H}{\partial q_{i}}}=\frac{dq_{i}}{\dfrac{%
\partial H}{\partial p_{i}}}=dt.  \tag{4.7}
\end{equation}

But these are just Hamilton's equations! Thus \ our assumption about the
(anti)collinearity of the Reeb and Hamiltonian vector fields is correct,
provided that both Reeb conditions hold. Since the equation $i_{\tilde{X}%
}d\alpha =0$ is the 2nd Reeb condition (e.g. see eq.(3.8)), we only need to
make sure that the 1st Reeb condition $i_{\tilde{X}}\alpha =1$ also holds.
Since according to eq.(3.4) $\alpha =i_{X}\omega ,$ where $X$ is the
Liouville field, we can write $i_{\tilde{X}}\alpha =i_{\tilde{X}}i_{X}\omega
=\omega (\tilde{X},X)=1.$ It remains now to check if such a condition always
holds. \ Since we are working on $S^{3},$ it is sufficient to check this
condition for $S^{3}.$The proof of the general case is given in the book by
Geiges [$14$], page 25. For $S^{3}$ the Reeb vector field is given by
eq.(3.11) while the Liouville vector field is given by eq.(3.5). Since the
symplectic 2-form is given by eq.(3.1), by direct computation we obtain $i_{%
\tilde{X}}i_{X}\omega =\omega (\tilde{X},X)=1$. Thus, we just obtained the
following correspondences of major importance: Beltrami vector fields $%
\Longleftrightarrow $ Reeb vector fields $\Longleftrightarrow $ Hamiltonian
vector fields. The obtained correspondence allows us now to utilize all
knotty results known for dynamical systems for the present case of Abelian
(Maxwellian) gauge fields. We begin our study of this topic in the next
section\bigskip

\textbf{5. \ \ From Weinstein conjecture to nonsingular Morse-Smale
flows\bigskip }

5.1. \ Some facts about Weinstein conjecture\bigskip

The Weinsten conjecture is just a mathematical restatement of the issue
about the existence of closed orbits on constant energy surfaces. These are
necessarily manifolds of contact type. Indeed, if $2n$ is the dimension of
the symplectic manifold, then the dimension of the constant energy surface
embedded in such a manifold is $2n-1$ which is the odd number. \ All odd
dimensional manifolds are contact manifolds [$6,14$]. Following Hofer [$40$%
], we now formulate\bigskip

\textbf{Conjecture 5.1a.} (Weinstein) \textit{Let }$W$\textit{\ be
symplectic manifold with 2-form }$\omega $\textit{. Let }$\mathit{H}$\textit{%
\ be a smooth Hamiltonian H\ so that }$M:=H^{-1}(E)$\textit{\ is compact
regular energy surface (for some prescribed energy }$E$\textit{). If there
exist a 1-form }$\lambda $\textit{\ on M such that }$\lambda (\mathbf{v}%
_{H}(x))\neq 0$\textit{\ }$\forall x\in M$\textit{\ and }$d\lambda =\omega
\mid _{M},$\textit{then there exists a periodic orbit on }$M$\textit{%
.\bigskip }

According to eq.(3.6) $d\alpha =\omega \mid _{M}$ and, surely, we can
replace $\lambda $ by $\alpha .$ Then, the condition $\lambda (\mathbf{v}%
_{H}(x))\neq 0$ is equivalent to the 1st Reeb condition, eq.(3.7), since $%
\mathbf{v}_{H}$ is equivalent to $\tilde{X}.$ The 2nd Reeb condition surely
holds too in view of the result $i_{\mathbf{v}_{H}}\omega =i_{\tilde{X}%
}d\alpha =i_{\tilde{X}}\omega =0$ obtained in the previous section. Thus,
the above conjecture can be restated as [$41$] \bigskip

\textbf{Conjecture 5.1b.} (Weinstein) \textit{Let }$M$\textit{\ be closed
oriented odd-dimensional manifold with a contact form }$\alpha .$ \textit{%
Then, the associated Reeb vector field has a closed orbit. \bigskip }

Hofer [$42$] using theory of pseudoholomorphic curves demonstrated that the
Weinstein conjecture is true for $S^{3}.$ Much later the same result was
obtained by Taubes, \ e.g. read [$41$] for a review, who used results of
Seiberg-Witten \ and Floer theories. Since some of Floer's results were
exploited in part I, our readers might be interested to know how further
development of Floer's ideas can be used for proving Weinstein's conjecture.
This can be found by reading Ginzburg's paper [$43$]. \ Thus, we now know
that \ on $S^{3}$ trajectories of the Reeb vector fields do contain closed \
orbits. This is surely true \ in the simplest case of Reeb orbits described
by eq.s(4.1) and (4.2). The question arises: Is there other Reeb orbits for
the Hamiltonian system described by eq.s(4.2)? \ We had provided some answer
in eq.s(4.5). \ Now the question arises: Is eq.s(4.5) exhaust all
possibilities? Surprisingly, the answer is "no"!. \ Hofer,Wysocki and
Zehnder [$44$] proved the following\bigskip

\textbf{Theorem 5.2.} (Hofer,Wysocki and Zehnder\textit{) Let the standard
contact form }$\alpha _{0}$\textit{\ on }$S^{3}$\textit{\ be given ether by
eq.(4.6) of part I or, equivalently, by eq.(4.3) above, then there should be
a smooth, positive function \ }$f$\textit{: }$S^{3}$\textit{\ }$\rightarrow
(0,\infty ),$\textit{\ such that if the Reeb vector field associated with
the contact form }$\alpha =f\alpha _{0}$\textit{\ possesses a knotted
periodic orbit, then it possesses infinitely many periodic orbits.\bigskip }

Here $\alpha _{0}$ is the standard contact form, e.g. that given by
eq.(4.3). Eq.(4.4) is an example of relation $\alpha =f\alpha _{0}.$If these
orbits are unknotted, then they are all equivalent. Thus, "infinitely many"
presupposes \ nonequivalence of closed orbits which is possible only if they
are knotted. Etnyre and Ghrist [$2$] proved that periodic orbits on $S^{3}$%
contain knots/links of all possible types simultaneously. Their proof is of
existence-type though since they were not able to find the function $f$
explicitly. In hydrodynamics, finding $f$ \ seemingly provides the
affirmative answer to the Moffatt conjecture [$3$]. Subsequently, Enciso and
Peralta-Salas \ [$1$] proved Moffatt's conjecture by different methods.

In this paper we \ also unable to find \ $f$ explicitly. To by pass this
difficulty, we employ arguments based on the established equivalence (at the
end of section 4) between the Beltrami and Hamiltonian \ vector flows, on
one hand, and on results obtained in paper by Zung and Fomenko [$15$], on
another. In it, the topological classification of non-degenerate Hamiltonian
flows on $S^{3}$ was developed$.$ Other methods of generation of knots/
links \ of all types will be discussed in section 7.\bigskip

5.2. \ From Weinstein and Hofer to Zung and Fomenko \bigskip

As we just stated, because of equivalence established at the end of section
4, it is convenient to adopt results of Zung-Fomenko (Z-F) paper [ ] for \
needed proofs.\ Specifically, the dynamical system whose equations of motion
are given by eq.s(4.2) fits perfectly into Zung-Fomenko general theory. In
it, in addition to the Hamiltonian $h=\frac{1}{2}%
(p_{1}^{2}+p_{2}^{2}+y_{1}^{2}+y_{2}^{2})$ there are several other integrals
of motion among which the integral $F=\frac{1}{2}(f_{2}-f_{1}),$ where $%
f_{1}=p_{1}^{2}+y_{1}^{2}$ and $f_{2}=p_{2}^{2}+y_{2}^{2}$ , is playing a
special role. In Z-F paper it is called the Bott integral for reasons which
will be explained shortly below. Full analysis of this dynamical system was
given in the paper by Jovanovi\v{c} \ [$34$]. From it, it follows that the
solution is made out of two interlocked circular trajectories describing the
Hopf link. This result is consistent with results obtained in part I and,
therefore, with results of Ranada. \ According to Z-F theory other, more
complicated knots/ links can be constructed from the Hopf link with help of
the following 3 topological operations.\bigskip

1. \textsl{\ }$A$ $connected$ $sum$ \# ;

2. \textsl{\ A toral winding} \ is described as follows. Let $K$ be a link, $%
K=\{S_{1},...,S_{k}\}.$

\ \ \ \ \ Select, say, $S_{i}$ and design a regular tubular neighborhood
(that is torus $T^{2}$)

\ \ \ \ \ around $S_{i}.$ Draw on $T^{2}$ a simple closed smooth curve $%
S_{i}(T).$ Then the operation

\ \ $\ \ \ K\rightarrow K\cup S_{i}(T^{2})$ is called \textit{toral winding%
\footnote{%
In knot-theoretic literature this operation is called "cabling operation".
It will be discussed in detail in the next section.}.}

3. \ \textsl{A special toral winding} is described as follows. Let $S_{i}\in
K$ and let $S_{i}(T^{2})$ be the toral

\ \ \ \ \ winding around $S_{i}$ of the type $(2,2l+1)$, $l\in \mathbf{Z}.%
\footnote{%
Here $(2,2l+1)$ is the standard notation for torus knots. In particular,
(2,3) denotes the trefoil knot.}$

\ \ \ \ \ Then, the operation $K\rightarrow K\cup S_{i}(T^{2})\setminus
S_{i} $

\ \ \ \ \ is called \ a \textit{special toral winding}.\bigskip

\ \ \ \ \ Zung and Fomenko [$15$] proved the following\bigskip

\textbf{Theorem 5.3.}\textit{\ Generalized iterated toral windings are
precisely all the possible links of stable periodic trajectories of
integrable systems on }$S^{3}.\bigskip $

\textbf{Corollary 5.4}. \textit{A generalized iterated torus knot is a knot
obtained from trivial knots by toral windings and connected sums.\bigskip
These are the only knots of stable periodic trajectories of integrable
systems on }$S^{3}.\medskip \bigskip $

\textbf{Remark 5.5. }\ Theorem 5.3. provides needed classification of \ all
knots/links which can be dynamically generated. It implies that not every
knot/link of stable periodic trajectories can be generated by integrable
dynamical system on $S^{3}.$\textbf{\ }For instance,\textbf{\ }there are no
\ dynamically generated knots/links containing figure eight knot and,
therefore, \ containing any other hyperbolic knot/link. This observation
immediately excludes from consideration results of Birman and Williams [$17$%
], of Etnyre and Ghrist [$2$] and of Enciso and Peralta-Salas [$1$]. Such an
exclusion is caused by the Hamiltonian nature of the dynamical flows on $%
S^{3}.$ Physically, it happens that such an exclusion is very plausible. It
lies at the heart of \ the particle-knot/link correspondence developed in
Kholodenko [$13$]. \ In support of this correspondence, in this paper we
shall discuss in detail conditions under which all knots/links could be
generated. \ Apparently, these conditions cannot be realized in high energy
physics.\bigskip \bigskip

5.3. \ From Zung and Fomenko to Morse-Smale \bigskip

It is very instructive to re interpret the discussed results in terms of
dynamics of the nonsingular Morse-Smale (NMS) flows. Morgan \ [$45$]
demonstrated that any iterated torus knot can be obtained as an attracting
closed orbit for some NMS flow.\textit{\ }He proved the following\medskip

\textbf{Theorem 5.6.}(Morgan\textbf{) }\textit{If }$K\subset S^{3}$\textit{\
is an attracting closed orbit for a NSM flow on }$S^{3},$\textit{\ then, }$K$%
\textit{\ is iterated torus knot.\medskip }

\textbf{Remark 5.7}. Clearly Theorem 5.3. and Theorem 5.6. produce the same
result. It was obtained by different methods though. These facts are in
agreement with the content of Remark 5.5. \bigskip

Appendix \ D \ contains basic results on Morse-Smale flows. Beginning from
works by Poincare$^{\prime },$ it has become clear that description of
dynamical flows on manifolds is nonseparable from the description of the
topology of the underlying manifolds. Morse theory brings this idea to
perfection by utilizing \ the gradient flows. The examples of gradient flows
are given in part I, e.g. see eq.s (2.22) and (2.33).The basics on gradient
flows can be found, for instance, in the classical book by Hirch and Smale [$%
46$]. The basics on Morse theory known to physicists, e.g. from Nash and Sen
[$47$] and Frankel [$48$], are not sufficient for understanding of the
present case. This is so because the standard Morse theory deals with \ the
nondegenerate and well separated critical points. In the present case we
need to discuss not critical points but critical (sub)manifolds. The
extension of Morse theory covering the case of critical (sub)manifolds was
made by Bott. \ His results and further developments are discussed in the
review paper by Guest [$49$]. In the context of evolution of dynamical
systems on manifolds this extension naturally emerges when one is trying to
provide answers to the following set of \ questions.

a) How is complete integrability of a Hamiltonian system related to the
topology of the

phase or configuration space of this system? It is well known that in the

action-angle variables \ the completely integrable system is decomposed into

Arnol'd -Liouville tori.

b) But what is relative arrangement of these tori in the phase space?

c) Can such\ arrangement of these tori\ result in them to be knotted?

These questions can be answered by studying already familiar dynamical
system described by eq.s(4.2). It has the Hamiltonian $h$ as the integral of
motion. But in addition, it has another (Bott) integral $F.$ To move
forward, we notice that on $S^{3}$ grad $h\neq 0$ \ while grad $F$ can be
zero. Since\ the Bott integral $F$ lives on $h,$ the equation grad $F=0$ is
in fact the equation for the critical submanifold. The theory developed by
Fomenko [$24$] is independent of the specific form of $F$ and requires only
its existence.\bigskip The following \bigskip theorem is crucial\bigskip

\textbf{Theorem 5.7. }\textit{Let F be the Bott integral on some
3-dimensional compact nonsingular \ isoenergic surface h. Then it can have
only 3 types of critical submanifolds: }$S^{1}$\textit{, }$T^{2}$\textit{\
and Klein bottles. \medskip \bigskip }

\textbf{Remark 5.8. }Without loss of generality we can exclude from
consideration the Klein bottles by working only with orientable manifolds.
In such a case we are left with $S^{1}$\textit{, }$T^{2}$ and these were the
only surfaces of Euler characteristic zero discussed in Theorem 4.1. of part
I. Thus, using just this observation we can establish the relationship
between the Morse-Smale and the Beltrami (force-free) flows.\bigskip

Alternatively, following Wada [$50$] \ we attach index 0 to the orbit when
it\ is an attractor, we attach index 1 to the orbit when it is a saddle and
2 when the orbit is a repeller. Both the repeller, Fig.1a), and the
attractor, Fig.1b), are circular orbits: $\alpha =S^{1}$. Fig.1. does not
include the saddle orbit. This orbit happens to be less important as
explained in Theorem 5.9. stated below

\begin{figure}[tbp]
\begin{center}
\includegraphics[scale=1.8]{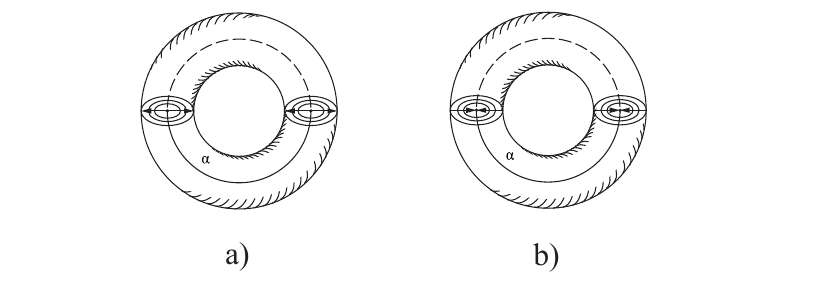}
\end{center}
\caption{ Index 2 (repeller) a), index 0 (attractor) b), $S^{1}-$type
orbits. }
\end{figure}

It will be demonstrated below that both $S^{1}$and\textit{\ }$T^{2}$ can
serve as attractors, repellers or saddles. For $T^{2}$ the analogous
situationis depicted in Fig.2.

\begin{figure}[tbp]
\begin{center}
\includegraphics[scale=1.8]{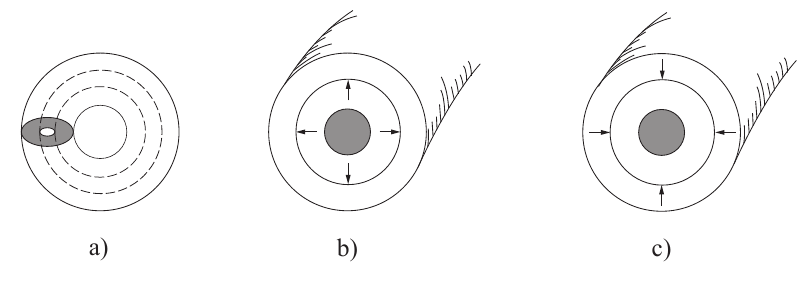}
\end{center}
\caption{The "orientable cylinder" whose boundaries are made out of two $%
T^{2\prime }s.$The repelling b) and the attracting c) tori $T^{2}.$}
\end{figure}

In the case of $T^{2}$ the following theorem by Wada is of importance
\medskip

\textbf{Theorem 5.9.} \textit{Every indexed link which consists of all
closed orbits of a NMS flow on S}$^{3}$\textit{\ is obtained from (0,2) Hopf
link by applying six operations. Convesely, every indexed link obtained from
(0,2) Hopf link by applying these six operations is the set of all the
closed orbits of some NMS flow on S}$^{3}\bigskip \medskip $

\bigskip These six Wada operations will be discussed in the next section. In
the meantime we notice the following. Since by definition the NMS flow does
not have fixed points on the underlying manifold $\mathit{M}$, use of the
Poincare$^{\prime }$-Hopf index theorem \ leads to $\chi (M)=0,$ where $\chi
(M)$ is Euler characteristic of $\mathit{M}$. \ It can be demonstrated that
Euler characteristic for odd dimensional manifolds without boundary is zero
\ [$51$]. The proof for $\mathit{S}^{3}$ is especially simple and is
provided below. Because of this, $\mathit{S}^{3}$ can sustain the NMS flow.
The full implications of this fact were investigated by Morgan [$45$] (e.g.
see Theorem 5.6.above). According to (now proven) geometrization conjecture
every 3-manifold can be decomposed into no more than 8 fundamental pieces [$%
52,53$]. Morgan proved that any 3-manifold which does not contain a
hyperbolic piece can sustain the NMS flows. This result is compatible with
the Remark 5.5.

For $\mathit{S}^{3}$ the above results can be proven using elementary
arguments. Specifically, let us\ notice that Euler characteristic of both $%
S^{1}$ and $T^{2}$ is zero. \ In addition, it is well known that $S^{3}$ can
be made\ out of two solid tori glued together. In fact, any 3-manifold
admits Heegaard splitting \ [$53$]. This means that it can be made by
appropriately gluing \ together two handlebodies whose surfaces are
Riemannian surfaces of genus $g$. In our case $T^{2}$ is the Riemannian
surface of genus one and the handlebody (solid torus) is $V=$ $D^{2}\times
S^{1}$. It has $T^{2}$ as its surface. Consider now the Hopf link. It is
made out of two interlocked circles. We can inflate these circles thus
making two solid tori out of them. These tori (toric handlebodies) can be
glued together. If the gluing \={h} is done correctly [$54,55$], we obtain $%
S^{3}$ as result. The following set-theoretic properties of Euler
characteristic $\chi (M\cup N)=\chi (M)+\chi (N)-\chi (M\cap N)$ \ and $\chi
(M\times N)=\chi (M)\cdot \chi (N)$\ can be used now to calculate the Euler
characteristic of $S^{3}.$ For the solid torus $V=$ $D^{2}\times S^{1}$ we
obtain: $\chi (D^{2}\times S^{1})=\chi (D^{2})\cdot \chi (S^{1})=0$ since $%
\chi (S^{1})=0$ . Next \ $S^{3}=(D_{1}^{2}\times S_{1}^{1})\cup _{\text{\={h}%
}}(D_{2}^{2}\times S_{2}^{1}),$ where \={h} is gluing homeomorphism \={h} $:$
$T_{1}^{2}\rightarrow T_{2}^{2}.$ Therefore $\chi (S^{3})=\chi (V_{1}\cup _{%
\text{\={h}}}V_{2})=\chi (V_{1})+\chi (V_{2})=0$ . Furthermore, the solid
torus $V$ is the trivial case of the Seifert fibered space (see appendix E).
Therefore $S^{3}$ is also Seifert fibered space [$56$]. According to Theorem
5.7. and Definition D.5 of Appendix D the NMS\ flows are made of finite
number of periodic orbits which are either circles or tori. Thus, \textsl{%
the NMS flows can take place on Seifert fibered manifolds or on graph
manifolds.}These are made of appropriately glued together Seifert fibered
spaces \ [$53$]. This conclusion is in accord with that obtained by Morgan [$%
45$] differently.

Let $(M)$ denote the class of all closed compact orientable 3-manifolds and $%
(H)$ denote the class of all closed compact orientable nonsingular
isoenergic surfaces of Hamiltonian systems that can be integrated with help
of the Bott integrals. Then, the following question arises: Is it always
true that $(H)\subset (M)?$

There is a number of ways to construct 3-manifolds from elementary (prime)
blocks \ [$52,53$]. In particular, to answer the above question Fomenko [$24$%
] introduces 5 building blocks to be assembled. Four out of 5 blocks are
associated with orientable manifolds. In this paper we shall discuss only
these types of manifolds. They can be described as follows.

\bigskip

1. \ The solid torus $D^{2}\times S^{1}$ whose boundary is $T^{2}.$

2. \ The orientable "cylinder" \ $T^{2}\times D^{1}$, e.g. see Fig.2a). Its
boundary of

\ \ \ \ \ is made out of two $T^{2}.$

3. \ An "orientable saddle" $N^{2}\times S^{1},$ e.g. see Fig.3a)

\begin{figure}[tbp]
\begin{center}
\includegraphics[scale=1.8]{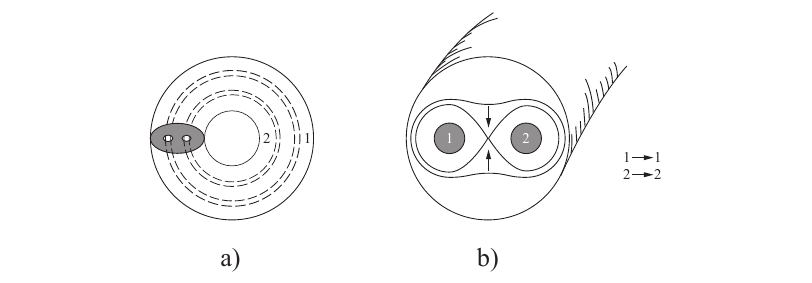}
\end{center}
\caption{Orientable saddle a) and its bifurcation b)}
\end{figure}

\ \ \ \ Its boundary is made out of three $T^{2}.$

4. A "non-orientable saddle", \ e.g. see Fig. 4a).

\begin{figure}[tbp]
\begin{center}
\includegraphics[scale=1.8]{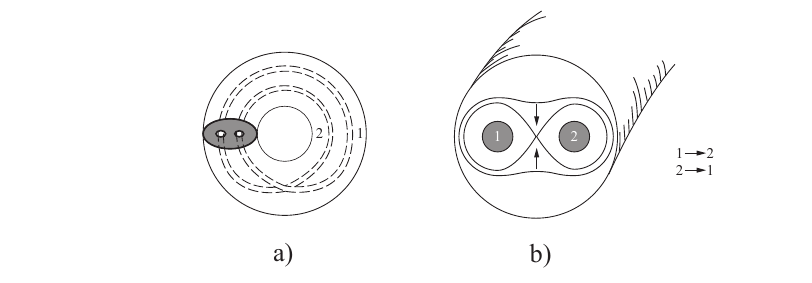}
\end{center}
\caption{A \ non-orientable saddle a) and its bifurcation b)}
\end{figure}

Its boundary is made out of two $T^{2}.$

Just by looking at these figures, it is clear that:

orientable cylinder= orientable saddle+solid torus \medskip

\textbf{Theorem 5.10}.\textit{\ Let }$H=const$\textit{\ be some isoenegy
surface of some Hamiltonian system }

\textit{integrable on H via Bott integral F. \ Then H can be made by gluing}

\textit{together a certain number of solid tori }$D^{2}\times S^{1}$ \textit{%
and orientable saddles}

$N^{2}\times S^{1},$\textit{\ that is }$H=\alpha (D^{2}\times S^{1})+\beta
(N^{2}\times S^{1})$\textit{.\quad }

\textit{The elementary manifolds are glued together via~\ diffeomorphisms of
their boundary tori}

\textit{It is symbolically denoted by the sign "+". Here }$\alpha $\textit{\
and }$\beta $\textit{\ are some \ nonnegative integers}

\allowbreak \smallskip \medskip

In 1988 Matveev and Fomenko [$57$] proved the following theorem (e.g. read
Theorem 3 of this reference) \medskip

\textbf{Theorem 5.11.} \textit{A compact orientable 3-manifold with }(%
\textit{possibly empty})\textit{\ torus-type }

\textit{boundary belongs to the class }(\textit{H})\textit{\ if and only if
\ its interior admits \ a canonical }

\textit{decomposition into pieces having geometries of the first seven
types. }

\textit{In particular, }(\textit{H})\textit{\ contains no hyperbolic
manifolds.\medskip \medskip }

\textbf{Remark 5.12.} According to Thurston's geometrization conjecture (now
proven by G. Perelman) every 3-manifold admits a \textit{canonical
decomposition} into 8 basic building blocks (or geometries). Out of these,
only one is hyperbolic, e.g. read Scott \ [$52$].This result is fully
consistent with earlier made Remark 5.5.\bigskip

\textbf{Corollary 5.13.} Theorem 5.11. \ provides an answer to the question
"Is it always true that $(H)\subset (M)?"\bigskip $

\bigskip In \ view of Remark 5.5., Theorem 5.11. is of central importance
for this paper. Because of this, we shall return to its content a number of
times in what follows. In the meantime, in preparing results for the next
section, using Fomenko's [$24$] results, we need to reinterpret Wada's
results, that is \ to describe how his results are related to bifurcations
of the Arnol'd-Liouville tori. Bifurcations of these tori are caused by
changes in the constant $c$ in the equation $F=c$ for the Bott integral.
These are depicted in Fig.26 of Fomenko's paper but, again, we do not need
all of them. Those which we need can be easily described by analogy with 4
basic blocks described above. Our presentation is facilitated by results of
Theorem 5.7. Using it, we conclude that in orientable case we have to deal
only with $S^{1}$ and $T^{2}.$ Stability or instability of these structures
cause us to consider along with them a nearby space foliated, say, by tori.
Specifically,

1. A torus $T^{2}$ is contracted to the axial circle $S^{1}=\alpha $ and
then, it may even vanish,

\ \ \ \ depending upon the value of $c$. Thus, we get $T^{2}\rightarrow
S^{1}\rightarrow 0.$ Naturally, the process

\ \ \ \ can go in opposite direction as well, e.g. see Fig. 1 a) and b).

2. Two tori $T^{2}$ move toward each other (that is \ they both flow into $%
T^{2})$ as depicted

\ \ \ \ in Fig.2 b) and c). Let, say, both \ the outer boundary and the
inner boundary be

\ \ \ \ unstable in the sense that there is a $T^{2}$ surface somewhere in
between the inner

\ \ \ \ and the outer $T^{2\prime }s$ . Such a torus is as an attractor
since both the inner and the

\ \ \ \ outer $T^{2\prime }s$ are being attracted \ to it. In this case we
may have the following

\ \ \ \ process: $2T^{2}\rightarrow T^{2}\rightarrow 0.$

\ \ \ \ Apparently, the process can go in reverse too. In such a case we are
dealing with

\ \ \ \ repeller.

3. Imagine now a pair of pants. Consider a succession of crossections for
such pants.

\ \ \ \ On one side, we will have $D^{2}($the waist). This configuration
will continue till it will

\ \ \ \ hit the fork- the place from where the pants begin . The fork
crossection

\ \ \ \ is made of figure 8. After passing that crossection we are entering

\ \ \ \ the pants. \ This process is depicted \ in Fig.3. We begin with the
configuration

\ \ \ \ of Fig.2a), then the bifurcation depicted in Fig.3b) is taking place
\ resulting in

\ \ \ \ configuration depicted in fig.3 a).

\ \ \ \ Thus, initially we had $T^{2}$ and finally we obtained $2$ $T^{2}$.
\ That is now we have :

\ \ $\ \ T^{2}\rightarrow 2T^{2}.$

4. A non-orientable saddle \ is obtained if for any crossection of $%
N^{2}\times S^{1}$ we can

\ \ \ \ swap the 1st $D_{1}^{2}$ $\subset T_{1}^{2}$ with $D_{2}^{2}$ $%
\subset T_{2}^{2}$ as depicted in Fig.4 a) and b). In such a case

\bigskip\ \ \ we obtain: $T^{2}\rightarrow T^{2}.$ Such description of
bifurcations is consistent with Theorem 5.7.\bigskip

\ \ It can be demonstrated that the bifurcations 2 and 4 can be reduced to 1
and 3.

\ \ This fact will \ be used in the next section.\bigskip

\textbf{Remark 5.14}. Just described processes are expected to play major
role in topological reinterpretation of scattering processes of high energy
physics advocated in [$13$]. \ Very likely, the already developed formalism
of topological quantum field theories (TQFT)\ and Frobenius algebras , e.g.
as discussed in [$58$], could be used for this purpose. Alternatively,
following ideas of Fomenko and Bolsinov book [$33$], and that by Manturov,
e.g. read chapter 8 of [$59$], it might be possible to develop topological
scattering theory using theory of virtual knots and links initiated by
Kauffman [$60$].

\bigskip

\bigskip

\textbf{6. \ \ \ \ Dynamical bifurcations and topological transitions
associated with them \medskip }

\bigskip

\bigskip 6.1. \ General remarks\medskip \bigskip

In the previous section we mentioned works by Wada, Morgan and Fomenko-Zung
related to dynamics of NMS flows. The focus of this paper however is not on
these flows as such but rather on descriptions of mechanisms of generation
of knotted/linked trajectories. Theorems 5.3. and 5.6. provide us with
guidance regarding the types of knots/links which can be generated by the
NMS flows. However, the above theorems do not explain how such knots/links \
are actually generated. Wada's results, summarized in Theorem 5.9., provide
a formal description of the sequence of topological moves producing the
iterated knots/links, beginning with the Hopf links. These moves are
depicted in Fig.s 2-7 of the paper by Campos et al [$61$]. These are, still,
just particular kinds of Kirby moves as explained in Appendix F. The
description of these moves is totally disconnected from the description of
dynamical bifurcations depicted in Fig.s 1-4 above. Fomenko [$24$] designed
a graphical method helpful for understanding of the sequence of topological
transitions. More details on this topic is given in the monograph by Fomenko
and Bolsinov [$33$]. Remark 5.14. provides us with suggestions for the
further development. Nevertheless, using the already obtained results, we
are now in the position to explore still other ways for connecting Wada's
results with dynamical bifurcations. They are discussed in this section. In
it, we develop our own approach to the description of topological
transitions between dynamically generated knots/links.\bigskip

\bigskip 6.2. \ \ Generating cable and iterated torus knots\medskip \bigskip

In view of Theorems 5.3 and 5.6. we need to provide more detailed
description of the iterated torus knots/links first. For this purpose,
following Menasco [$62$] we introduce \ an \textit{oriented \ }knot $%
S^{1}\rightarrow $ $\mathit{K}\subset S^{3}.$ Let then $V_{K}$ be a solid
torus neighborhood of $K$. Let $\partial V_{K}$ $=T_{K}^{2}\subset S^{3}.$
As in appendix E, we write 
\begin{equation}
J\sim \nu \mathfrak{M}+\mu \mathfrak{L.}  \tag{6.1}
\end{equation}%
This is the same equation as eq.(E.1a) describing a simple oriented closed
curve $J$ \ going $\nu $ times around the meridian $\mathfrak{M}$ and $\mu $
times around the longitude $\mathfrak{L}$ of $T_{K}^{2}.$ $J$ belongs to the
homotopy class $\pi _{1}(T_{K}^{2})$ if and only if either $\nu =\mu =0$ or $%
g.c.d(\mu ,\nu )=1$ (Rolfsen [$55]$).\bigskip

\textbf{Definition 6.1. }When $K$ is unknot\textbf{, }$K=K_{0}$\textbf{, }%
the curve $J$ is called ($\mu ,\nu )$ a \textit{cable }of $K$. \ Thus, $K$($%
\mu ,\nu )$ \textit{torus} \textit{knot} is a cable of the unknot. The 
\textit{cabling operation }leading to the\textit{\ }formation of\textit{\ }$%
\mathit{K}$($\mu ,\nu )$\textit{\ }torus knot from now on will be denoted as%
\textit{\ }\textbf{C}($K_{0}$, ($\mu ,\nu ))$\textit{. \bigskip }

\ The above definition\textit{\ }allows us\textit{\ }to generate the
iterated torus knot inductively. Beginning with some unknot $K_{0},$ we
select a sequence of co-prime 2-tuples of integers $(P,Q)=\{(p_{1},q_{1})$,$%
(p_{2},q_{2}),\cdot \cdot \cdot ,(p_{n},q_{n})\}$, where $p_{1}<q_{1}.$ This
information allows us to construct the oriented \textit{iterated} \textit{%
torus knot}%
\begin{equation}
\mathit{K(P,Q)}=\mathbf{C}(\mathbf{C}(\cdot \cdot \cdot \mathbf{C}(\mathbf{C(%
}K_{0},(p_{1},q_{1}))(p_{2},q_{2}))\cdot \cdot \cdot
,(p_{n-1},q_{n-1})),(p_{n},q_{n})).  \tag{6.2}
\end{equation}%
No restrictions on the relative magnitudes of \ $p_{i}$ and $q_{i}$ are
expected to be imposed for $n>1,$[$63$].

These general rules \ can be illustrated using the trefoil knot $K(2,3)$ as
an example$.$ It is a cable of the unknot. Since the trefoil is the
(simplest) torus knot \ it can be placed on the surface of the solid torus.
This torus has an unknot $K_{0}$ as the core. In Fig.5

\bigskip

\begin{figure}[tbp]
\begin{center}
\includegraphics[scale=1.8]{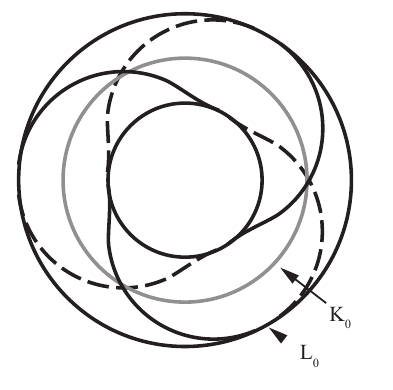}
\end{center}
\caption{One of the ways to depict the trefoil knot }
\end{figure}

the core $K_{0}$ is depicted as a circle while the longitude of the solid
torus around $K_{0}$ is labeled by $\mathit{L}_{0}.$ Without loss of
generality both circles $K_{0}$ and $\mathit{L}_{0}$ are placed on the same
plane $z=0$. \ \ The projection of the knot $K(2,3)$ into the same z-plane
intersects $L_{0}$ in $q=3$ points. Instead of \ Fig.5,\ the same \
configuration can be interpreted in terms of closed braids. For this
purpose, following Murasugi [$64$], \ the representation of torus knot $%
K(q,p)$ in terms of closed braids is depicted in Fig.6.

\begin{figure}[tbp]
\begin{center}
\includegraphics[scale=1.6]{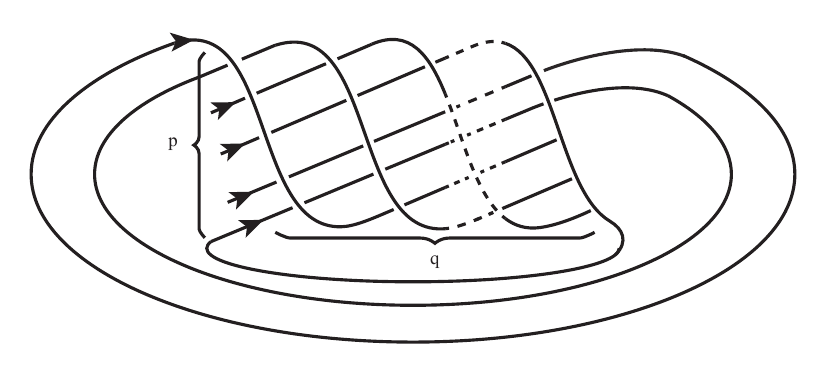}
\end{center}
\caption{Braid-type presentation of the torus knot K(p,q)}
\end{figure}

Such a representation is not unique. It is so because $K(q,p)=K(p,q).$
Furthermore, $K(-q,p)$ is the mirror image of $K(q,p).$ If \ $g.c.d(p,q)=1$,
then $K(-q,-p)$ is the same torus knot but with the reverse orientation.
Being armed with these results, we need to recall the presentation of the
braid group B$_{n}$ made out of $n$ strands. Its generators and relations are%
\begin{equation}
\text{B}_{n}=\left( \sigma _{1},...,\sigma _{n-1}\mid 
\begin{array}{c}
\sigma _{i}\sigma _{j}=\sigma _{j}\sigma _{i},\left\vert i-j\right\vert \geq
2 \\ 
\sigma _{i}\sigma _{i+1}\sigma _{i}=\sigma _{i+1}\sigma _{i}\sigma
_{i+1},i=1,2,...,n-2%
\end{array}%
\right) .  \tag{6.3}
\end{equation}%
The connection between the $K(q,p)$ and its braid analog\ depicted in Fig.6.
can be also established analytically via 
\begin{equation}
K(q,p)\Rightarrow (\sigma _{p-1}\sigma _{p-2}\cdot \cdot \cdot \sigma
_{2}\sigma _{1})^{eq},  \tag{6.4.}
\end{equation}%
where $e=\pm 1$ depending on knot orientation$.$ Here the symbol $%
\Rightarrow $ means "closure of the braid" -an operation converting braids
into knots/links [$64$]. These general results adopted for the trefoil knot
\ are depicted in Fig.7.\ 

\begin{figure}[tbp]
\begin{center}
\includegraphics[scale=1.8]{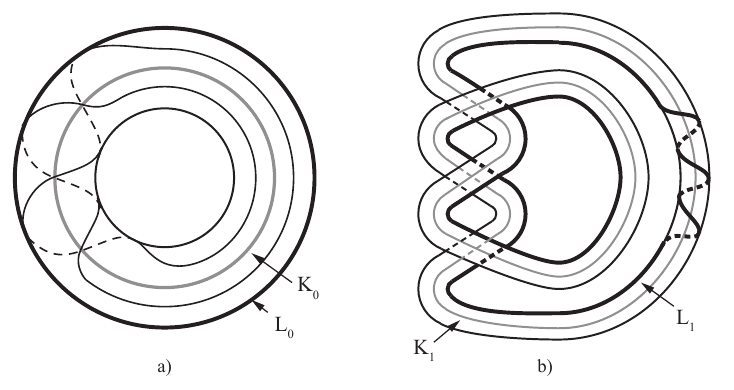}
\end{center}
\caption{Braid-type presentation of the torus knot $K(2,3)$. In a) we use
the same notations as in Fig.5 while in b) the core $K_{1}$ is in fact \ our
trefoil knot $K(2.3)$. Such braid-type presentation is the most convinient
for developing the iteration process generating all iterated torus knots
associated with K(2,3)}
\end{figure}

By design, the notations on this figure are meant to facilitate
visualization of the iteration process. It is formalized in the
following\bigskip

\textbf{Definition 6.2. }\textit{Cabling operation.} Let $K_{1}$ be an
arbitrary oriented knot in $S^{3}$ and $N(K_{1})$ is its solid torus tubular
neighborhood. Let furthermore $L_{K_{1}}$ be a longitude for $K_{1}$. It is
a simple closed curve on $\partial N(K_{1})$ homologous to $K_{1}$ in $%
N(K_{1})$ and null- homologous in $S^{3}\smallsetminus K_{1}.$ \ Consider
now a homeomorphism $h:N(K_{0})\rightarrow N(K_{1})$ \ which is also mapping 
$L_{0}$ into $L_{K_{1}}.$ By relabeling : $K_{0}\rightarrow K_{i}$ and $%
K_{1}\rightarrow K_{i+1}$ , the \textit{cabling operation} \textbf{C }can be
formally defined now as $h:N(K_{i})\rightarrow N(K_{i+1}),$ with $L_{i}$
being mapped into $L_{i+1}.\bigskip $

\textbf{Definition 6.3}. \textit{A cable space} C is a Seifert fibered
manifold obtainable from the solid torus $S^{1}\times D^{2}$ by removing $%
N(K_{1})\subset S^{1}\times \mathring{D}^{2}$ from $S^{1}\times \mathring{D}%
^{2},\mathring{D}^{2}=D^{2}$ $\smallsetminus \partial D^{2}$. Thus, in
accord with results of Appendix E, it is a Seifert fibered manifold having
no exceptional fibers. Alternatively, following Jaco and Shalen [$65$], page
182, a cable space C can be defined as follows. Let $S$ be a Seifert fibered
space over a disc $D^{2}$ with one exceptional fiber, then \ the complement
in $S$ of an open regular neighborhood of a \textit{regular fiber} is a
cable space C.\bigskip

The validity of the above definition is based on the following \medskip

\textbf{Theorem 6.4.} (Hempel [$66$]) \textit{Let }$V_{K}\subset S^{1}\times
D^{2}$\textit{\ be a tubular neighborhood of the torus knot }$K$\textit{\ }(%
\textit{including the unknot})\textit{\ in }$S^{3}$ \textit{and let }$M$%
\textit{\ be \ a simply connected 3-manifold containing a solid torus }$V$%
\textit{. Suppose that there is a homeomorphism }$h$\textit{\ of }$%
S^{3}\smallsetminus $\textit{\ }$\mathring{V}_{K}$\textit{\ onto }$%
M\smallsetminus \mathring{V},$ \textit{then }$M=S^{3}.$\textit{\ \medskip }

\textbf{Corollary 6.5. }Due to the Heegaard decomposition, $S^{3}$ is always
decomposable into two solid tori. Therefore, in view of the above
homeomorphism, it is sufficient to replace $S^{3}$ by ( that is work with)
solid torus $V=S^{1}\times D^{2}$ \ only. This provides a justification of
the operations inside the solid torus depicted in Fig.12 (appendix
F).\bigskip

The information \ we have accumulated allows us now to reobtain one of Zung
and Fomenko's results. Specifically, we have in mind the special toral
windings leading to\ torus knots of the type $K(2,2l+1)$. \ Now we are in
the position \ enabling us to explain the meaning of this result.

Since $K(q,p)=K(p,q)$, we obtain: $K(2,2l+1)=$ $K(2l+1,2)$ implying that we
are dealing with torus knots made out of just 2 braids twisted $n=2l+1$
times followed by the closure operation making a knot out of them. \ Adopted
for such a case eq.(6.4) now reads: $K(n,2)\rightleftarrows (\sigma
_{1})^{n}.$ Switching of just one crossing in the knot projection, e.g. see
Fig.6, \ makes $(\sigma _{1})^{n}$ to be replaced by $\sigma _{1}^{i}\sigma
_{1}^{-1}\sigma _{1}^{j}$ so that $i+j-1=m$ . To find $m$ it is sufficient
to notice that initially (that is before switching) we had $i+j+1=n.$ Thus, $%
i+j=n-1$ and, therefore, $n-2=m.$ In view of the property $K(q,p)\simeq
K(-q,-p)$ it is sufficient to discuss only the case $m\geq 0$ leading to $%
n\geq 2.$ Since $n=2l+1$ we obtain $n=3$ when $l=1$ that is we are dealing
with the trefoil $K(2,3)$. When $l=0$ we are dealing with the unknot and so
on. The obtained result is consistent with the Kirby move depicted in Fig.17
(appendix F). Indeed, \ we can always place the unknot into solid torus
(Corollary 6.5). If we begin with the Hopf link, which is $\pm 1$ framed
unknot, and fix our attention at one of the rings which is unknot, then
another ring can be looked upon as framed with framing $\pm 1.$ Such type of
framing converts another unknotted ring into the Hopf link again.

\begin{figure}[tbp]
\begin{center}
\includegraphics[scale=1.6]{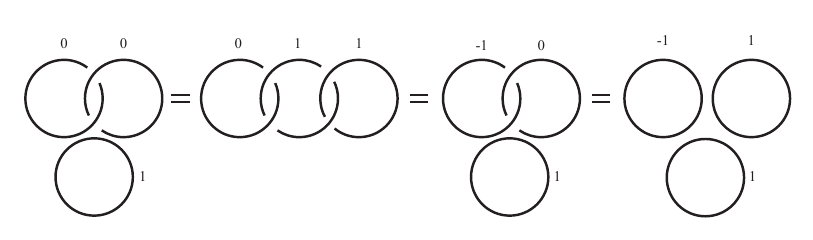}
\end{center}
\caption{Physics behind this picture is charge conservation. It will
explained later in the text}
\end{figure}

This situation is depicted in Fig.8.\ The extra ring can always be found in
the spirit of Wada's (1989) paper [$50$]. Say, we can \ take the meridian of
the solid torus into which the first unknotted ring of the Hopf ring was
enclosed as an extra ring. In fact, this is the content of Theorem 2 by
Menasco [$62$]. Thus, the Kirby moves depicted in Fig. 15 generate all
special toral windings obtained in Fomenko and Zung paper.

With this result in our hands, we still have to uncover the topological
mechanism by which the iterated torus knots are generated in order to
recover the rest of Zung-Fomenko results and those obtained by Morgan (e.g.
see Theorems 5.3.and 5.6.above). It is important to notice at this stage
that torus knots are allowed in the NMS dynamics just because the
Kirby-Fen-Rourke moves allow such knots to exist. These moves are not
sufficient though for generation of the iterated torus knots/links.
Following works by Milnor [$67$] and Eisenbud and Neumann [$68$] it is
possible using \ methods of algebraic geometry \ to develop graphical
calculus generating all iterated torus knots and links. Incidentally, in
current physics literature\ one can find proposals for generating iterated
torus knots/links via methods of algebraic geometry just cited. E.g. read
Dennis et al \ [$69$] or Machon and Alexander [$70$]. While methods of
algebraic geometry are very effective for depicting knots/links, they are
rather formal because they are detached from the topological
content/mechanism of dynamical bifurcations generating various iterated
torus knots. Thus, we are going to proceed with the topological treatment of
dynamical bifurcations. For this purpose, following \ Jaco (1980) [$71$], we
begin with a couple of\bigskip

\textbf{Definition 6.6.} \ In accord with Theorem 6.4. a complement of a
torus knot $K(p,q)$ in $S^{3}$ is the \textit{torus knot space}.\bigskip

\textbf{Definition 6.7. }An (\textit{n-fold})\textit{\ composing space}%
\textbf{\ }is a compact 3-manifolds homeomorphic to $W(n)\times S^{1},$
where $W(n)$ the disk with n-holes\bigskip

\textbf{Remark 6.8}. Evidently, previously defined cable space is just a
special case of the composing space.\bigskip

The complement of a link in $S^{3}$ made of a composition of $n$ torus-type
knots (including the unknot(s)) is an n-fold composing space. An n-fold
composing space is the Seifert\textit{\ }fibered space in which there are no
exceptional fibers (appendix E)\ and the base (the orbit space) is a disc
with $n$ holes\textit{. }Such a fibration of composing space is called%
\textit{\ standard. }The following theorem summarizes what had been achieved
thus far\medskip

\textbf{Theorem 6.9.} (Jaco and Shalen [$65$]), Lemma 6.3.4. \textit{A \
Seifert-fibered 3-manifold }$M(K)=S^{3}\smallsetminus \mathring{N}(K)$%
\textit{\ with incompressible boundary }$\partial M(K)=N(K)\smallsetminus 
\mathring{N}(K)$\textit{\ is either a torus knot space, a cable space or a
composing space.\medskip }

Following Jaco [$71$], page 32, the incompressibility can be defined as
follows. Set $M(K)=S^{3}\smallsetminus \mathring{N}(K),$ then $\partial
M(K)=S^{1}\times S^{1}$ is \textit{incompressible} in $M(K)$ iff $K$ is not
an unknot. \ A complement of \ a cabled knot \ in $S^{3}$ always contains a
cabled space with incompressible boundary [$65$], page 182.

At this point we are having all the ingredients \ needed for description of
the bifurcation cascade creating iterated torus knots/links. The process can
be described inductively. We begin with the seed- the cable space depicted
in Fig.2a). The first bifurcation is depicted in Fig.s 3 a),b). It is
producing the composing nonsingular Seifert fibered space whose orbit space
is the disc $D^{2}$ with two holes. The homeomorphism depicted in Fig.12
allows us to twist two strands \ as many times as needed. Thus, many (but
surely not all!) iterated torus knots are going to have the same complements
in $S^{3}.$ Clearly, the first in line of such type of knots is the trefoil
knot $K(2,3)$ depicted in Fig.7a). It plays the centarl role in
particle-knot correspondence [$13$]. Evidently, other knots of the type $%
K(p,q)$ are also permissible. Their complement is still going to be the
two-hole composing space depicted in Fig.3.a). The three-hole composing
space is generated now as follows. Begin with the composing space depicted
in Fig.3.a). Use the bifurcation process depicted in Fig.3.b) and apply it
to one of the two holes in Fig.3.a). As result, we obtain the three-hole
composing space. Again, we can use the homeomorphisms to entangle the
corresponding strands with each other.

The bifurcation sequence leading to creation of all types of iterated torus
knots is made of steps just described. \ It suffers from several
deficiencies. \ The first among them is a regrettable absence of the one-to
one correspondence between the links and their complements, e.g. see Fig.12.
The famous theorem by Gordon and Luecke [$72$] seemingly guarantees that the
degeneracy is removed for the case of knots since it states that knots are
being determined by their complements. This happens not always to be the
case. Details are given in the next section. The second deficiency becomes
evident already at the level of the trefoil knot which is the cable of the
unknot, e.g. see Fig.7. In Fig.7a) we see that it is permissible to
associate one strand of the two-holed composing space with the unknot while
another-with the trefoil knot. At the same time, if we want to use braids,
as depicted in Fig.7b), then we obtain 3 braids instead of two strands.
Thus, we are coming to the following

\textbf{Problem: }Is there a description of\textbf{\ }the iterated torus
knots in terms of braids as it is done, say, for the torus knots in Fig.6 ?

A connection between closed braids and knots/links is known for a long time.
It is of little use though if we are interested in providing a \textsl{%
constructive} solution to the problem we had just formulated. Surprisingly,
the solution of this problem is very difficult. It was given by Schubert [$%
73 $]. Recently, Birman and Wrinkle [$63$] found an interesting
interrelationship between the iterated torus knots, braids and contact
geometry.This interrelationship happen to be of profound importance in
studying of scattering processes in terms of particle-knot/link
correspondence discussed in [$13$], Kholodenko (2015b). In view of this, we
would like to discuss results of these authors\ in some detail in the next
subsection.\medskip

6.3. Back to contact geometry/topology. Remarkable interrelationship

\ \ \ \ \ \ \ between the iterated torus \ and transversely simple
knots/links\medskip \bigskip

6.3.1. Basics on Legendrian and transverse knots/links\bigskip

In sections 5 and 6 results of contact geometry/topology obtained in
sections 2-4 were used \ but without development. \ In this subsection we
would like to correct this deficiency. For this purpose we need to introduce
the notions of the Legendrian and transverse knots. As before, our readers
are encouraged to consult books by Geiges [$14$] and Kholodenko [$6$] for
details.

We begin with some comments on "optical knots" which were defined in the
Introduction section of part I. The term "optical knots" was invented by
Arnol'd [$74$] in connection with the following problem. \ 

Solutions of the eikonal equation $\left( \partial S/\partial q\right)
^{2}=1 $ determine the optical Lagrangian submanifold $p=\partial S/\partial
q$ \ belonging to the hypersurface $p^{2}=1.$ Every stable Lagrangian
singularity is revealing itself \ in the projection of the Lagrangian
submanifold into the base ($q$-space), e.g. read Appendix 12 of Arnol'd book[%
$31$]. If these results are used in \textbf{R}$^{3},$ then the base is just
the whole or part of \textbf{R}$^{2}.$ In such a case we can introduce the
notion of a \textit{Legendrian} \textit{knot}. It originates from the
equation $p=\partial S/\partial q$ written as $dS-pdq=0$ which we had
already encountered in part I, section 4. This time to comply with
literature on Legendrian knots \ we are going to relabel the entries in the
previous equation as follows 
\begin{equation}
dz-ydx=0.  \tag{6.5a}
\end{equation}%
The minus sign in front of $y$ is determined by the orientation of \textbf{R}%
$^{3}.$ Change in orientation causes change in sign. The standard contact
structure $\xi $ in the oriented 3-space $\mathbf{R}^{3}=(r,\phi ,z),$ that
is in cylindrical coordinates, is determined by the kernel ($\xi =\ker
\alpha ,$ that is by the condition $\alpha =0)$ of the 1-form $\alpha ,$ 
\begin{equation}
\alpha =r^{2}d\phi +dz.  \tag{6.5b}
\end{equation}%
(compare this result against the eq.(4.3)\footnote{%
In fact, we can always use the contactomorphic transformation to replace
locally $\alpha =\frac{1}{2}(r_{1}^{2}d\phi _{1}+r_{2}^{2}d\phi _{2})$ by $%
\tilde{\alpha}=\frac{r_{1}^{2}}{r_{2}^{2}}d\phi _{1}+d\phi _{2}$})\bigskip

\textbf{Definition 6.11. }\textit{A Legendrian knot} $K_{L}$ in an oriented
contact manifold $(M,\xi )$ is a circle $S^{1}$ embedded in $M$ in such a
way that it is always tangent to $\xi .$ Let the indeterminate $\tau $
parametrize $S^{1}$ \ and choose $\mathbf{R}^{3}$ as $M$, then the embedding
is defined by the map $f:$ $S^{1}\rightarrow \mathbf{R}^{3}$ specified
either by $\tau \rightarrow (x(\tau ),y(\tau ),z(\tau ))$ or by $f:\tau
\rightarrow (r(\tau ),\phi (\tau ),z(\tau )).$ The coordinates $r(\tau
),\phi (\tau ),z(\tau )$ are real-valued periodic functions with period,
say, $2\pi .$The tangency condition is being enforced by the equation 
\begin{equation}
\frac{dz}{d\tau }=y(\tau )\frac{dx}{d\tau }.  \tag{6.6}
\end{equation}%
Thus, whenever $\dfrac{dx}{d\tau }$ vanishes, $\dfrac{dz}{d\tau }$ must
vanish as well.\bigskip

\textbf{Definition 6.12. }In terms of $x,y,z$ coordinates it is possible to
define either the \textit{front or the} \textit{Lagrangian projection} 
\textit{of the Legendrian knot}. \textit{The front projection} $\Pi $ is
defined by 
\begin{equation}
\Pi :\mathbf{R}^{3}\rightarrow \mathbf{R}^{2}:(x,y,z)\rightarrow (x,z) 
\tag{6.7}
\end{equation}%
The image $\Pi (K_{L})$ under the map $\Pi $ is called the \textit{front
projection of} $K_{L}.$ The condition, eq.(6.6), is causing the front
projection not to contain the vertical tangencies to the $K_{L}$ projection.
\ Because of this, the front projection of the $K_{L}$ is made of \ a
collection of cusp-like pieces (with all cusps arranged in such a way that
the cusp axis of symmetry is parallel to $x$-axis) joined between each
other. The front projection is always having $2m$ cusps, $m\geq 1.$ It is
important that these cusps exist only in the x-z plane, that is \textsl{not}
in 3-space. \textit{The} \textit{Lagrangian} \textit{projection} $\pi $ of
the Legendrian knot $K_{L}$ is defined by 
\begin{equation}
\pi :\mathbf{R}^{3}\rightarrow \mathbf{R}^{2}:(x,y,z)\rightarrow (x,y). 
\tag{6.8}
\end{equation}%
\bigskip For the sake of space, we shall not discuss details related to the
Lagrangian projection. They can be found in Geiges [$14$].\bigskip

\textbf{Remark 6.13}. Arnol'd \ \textit{optical knots} are Legendrian knots.
In geometrical optics such knots are also known as \textit{plane wavefronts}%
. They obey the Hugens principle.\bigskip

In addition to the Legendrian knots with their two types of projections \
there are also \textit{transverse } knots. They are immediately relevant to
this paper. It can be shown that the \textit{topological} knots/links can be
converted \ both to the Legendrian and to the transverse knots so that the
transverse knots can be obtained from the Legendrian ones and vice versa [$%
14 $], page 103. Since the Legendrian knots are also known as optical knots,
this fact provides a justification for the titles of both parts I and II of
this work.\bigskip\ 

Going back to the description of transverse knots, we begin with the\bigskip

\textbf{Definition 6.14}. \textit{A transverse knot} $K_{T}$ in contact
manifold $(\mathbf{R}^{3},\xi )$ is a circle $S^{1}$ embedded in $\mathbf{R}%
^{3}$ in such a way that it is always transverse to $\xi .$ The transverse
knots are always oriented. The front projection $\Pi (K_{T})$ must satisfy
two conditions [$14,75$] graphically depicted in Fig.9

\begin{figure}[tbp]
\begin{center}
\includegraphics[scale=1.5]{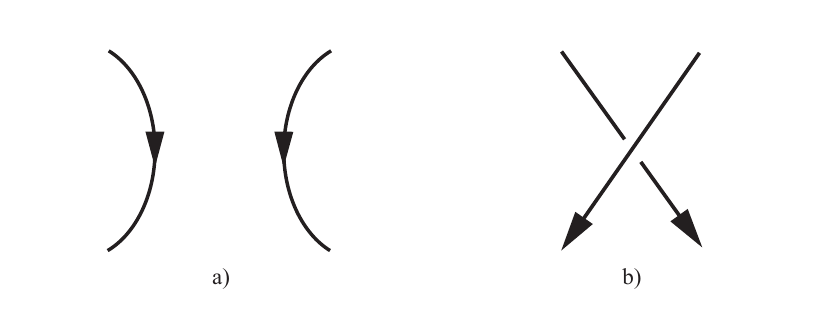}
\end{center}
\caption{Forbidden configurations for the front projection of transverse
knots}
\end{figure}

To formulate the meaning of the notion of transversality and of these
conditions analytically, consider a mapping $\ f:$ $S^{1}\rightarrow \mathbf{%
R}^{3}:$ $\tau \rightarrow (x(\tau ),y(\tau ),z(\tau ))$ with $x(\tau
),y(\tau )$ and $z(\tau )$ being some periodic functions of $\tau $. The
transversality condition now reads%
\begin{equation}
\frac{dz}{d\tau }-y(\tau )\frac{dx}{d\tau }>0.  \tag{6.9a}
\end{equation}%
This inequality defines the canonical orientation on $K_{T}.$ The front
projection $\Pi (K_{T})$ is parametrized in terms of the pair $(x(\tau
),z(\tau )).$ At the vertical tangency pointing down we should have $\frac{dx%
}{d\tau }=0$ and $\frac{dz}{d\tau }<0.$ This contradicts eq.(6.9a) thus
establishing the condition a) in Fig.9. \ The condition b) on the same
figure is established by using eq.(6.9a) written in the form 
\begin{equation}
y(\tau )<\frac{z^{\prime }}{x^{\prime }}.  \tag{6.9b}
\end{equation}%
This inequality implies that y-coordinate is bounded by the slope $dz/dx$ in
the x-z plane (the positive y-axis is pointing into the page). Etnyre [$75$]
argues that this observation is sufficient for proving that the fragment
depicted in Fig.9b) cannot belong to the fragment of the projection of $%
K_{T}.\bigskip $

The importance of transverse knots for this paper\ is coming from their
connection with closed braids studied in previous subsection. To describe
this connection mathematically, \ it is useful to introduce the cylindrical
system of coordinates $(r,\phi ,z)$ so that any closed braid can be looked
upon as a map $f:S^{1}\rightarrow \mathbf{R}^{3}:\tau \rightarrow (r(\tau
),\phi (\tau ),z(\tau ))$ for which $r(\tau )\neq 0$ and $\phi ^{\prime
}(\tau )>0$ $\forall \tau $ [$76$].\bigskip

\textbf{Definition 6.15. }A link\textbf{\ }$L$ is $transverse$ if\textbf{\ }%
the restriction of\textbf{\ }$\alpha =r^{2}d\phi +dz$ to $L$ \ nowhere
vanishes. Any conjugacy class in B$_{n}$ defines a \textit{transverse
isotopy class} of transversal links/knots. Bennequin [$77]$ proved that 
\textit{any transverse knot/link is transversely isotopic to a} \textit{%
closed braid}.\bigskip

Any knot $K$ $\subset \mathbf{R}^{3}$ belongs to its \textit{topological}
type $\mathcal{K}$, that is to the equivalence class under isotopy of the
pair $(K$,$\mathbf{R}^{3}).$ In the case of transverse knots/links, \ one
can define the transverse knot type $\mathcal{T(K)}$. It is determined by
the requirement $\frac{z^{\prime }(\tau )}{\phi ^{\prime }(\tau )}%
+r^{2}(\tau )>0$ at every stage of the isotopy and at every point which
belongs to the knot $K_{T}.$

In the standard knot theory knots are described with help of topological
invariants, e.g. by the Alexander or Jones polynomials, etc. \ Every
transverse knot belongs to a given topological type $\mathcal{K}$. This
means that the knot/link invariants such as Alexander or Jones polynomials
can be applied. In addition, though, the transverse knots have their own
invariant $\mathcal{T(K)}$ implying that all invariants for topological
knots, both the Legendrian and transverse, should be now supplemented by the
additional invariants. For the transverse knots in addition to $\mathcal{T(K)%
}$ one also has to use the \textit{\ Bennequin} (the self-linking)\textit{\
number} $\beta (\mathcal{T(K))}$. To define this number, following Bennequin
[$77$], we begin with a couple of definitions.\bigskip

\textbf{Definition 6.16}.a) The $braid$ $index$ $n=n(K)$ of a closed braid $%
K $ is the number of strands in the braid\bigskip

\textbf{Definition 6.17.}a\textbf{)} The \textit{algebraic length e}(\textit{%
K})\textit{\ of the braid b }%
\begin{equation}
b=\sigma _{i_{1}}^{\varepsilon _{1}}\cdot \cdot \cdot \sigma
_{i_{k}}^{\varepsilon _{k}}\in \text{B}_{n}  \tag{6.10a}
\end{equation}%
\textit{prior to its closure resulting in knot \ K \ is defined \ as}%
\begin{equation}
\mathit{e}(\mathit{K})=\varepsilon _{1}+\cdot \cdot \cdot +\varepsilon
_{k}\in \mathbb{Z}\text{ .}  \tag{6.10b}
\end{equation}

\textbf{Definition 6.18.} The Bennequin number $\beta (\mathcal{T(K))}$ is
defined as 
\begin{equation}
\beta (\mathcal{T(K))=}e(K)-n(K).  \tag{6.11}
\end{equation}%
\bigskip Further analysis \ [$63$] of \ the results obtained by Bennequin
ended in alternative definitions of the braid index and the algebraic
length. Specifically, these authors came up with the following\bigskip

\textbf{Definition 6.16}.b) The $braid$ $index$ $n=n(K)$ of a closed braid $%
K $ is the linking number of $K$ with the oriented z-axis\footnote{%
Recall that we are using the cylindrical system of coordinates for
description of braids.}\bigskip

\textbf{Definition 6.17.}b\textbf{)} The \textit{algebraic crossing number
e=e}(\textit{K}) of the closed braid is the sum of the signed crossings in
the closed braid projection using the sign convention depicted in
Fig.13.\bigskip

A generic (front) projection of $K_{T}$ onto $r-\phi $ plane is called 
\textit{closed braid projection}. Since the braid is oriented, the
projection is also oriented in such a way that moving in the positive
direction following the braid strand (along the z-axis direction) increases $%
\phi $ in the projection$.$ To use these definitions effectively, we need to
recall the definition of a writhe.\bigskip

\textbf{Definition 6.19. }\textit{The writhe} $w(K)$ \ of the \textit{knot
diagram} $D(K)$ (knot projection into plane \textbf{R}$^{2})$ of an oriented
knot is the sum of signs of crossings of $D(K)$ using the sign convention
depicted in Fig.13.\bigskip

From here it follows that if the z-axis is perpendicular to the plane 
\textbf{R}$^{2}$ we obtain $n(K)=0$ and 
\begin{equation}
\beta (\mathcal{T(K))=}w(K).  \tag{6.12}
\end{equation}%
This result is in accord with that listed in Etnyre [$75$] and Geiges [$14$%
], page 127, where it was obtained differently.\bigskip

\bigskip 6.3.2. \ Computation of writhe\bigskip

For reasons which will become obvious upon reading and in view of eq.(6.12)
we would like to evaluate $w(K)$ now$.$ To do so, \ choose the point $%
z=(z_{1},z_{2})=(x_{1}+iy_{1},x_{2}$ $+iy_{2})\in S^{3}$ and, at the same
time, $z\in K_{T}.$ It is permissible to think about $z$ also as a point in 
\textbf{R}$^{4}.$ Because of this, it is possible to introduce physically
appropriate coordinate system in \textbf{R}$^{4}$ as follows. Using eq.(3.5)
we select the components of the Liouville vector field \textbf{X} , that is $%
\mathbf{X=(}x_{1}\mathbf{,}y_{1}\mathbf{,}x_{2}\mathbf{,}y_{2}\mathbf{),}$as
the initial reference direction. Since $S^{3}$ is determined by the equation 
$x_{1}^{2}+y_{1}^{2}+x_{2}^{2}+y_{2}^{2}=1,$ or $\left\vert z_{1}\right\vert
^{2}+\left\vert z_{2}\right\vert ^{2}=1,$ we obtain $%
z_{1}dz_{1}+z_{2}dz_{2}=0.$ This is an equation for a hyperplane $\xi $ ($%
\xi =\ker \alpha ,$ where $\alpha $ is defined by eq.(4.3)). \ 

Recall that, say, in \textbf{R}$^{3}$ the plane $\mathcal{P}$ is defined as
follows. Let $X_{0}=(x_{0},y_{0},z_{0})$ $\in \mathcal{P}$. Let the normal N
to $\mathcal{P}$ is given by N$=(a,b,c)$ where both $X_{0}$ and N vectors
are determined with respect to the common origin. Then, $\forall $ $%
X=(x,y,z)\in \mathcal{P}$ the equation for the plane is \ \textbf{N}$\cdot (%
\mathbf{X}-\mathbf{X}_{0})\equiv $N$\cdot d\mathbf{X}=0$. To relate the
complex and real cases, we rewrite $z_{1}dz_{1}+z_{2}dz_{2}=0$ as $z\cdot
dz=0$. This result is not changed if we replace $dz$ by $idz$. In such a
case, the vector $\mathbf{X}$ is being replaced by $\mathbf{\tilde{X}}=$ $%
\mathbf{(-}y_{1}\mathbf{,}x_{1}\mathbf{,-}y_{2}\mathbf{,}x_{2}\mathbf{).}$
Eq.s (3.5) and (3.11) \ help us \ to recognize in $\mathbf{\tilde{X}}$%
\textbf{\ }the Reeb vector field\textbf{.} Evidently, the equation $%
z_{1}dz_{1}+z_{2}dz_{2}=0$ defining the contact structure $\xi $ ($\xi =\ker
\alpha )$ is compatible now with the condition of orthogonality $\mathbf{X}%
\cdot \mathbf{\tilde{X}}=0$ in \textbf{R}$^{4}$.

In section 4 we established that the Reeb vector field is proportional to a)
the Hamiltonian vector field and to b) the Beltrami vector field. Therefore,
the knot/link transversality requires us to find a plane $\mathcal{\tilde{P}}
$ such that the Beltrami-Reeb vector field \textbf{\~{X} }is pointed in the
direction orthogonal/transversal to the contact plane $\mathcal{\tilde{P}}$.
\ To find this plane unambiguously, we have to find a set of mutually
orthogonal vectors \textbf{X}, \textbf{\~{X},\ \u{X} }and\textbf{\ }$\mathbf{%
\check{X}}$ which span \textbf{R}$^{4}$. By keeping in mind that \textbf{R}$%
^{4}$ is a symplectic manifold into which the contact manifold $S^{3}$ is
embedded, we then should adopt these vectors to $S^{3}$. Taking into account
that the Liouville field \textbf{X} is orthogonal to the surface $\left\vert
z_{1}\right\vert ^{2}+\left\vert z_{2}\right\vert ^{2}=1$ we can exclude it
from consideration. Then, the velocity $\mathbf{\dot{x}}(\tau )$ of any
curve $\mathbf{x}(\tau )=(x_{1}(\tau ),y_{1}(\tau ),x_{2}(\tau ),y_{2}(\tau
))$ in $S^{3}$ admits the following decomposition [$35$]

\begin{equation}
\mathbf{\dot{x}}(\tau )=a(\tau )\mathbf{\breve{X}+}b(\tau )\mathbf{\check{X}+%
}c(\tau )\mathbf{\tilde{X}.}  \tag{6.13}
\end{equation}%
To understand the true meaning of this result, we shall borrow some results
from our book [$6$]. In it we emphasized that contact geometry and topology
is known under different names in different disciplines. In particular, we
explained that the basic objects of study in sub-Riemannian and contact
geometries coincide. This gives us \ a permission to re interpret the
obtained results in the language of sub-Riemannian geometry.

By means of contactomorphism: $(x,y,z)\rightarrow (x,y,\frac{1}{2}xy-z),$
the standard 1-form of contact geometry $\alpha =dz+xdy$ after subsequent
replacement of $z$ by $t/4$ \ acquires the following look%
\begin{equation}
\alpha =-\frac{1}{4}dt+\frac{1}{2}(ydx-xdy).  \tag{6.14a}
\end{equation}%
This form vanishes on two \textit{horizontal} vector fields%
\begin{equation}
\mathbf{X}_{1}=\partial _{x}+2y\partial _{t}\text{ and }\mathbf{X}%
_{2}=\partial _{x}-2y\partial _{t}.  \tag{6.14b}
\end{equation}%
Mathematically, the \textit{condition of horizontality} is expressed as%
\begin{equation}
\alpha (\mathbf{X}_{i})=0,i=1,2.  \tag{6.14c}
\end{equation}%
The Reeb vector field $\mathbf{R}$ is obtained in this formalism as a
commutator:%
\begin{equation}
\lbrack \mathbf{X}_{1},\mathbf{X}_{2}]=-4\partial _{t}\equiv \mathbf{R}. 
\tag{6.14d}
\end{equation}%
Clearly, this commutator is sufficient for determination of $\mathbf{R}$. It
was demonstrated in [$6,35$] that the above commutator is equivalent to the
familiar quantization \ postulate $[q,p]=i\hbar I.$ That is, the commutator,
eq.,(6.14d) defines the Lie algebra for the Heisenberg group. This group has
3 real parameters so that the Euclidean space \textbf{R}$^{3}$ can be mapped
into the space of Heisenberg group. Because of the commutator, eq.(6.14d),
it follows that the motion in the 3rd (vertical) dimension is determined by
the motion in the remaining two (horizontal) dimensions so that \textbf{R}$%
^{3}=\mathbf{R}^{2}\times \mathbf{R}^{1}$. \bigskip

\textbf{Remark 6.20.} This is the simplest form of the \textit{Holographic} 
\textit{principle} used in high energy physics. It also lies at the
foundation of the sub-Riemannian geometry.\bigskip

Specifically, let us suppose that we are having a curve $\mathbf{x}%
(s)=(x(s),y(s),t(s))$ in \textbf{R}$^{3}.$ Its velocity vector $\mathbf{\dot{%
x}}(s)$ can be decomposed as%
\begin{eqnarray}
\mathbf{\dot{x}}(s) &=&\dot{x}\partial _{x}+\dot{y}\partial _{y}+\dot{t}%
\partial _{t}=\dot{x}(\partial _{x}+2y\partial _{t})-2\dot{x}y\partial _{t} 
\notag \\
&&+\dot{y}(\partial _{x}-2y\partial _{t})+2\dot{x}y\partial _{t}+\dot{t}%
\partial _{t}  \notag \\
&=&\dot{x}\mathbf{X}_{1}+\dot{y}\mathbf{X}_{2}-\frac{1}{4}(\dot{t}+2x\dot{y}%
-2y\dot{x})\mathbf{R}  \TCItag{6.15}
\end{eqnarray}%
so that the curve $\mathbf{x}(s)$ is \textit{horizontal} if $\ \dot{t}+2x%
\dot{y}-2y\dot{x}=0.$ This condition is equivalent to the condition $\xi
=\ker \alpha $ introduced before, e.g. see eq.(6.6), in connection with the
Legendrian knots/links. Thus, the decomposition given by eq.(6.15) for the
contact manifold \textbf{R}$^{3}$ should be replaced now by the
decomposition given by eq.(6.13) for $S^{3}.$ In spite of the apparent
differences in appearance between these two results, they can be brought
into correspondence with each other. For this purpose, following Geiges [$14$%
], pages 76 and 95, \ we need to construct the \textit{neighborhood of a} 
\textit{transverse knot/link}. Since any knot $K$ is just an embedding of $%
S^{1}$ into $S^{3}($or \textbf{R}$^{3}),$ locally we can imagine $S^{1}$
piercing a plane $\mathcal{P}$ perpendicularly. Such a plane can be spanned,
say, by the vectors $\mathbf{X}_{1}$ and $\mathbf{X}_{2}$ we just had
described. Clearly, this makes our knot/link transverse and the neighborhood
of $K_{T}$ is described by the two conditions 
\begin{equation}
a)\text{ }\gamma _{T}:=(\theta ,x=0,y=0)\text{ and }\theta \in S^{1};\text{
\ b) }d\theta +xdy-ydx=0.  \tag{6.16}
\end{equation}%
The contact structure $\xi =\ker \alpha $ is determined by eq.(6.14a)) in
which the "time $t$" coordinate is compactified to a circle $S^{1}$. Once
this is done we immediately recognize that:

a) The contact structure in the present case is exactly the same as can be
obtained from the one-form $\tilde{\alpha}=\frac{r_{1}^{2}}{r_{2}^{2}}d\phi
_{1}+d\phi _{2}$ we have introduced earlier.

b) The locality of the result, eq.(6.16), makes it not sensitive to the
specific nature of knot/link. In particular, it remains valid for the Hopf
link too.

These observations were proven previously by Bennequin [$77$] who derived
them differently and formulated them in the form of the \medskip

\textbf{Theorem 6.21.} (Bennequin [$77$], Theorem 10) \textit{Every link
transversal to standard contact structure on }$S^{3}($\textit{given by our
eq.}(\textit{4.3}))\textit{, transversally isotopic to a link L whose
tangent }$TL$\textit{\ at any point which belongs to }$L$\textit{\ is
arbitrarily close to the tangent to the fibre of the Hopf fibration.\medskip 
}

To take full advantage of this result we have to make several additional
steps. For this purpose we have to use the quaternions. Recall, that the
quaternion $q$ can be represented as $q=z_{1}+j\bar{z}_{2}=(z_{1},\bar{z}%
_{2}),$ where $j$ is another complex number, $j^{2}=-1,$ such that $ij=-ji$
. Formally, there is also the third complex number $k$, $k^{2}=-1.$ In view
of the definition of $q$ and the commutation \ relation $ij=-ji$ its use can
be by-passed. In view of just defined rules, $jq=jz_{1}+j^{2}\bar{z}_{2}=-%
\bar{z}_{2}+jz_{1}=(-\bar{z}_{2},\bar{z}_{1})$. \ The obtained result we use
to encode yet another vector $\mathbf{\breve{X}}=$ $\mathbf{(-}x_{2}\mathbf{,%
}y_{2}\mathbf{,}x_{1}\mathbf{,-}y_{1}\mathbf{)}$ in \textbf{R}$^{4}.$ \ This
vector was introduced by Bennequin without derivation. With the vectors 
\textbf{X}, \textbf{\~{X} }and\textbf{\ \u{X} }just\textbf{\ }defined, we
notice that they are mutually orthogonal in \textbf{R}$^{4}$ by design.
Instead of 3 tangent vectors \textbf{X}$_{1}$,\textbf{X}$_{2}$ and \textbf{R}
which span \textbf{R}$^{3}$ now we have 4 vectors which span \textbf{R}$^{4}$%
. These are \textbf{X}, \textbf{\~{X}, \u{X} }and\textbf{\ }$\mathbf{\check{X%
}=(-}y_{2},-x_{2},y_{1},x_{1}).$ The last vector is written in accord with
that used by Hurtado and Rosales [$78$]. In this paper the remaining three
vectors also coincide with ours. When adopted to $S^{3},$ three vectors are
tangent to $S^{3}$ and one, that is X, is normal to $S^{3}.$ Thus, it is
sufficient to consider only the vectors which span the contact manifold $%
S^{3}.$ In such a case the analog of the commutator, eq.(6.14d), is given by 
\begin{equation}
\lbrack \mathbf{\breve{X}},\mathbf{\check{X}]=-}2\mathbf{\tilde{X}.} 
\tag{6.17}
\end{equation}%
In accord with eq.(6.14d)), $\mathbf{\tilde{X}}$ is the Reeb vector field.
The associated contact 1-form was defined in part I, eq.(4.14), as\footnote{%
For convenience of our readers we reproduce it again using current
numeration.} 
\begin{equation}
\alpha =-y_{1}dx_{1}+x_{1}dy_{1}-y_{2}dx_{2}+x_{2}dy_{2}.  \tag{6.18}
\end{equation}%
In the present case the horizontality conditions $\ $are: $\alpha ($\textbf{%
\v{X}}$)=\alpha (\mathbf{\breve{X}})=0.$ Obtained information is sufficient
for explanation of the expansion given in eq.(6.13) (to be compared with
eq.(6.15)) and, hence, for utilization of Theorem 6.21. Results of this
theorem are in accord with the results obtained by Wada (Theorem 5.9.)
stating the the standard Hopf link serves as the seed generating \ all
oriented iterated torus knots. Being armed with these results, now we are in
the position to evaluate (to estimate) the writhe in eq.(6.12). Geiges [$14$%
] page 128, proved the following\medskip

\textbf{Theorem 6.22}. \textit{Every integer can be realized as the
self-linking number }(\textit{that is writhe}) \textit{of some transverse
link\medskip \medskip }

This theorem is providing only a guidance to what to expect. More useful is
the concept of transversal simplicity\bigskip

\textbf{Definition 6.23.} A transverse knot is \textit{transversely simple}
if it is characterized (up to transversal isotopy) by its topological knot
type $\mathcal{K}$ and by $\beta (\mathcal{T(K))}$.\bigskip

By definition, in the case of a link $\beta (\mathcal{T(K))}$ is the sum of $%
\beta (\mathcal{T(K))}$ for each component of the link. Thus, the
m-component unlink is transversely simple since the unknot is transversely
simple. To go beyond these obvious results requires introduction of the
concept of \textit{exchange reducibility}. Leaving details aside (e.g. read
Birman and Wrinkle [$63$], it is still helpful to notice the following.

Closed n-braid isotopy classes are in one-to-one correspondence with the
conjugacy classes of the braid group B$_{n}$. This result survives under the
transverse isotopy. Such isotopy preserves both $n(K)$ and $e(K).$ In
addition to the isotopy moves, there are positive/negative ($\pm )$ \textit{%
destabilization moves}. The destabilization move reduces the braid index
from $n$ to $n-1$ by removing a trivial loop. If such a loop contains a
positive crossing, the move is called positive (+). The move \textsl{reduces}
$e(K)$ \textsl{by one} and thus \textsl{preserves} $\beta (\mathcal{T(K))}$.
The negative destabilization (-) increases $\beta (\mathcal{T(K))}$ by 2. \
Finally, there is an \textsl{exchange} move. It is similar to the 2nd
Reidemeister move. \textsl{It changes the conjugacy class} and, therefore,
it cannot be replaced by the braid isotopy. Nevertheless, an exchange move
preserves \ both $n$ and $e$ and, therefore, also $\beta (\mathcal{T(K))}$.
Clearly, these moves allow us to untangle the braid of $n$ strands and to
obtain the unlink whose projection is made of $m$ unknots.\bigskip

\textbf{Definition 6.24.} A knot of type $\mathcal{K}$ is \textsl{exchange
reducible} if a closed n-braid representing this knot \ can be changed by
the finite sequence of braid isotopies, exchange moves and $\pm $
destabilizations to the minimal m-component unlink $m=n_{\min }(\mathcal{K}%
). $ In such a case max$\beta (\mathcal{T(K))}$ is determined by $n_{\min }(%
\mathcal{K}).$ For the iterated torus knots the exact value of max$\beta (%
\mathcal{T(K))}$ , that is of $n_{\min }(\mathcal{K}),$is known\footnote{%
E.g. read Birman and Wrinkle \ [$63$], Corollary 3.}. In view of eq.(6.12)
we finally obtain 
\begin{equation}
max\beta (\mathcal{T(K))=}n_{\min }(\mathcal{K})=w(K).  \tag{6.19}
\end{equation}

6.3.3. Implications for the interated torus knots\bigskip

Based on just obtained result, it \ is possible to prove the following
\medskip

\textbf{Theorem 6.25. }\textit{If }$\mathit{K}\in \mathcal{T(K)}$\textit{\
such that }$K$\textit{\ is exchange reducible, then \ it is transversely
simple.\medskip }

Proof of this theorem allows to prove the theorem of central importance
\medskip

\textbf{Theorem 6.26}. \textit{The} \textit{oriented iterated torus knots
are exchange reducible. Thus they are transversely simple.\medskip }

\textbf{Question: }Are all knot types $\mathcal{K}$, when converted to
transverse knot types $\mathcal{T}$($\mathcal{K}$) transversely
simple?\bigskip

\textbf{Answer}: No! \ \ \ \ \ \ \ \ \ \ \ \ \ \ \ \ \ \ \ \ \ \ \ \ \ \ \ \
\ \ \ \ \ \ \ \ \ \ \ \ \ \ \ \ \ \ \ \ \ \ \ \ \ \ \ \ \ \ \ \ \ \ \ \ \ \
\ \ \ \ \ \ \ \ \ \ \ \ \ \ \ \ \ \ \ \ \ \ \ \ \ \ \ \ \ \ \ \ \ \ \ \ \ \
\ \ \ \ \ \ \ \ \ \ \ \ \ \ \ \ \ \ \ \ \ \ \ \ \ \ \ \ \ \ \ \ \ \ \ \ \ \
\ \ \ \ \ \ \ \ \ \ \ \ \ \ \ \ \ \ \ \ \ \ \ \ \ \ \ \ \ \ \ \ \ \ \ \ \ \
\ \ \ \ \ \ \ \ \ \ \ \ \ \ \ \ \ \ \ \ \ \ \ \ \ \ \ \ \ \ \ \ \ \ \ \ \ \
\ \ \ \ \ \ \ \ \ \ \ \ \ \ \ \ \ \ \ \ \ \ \ \ \ \ \ \ \ \ \ \ \ \ \ \ \ \
\ \ \ \ \ \ \ \ \ \ \ \ \ \ \ \ \ \ \ \ \ \ \ \ \ \ \ \ \ \ \ \ \ \ \ \ \ \
\ \ \ \ \ \ \ \ \ \ \ \ \ \ \ \ \ \ \ \ \ \ \ \ \ \ \ \ \ \ \ \ \ \ \ \ \ \
\ \ \ \ \ \ \ \ \ \ \ \ \ \ \ \ \ \ \ \ \ \ \ \ \ \ \ \ \ \ \ \ \ \ \ \ \ \
\ \ \ \ \ \ \ \ \ \ \ \ \ \ \ \ \ \ \ \ \ \ \ \ \ \ \ \ \ \ \ \ \ \ \ \ \ \
\ \ \ \ \ \ \ \ \ \ \ \ \ \ \ \ \ \ \ \ \ \ \ \ \ \ \ \ \ \ \ \ \ \ \ \ \ \
\ \ \ \ \ \ \ \ \ \ \ \ \ \ \ \ \ \ \ \ \ \ \ \ \ \ \ \ \ \ \ \ \ \ \ \ \ \
\ \ \ \ \ \ \ \ \ \ \ \ \ \ \ \ \ \ \ \ \ \ \ \ \ \ \ \ \ \ \ \ \ \ \ \ \ \
\ \ \ \ \ \ \ \ \ \ \ \ \ \ \ \ \ \ \ \ \ \ \ \ \ \ \ \ \ \ \ \ \ \ \ \ \ \
\ \ \ \ \ \ \ \ \ \ \ \ \ \ \ \ \ \ \ \ \ \ \ \ \ \ \ \ \ \ \ \ \ \ \ \ \ \
\ \ \ \ \ \ \ \ \ \ \ \ \ \ \ \ \ \ \ \ \ \ \ \ \ \ \ \ \ \ \ \ \ \ \ \ \ \
\ \ \ \ \ \ \ \ \ \ \ \ \ \ \ \ \ \ \ \ \ \ \ \ \ \ \ \ \ \ \ \ \ \ \ \ \ \
\ \ \ \ \ \ \ \ \ \ \ \ \ \ \ \ \ \ \ \ \ \ \ \ \ \ \ \ \ \ \ \ \ \ \ \ \ \
\ \ \ \ \ \ \ \ \ \ \ \ \ \ \ \ \ \ \ \ \ \ \ \ \ \ \ \ \ \ \ \ \ \ \ \ \ \
\ \ \ \ \ \ \ \ \ \ \ \ \ \ \ \ \ \ \ \ \ \ \ \ \ \ \ \ \ \ \ \ \ \ \ \ \ \
\ \ \ \ \ \ \ \ \ \ \ \ \ \ \ \ \ \ \ \ \ \ \ \ \ \ \ \ \ \ \ \ \ \ \ \ \ \
\ \ \ \ \ \ \ \ \ \ \ \ \ \ \ \ \ \ \ \ \ \ \ \ \ \ \ \ \ \ \ \ \ \ \ \ \ \
\ \ \ \ \ \ \ \ \ \ \ \ \ \ \ \ \ \ \ \ \ \ \ \ \ \ \ \ \ \ \ \ \ \ \ \ \ \
\ \ \ \ \ \ \ \ \ \ \ \ \ \ \ \ \ \ \ \ \ \ \ \ \ \ \ \ \ \ \ \ \ \ \ \ \ \
\ \ \ \ \ \ \ \ \ \ \ \ \ \ \ \ \ \ \ \ \ \ \ \ \ \ \ \ \ \ \ \ \ \ \ \ \ \
\ \ \ \ \ \ \ \ \ \ \ \ \ \ \ \ \ \ \ \ \ \ \ \ \ \ \ \ \ \ \ \ \ \ \ \ \ \
\ \ \ \ \ \ \ \ \ \ \ \ \ \ \ \ \ \ \ \ \ \ \ \ \ \ \ \ \ \ \ \ \ \ \ \ \ \
\ \ \ \ \ \ \ \ \ \ \ \ \ \ \ \ \ \ \ \ \ \ \ \ \ \ \ \ \ \ \ \ \ \ \ \ \ \
\ \ \ \ \ \ \ \ \ \ \ \ \ \ \ \ \ \ \ \ \ \ \ \ \ \ \ \ \ \ \ \ \ \ \ \ \ \
\ \ \ \ \ \ \ \ \ \ \ \ \ \ \ \ \ \ \ \ \ \ \ \ \ \ \ \ \ \ \ \ \ \ \ \ \ \
\ \ \ \ \ \ \ \ \ \ \ \ \ \ \ \ \ \ \ \ \ \ \ \ \ \ \ \ \ \ \ \ \ \ \ \ \ \
\ \ \ \ \ \ \ \ \ \ \ \ \ \ \ \ \ \ \ \ \ \ \ \ \ \ \ \ \ \ \ \ \ \ \ \ \ \
\ \ \ \ \ \ \ \ \ \ \ \ \ \ \ \ \ \ \ \ \ \ \ \ \ \ \ \ \ \ \ \ \ \ \ \ \ \
\ \ \ \ \ \ \ \ \ \ \ \ \ \ \ \ \ \ \ \ \ \ \ \ \ \ \ \ \ \ \ \ \ \ \ \ \ \
\ \ \ \ \ \ \ \ \ \ \ \ \ \ \ \ \ \ \ \ \ \ \ \ \ \ \ \ \ \ \ \ \ \ \ \ \ \ 

It happens, Etnyre [$75$], that transversely simple knots comprise a
relatively small subset of knots/links. It is being made of:

a) the unknot;

b) the torus and iterated torus knots;

c) the figure eight knot.

From the results we obtained thus far we know that the figure eight knot
cannot be

dynamically generated (Theorem 5.11). Thus, the transversal simplicity is 
\textbf{not} equivalent to

the previously obtained correspondence for flows: Beltrami$\rightarrow $Reeb$%
\rightarrow $Hamiltonian.\bigskip

\textbf{Question:} Is this fact preventing generation of hyperbolic optical
knots of which the figure eight knot is the simplest representative ?\bigskip

\textbf{Answer}: No! The explanation is provided in the next
section.\bigskip \medskip

\textbf{Remark 6.27}. Theorem 6.26 allows to develop particle- knot/link
correspondence constructively when superimposed with the Dehn surgery $-$
crosssing change correspondence discussed in Kholodenko (2015b).\bigskip

\textbf{7. \ \ \ \ From Lorenz equations to cosmetic knots/links \medskip
\bigskip }

\bigskip In the previous section we described in detail \ the cascade of
bifurcations \ generating \ all possible torus and iterated torus knots and
links dynamically. These results are sufficient for their use in high energy
physics applications \ described in Kholodenko (2015b). In this section we
would like to discuss, also in detail, how other types of knots/links can be
generated \ and why these other types should not be considered in potential
high energy physics applications. We begin with discussion of the classical
works by Birman, Williams and Ghys. \bigskip

7.1.\ \textbf{\ }Birman-Williams treatment of the Lorenz equations\textbf{\
\ \ \medskip\ \ \ \ \ \ \ \ }

Using methods of topological and symbolic dynamics Birman and Williams
(BWa)) [$16$] explained the fact that flows generated by eq.s(4.5b) are
special cases of those which follow from the description of periodic orbits
of Lorenz equations. These \ equations were discovered by Lorenz \ in 1963
who obtained them as finite-dimensional reduction of the Navier-Stokes
equation. Lorenz equations are \ made of a coupled system of three ordinary
differential equations.%
\begin{eqnarray}
\dot{x} &=&-10x+10y,  \notag \\
\dot{y} &=&rx-y-xz,  \TCItag{7.1} \\
\dot{z} &=&-8/3z+xy.  \notag
\end{eqnarray}%
Here $r$ is real parameter, the Rayleigh number. It is typically taken to be
about 24. The equations are not of the Hamiltonian-type since they contain
dissipative term. Accordingly, the results of previous sections cannot be
immediately applied. Nevertheless, the results of BWa) paper are worth
discussing \ because of the following key properties of these equations.
They can be summarized as follows.\bigskip

1. \ There are infinitely many non equivalent Lorenz knots/links generated by

\ \ \ \ \ closed/periodic orbits of eq.s(7.1).

2. \ Every Lorenz knot is fibered\footnote{%
This concept will be explained below}.

3. \ Every algebraic knot is a Lorenz knot.

\ \ \ \ \ From the previous section it follows that all iterated torus
knots/links are of

\ \ \ \ \ Lorenz-type since they are algebraic.

4. \ There are Lorenz knots which are not iterated torus knots\footnote{%
This means that such knots/links cannot be generated via mechanism of
previous section. For example, these are some hypebolic knots/links
described on page 72 of BW a).}; there are

\ \ \ \ \ iterated torus\ knots which are Lorenz but not algebraic\footnote{%
This is proven in Theorem 6.5. of BW a). It is compatible with results of
previous sections.}.

5. \ Every Lorenz link is a closed positive braid\footnote{%
That is all $\varepsilon _{i}>0$ in eq.(6.10a).}, however there are closed

\ \ \ \ positive braids which are not Lorenz.\medskip \medskip

\bigskip 6. \ Every non-triviual Lorenz link which has 2 or more components
is non splittable.

7. \ Non-trivial Lorenz links \ are non-amphicheiral. The link $L$ is
amphicherial

\ \ \ \ \ if there is an orientation-reversing homeomorphism H: $%
(S^{3},L)\rightarrow (S^{3},L)$ reversing

\ \ \ \ \ orientation of $L$. \ Because the Lorenz links are
non-amphicheiral,

\ \ \ \ they are oriented links. From the Definition 6.14 we know that all
transverse

\ \ \ \ knots/liks are oriented. Thus,\textsl{\ }all\textsl{\ }Lorenz
knots/links are subsets of transverse

\ \ \ \ \ knots/links

8. \ Non-trivial Lorenz links have positive signature. This technical
concept is explained

\ \ \ \ \ in Murasugi [$64$]. In Kholodenko (2015b) the the signature is
given

\ \ \ \ \ physical interpretation.\bigskip \bigskip

7.2. \ \ Birman-Williams (1983b)) paper $[17$]. The universal template and
the

\ \ \ \ \ \ \ \ \ Moffatt conjecture\medskip \bigskip

BWa) study of periodic orbits was greatly facilitated by the template
construction invented in their work. See also Ghrist et al [$37$] ,

Gilmore and Lefranc [$38$] and Ghrist [$79$] for many detailed
examples.\bigskip

\textbf{Definition 7.1.} \textit{A template }$\mathcal{T}$\textit{\ \ is a
compact branched two-manifold with boundary built from finite number of
branch line charts.} \bigskip

The key justification for introduction of templates can be explained as
follows. Upon embedding of $\mathcal{T}$ \ into $S^{3}$ the periodic orbits
of (e.g. Lorenz) semiflow tend

to form knots/links. In BWa) we find a\bigskip

\textbf{Conjecture 7.2.} \textit{There does not exist an embedded }$\mathcal{%
T}$\textit{\ supporting all (tame) knots/links as periodic orbits, that is
the are no universal template.\bigskip }

\textbf{Definition 7.3.} \textit{An universal template is an embedded
template }$\mathcal{T}\subset S^{3}$\textit{\ among whose close orbits can
be found knots of every type}.\bigskip

In 1995 guided by Birman and Williams (BWb) [$17$] Ghrist designed an
universal template thus disproving the Birman-Williams conjecture. The full
proof was subsequently published in [$18$]. Etnyre and Ghrist [$2$] took the
full advantage of this fact and came up with the existence-type proof of the
Moffatt conjecture \ formulated by Moffatt in 1985 [$3$] . Recall, that this
conjecture is claiming that in steady Euler flows there could be knots of
any types. From results presented thus far in parts I and II it should be
clear that the same conjecture is applicable to optical knots. More
recently, Enciso and Peralta-Salas [$23$] produced also the existence-type
proof of the Moffatt conjecture for Eulerian steady flows. These
existence-type results are incompatible with the Beltrami$\rightarrow $Reeb$%
\rightarrow $Hamiltonian flows correspondence \footnote{%
Incidentally, this correspondence was \ also established by Etnyre and
Ghrist [$27$] by different methods} and, accordingly, with the Theorem 5.3.
and its numerous corollaries.

\textbf{Question}: Is there way out of the existing controversy?

\textbf{Answer}: Yes, there is. It is based on the discussion of specific
experimentally realizable set ups for generating knots/links of different
types.

The first example of such a set up was discussed in great detail already in
the B-Wb) paper in which the following experiment was considered. Suppose we
are given a knotted piece

of wire with steady current flowing through the wire. The task lies in
describing the magnetic field in the complement of such knot.\bigskip
\medskip

7.3. \ Physics and mathematics of experimental design\medskip \bigskip

In BWb) paper the authors discussed knotted magnetic field configurations
surrounding a piece of wire coiled in the shape of figure 8 knot. From knot
theory it is known that the Seifert surface of figure 8 knot \ is punctured
torus. Remarkably, but the punctured torus is also Seifert surface for the
trefoil knot. The trefoil knot is a torus knot \ and, therefore, it can be
generated mechanically as was discussed in previous sections. Thus the
discussion of similarities and differences between these knots will help us
to resolve the controversy described in the previous subsection. In fact,
discussions related to the trefoil knot will also be helpful for alternative
description of the Lorenz knots which happens to be very illuminating.

The experimental set up not necessarily should involve compliance with
eq.s(2.1a,b). For instance, if we use wires which are ordinary conductors,
then we immediately loose both eq.s(2.1a,b) and all machinery we discussed
thus far is going to be lost. This fact is also consistent with Theorem 5.3.
The correspondence between physics of incompressible ideal Euler fluids and
physics of superconductors was noticed in the paper by Fr\"{o}hlich [$80$]
but, apparently, was left non appreciated till publication of our book \ [$6$%
] . The first 3 chapters of the book provide needed background for
recognition of the fact that the correct set up realizing the requirements
given by eq.s (2.1a,b) is possible only if we are using superconducting
wires. In part I and at the beginning of part II we discussed the Hopf links
\ made of two linked unknotted rings. If one of the rings is made out of
supercoducting wire, then the magnetic field in this wire will be collinear
with the direction of the superconducting current. The superconducting
current, in turn, will create yet another magnetic field and, since the
magnetic field does not have sources or sinks, we end up with the Hopf link.
In the present case, we have instead of superconducting circle (that is of
an unknot) a superconducting coil $K$ in the form of the trefoil or figure 8
knot. The magnetic field collinear with the current in the circuit obeys
eq.s (2.1a,b), as required, \ while the surrounding magnetic field will live
in $S^{3}\smallsetminus K$ space and should not be consistent with eq.s
(2.1a,b). Thus, we are left with the problem of describing all
knotted/linked structures in the space $S^{3}\smallsetminus K$ . We begin
this description in the next subsection.\bigskip

7.4. \ Some facts about fibered knots\medskip \bigskip

Both the trefoil and figure eight knots are the simplest examples of fibered
knots. We discussed both of them in detail in[81] in the context of dynamics
of 2+1 gravity. This circumstance greatly simplifies our tasks now. From
previous sections we know that the trefoil knot can be generated \ with help
of \ methods of Hamiltonian and contact dynamics while the figure eight knot
cannot. Nevertheless, both of them are having \textsl{the same} Seifert
surface $S-$the punctured/holed torus. An orientation-preserving surface
homeomorphisms $h:$ $S\rightarrow S$ of the holed torus should respect the
presence of a hole. This is so because the circumference of this hole is our
knot (trefoil or figure eight).

The \textit{mapping torus} fiber bundle $T_{h}$ can be constructed as
follows. Begin \ with the products $S\times 0$ (the initial state) and $%
S_{h}\times 1$ (the final state) so that for each point $x\in S$ we have $%
(x,0)$ and $(h(x),1)$ respectively. \ After this, the interval $I=(0,1)$ can
be made closed (to form a circle $S^{1}).$This is achieved by identifying $0$
with $1$ causing the identification $h:$ $(x,0)=(h(x),1).$ The fiber bundle $%
T_{h}$ \ is constructed now as a quotient 
\begin{equation}
T_{h}=\frac{S\times I}{h}.  \tag{7.2}
\end{equation}%
It is a 3-manifold which fibers over the circle $S^{1}$. Such constructed
3-manifold in $S^{3}$ is complimentary to the (fibered) knot, in our case
the figure eight or trefoil knot.

It happens that every closed oriented 3-manifold $M$ admits an \textit{open
book decomposition}. This means the following. An oriented link $L\subset M,$
called \textit{binding }can be associated with locally trivial bundle 
\textit{p}: $M\smallsetminus L\rightarrow S^{1}$ whose fibers are open
(Seifert) surfaces $F_{S},$called \textit{pages}. Technically speaking, $L$
is also required to have \ a tubular neighborhood $L\times D^{2}$ so that
the restricted map \textit{p}: $L\times (D^{2}\smallsetminus
\{0\})\rightarrow S^{1}$ is of the form $(x,y)\rightarrow y/\left\vert
y\right\vert .$ The closure of each page $F_{S}$ is then \ a connected
compact orientable surface with boundary $L$. Any oriented link $L\subset M$
that serves as a binding of an open book decomposition of $M$ is called 
\textit{fibered link/knot.}

There is a deep relationship between the Alexander polynomial $\Delta
_{K}(t) $ for the fibered knot $K$ and the associated with it 3-manifold $%
\mathit{M}$. If $V$ be the Seifert matrix of linking coefficients for $K$,
then $\Delta _{K}(t)=\det (V^{T}-tV).$ If the knot $K$ is fibered, the
polynomial $\Delta _{K}(t)$ possess additional symmetries. Specifically, in
such a case we obtain 
\begin{equation}
\Delta _{K}(t)=\dsum\limits_{i=0}^{2g}a_{i}t^{i}  \tag{7.3}
\end{equation}%
and it is known that $\Delta _{K}(0)=a_{0}=a_{2g}=\pm 1.$ Here $g$ is the
genus of the associated Seifert surface. Thus, fibered knots can be
recognized by analyzing their Alexander polynomial. It is monic for fibered
knots. Furthermore, using just described properties of $\Delta _{K}(t),$ we
obtain as well $\Delta _{K}(0)=\det (V^{T})=\det (V)=\pm 1.$ Because of
this, we obtain:%
\begin{equation}
\Delta _{K}(t)=\det (V^{-1}V^{T}-tE)\equiv \det (M-tE).  \tag{7.4}
\end{equation}%
Here $E$ is the unit matrix while $M=V^{-1}V^{T}$ is the monodromy matrix
responsible for the surface homeomorphisms $h:$ $S\rightarrow S.$ In the
case of figure 8 knot BWb) use 
\begin{equation}
M_{a}=\left( 
\begin{array}{cc}
2 & 1 \\ 
1 & 1%
\end{array}%
\right)  \tag{7.5a}
\end{equation}%
while below, in eq.(7.20), it is shown that there is yet another matrix

\bigskip 
\begin{equation}
M_{b}=\left( 
\begin{array}{cc}
2 & -1 \\ 
-1 & 1%
\end{array}%
\right)  \tag{7.5b}
\end{equation}%
participating in these homeomorphisms. Both matrices produce $\Delta
_{K_{8}}(t)=t^{2}-3t+1.$ The matrix $M$ \ for the trefoil knot T\ is
different. It is either 
\begin{equation}
M_{a}=\left( 
\begin{array}{cc}
1 & 1 \\ 
-1 & 0%
\end{array}%
\right) \text{ or }M_{b}=\left( 
\begin{array}{cc}
1 & -1 \\ 
1 & 0%
\end{array}%
\right)  \tag{7.6a,b}
\end{equation}%
producing $\Delta _{K_{\text{T}}}(t)=t^{2}-t+1.$ The 3-manifolds $T_{h}$
associated \ with the figure 8 and trefoil knots are different because of
the difference in the respective monodromy matrices. \ The 3-manifold for
the figure 8 knot is hyperbolic while for the trefoil it is Seifert-fibered.

Since locally, the complement of the figure 8 knot is behaving as the space
of constant negative curvature the figure 8 knot is hyperbiolic knot. In
hyperbolic space the nearby dynamical trajectories diverge. Nevertheless,
Ghrist [18] \ using BWb) was able to prove the following\bigskip

\textbf{Theorem 7.4. (}Ghrist\textbf{) }\textit{Any fibration of the
complement of the figure -eight knot in }$S^{3}$ over $S^{1}$ \textit{%
induces a flow on }$S^{3}$ \textit{containing every tame knot and link as
closed orbits}.\bigskip

The proof of this result involves explicit design of the universal template
which was used subsequently by Etnyre and Ghrist [$2$] in proving the
Moffatt conjecture. These knots and links are not generated as Hamiltonian
flows though. Nevertheless, in view of the above theorem the complement in $%
S^{3}$ of \ the superconducting current carrying wire coiled in the shape of
figure 8 knot \ is expected to contain magnetic lines \ of any (tame) knot
and link.

In developing their proof Etnyre and Ghrist [2] (2000a) needed to combine
the results of [$18$] with those coming from contact geometry. The specific
contactomorphisms described in \ section 4 are sufficient only for
generation of torus and the iterated torus knots. The experimental set up
discussed in previous subsection provides an opportunity to enlarge the
collection of knots. In the following subsections we provide some details \
of how this can be achieved.\bigskip

7.5. \ Lorenz knot/links \ living in the complement of the trefoil knot
\bigskip \medskip

We begin by noticing that the trefoil knot can be generated dynamically, on
one hand, and can be created by a superconducting piece of wire coiled in
the form of the trefoil knot, on another hand. In both cases \ the
complements of the trefoil in $S^{3}$ are Seifert-fibered 3-manifolds [$71$%
]. The well known theorem by Gordon and Luecke [$72$] is telling us that
knots are determined by their complements. This restriction is not
extendable to links though as discussed in Appendix F. Following Ghys [$27$%
], and Ghys and Leys [$82$], we shall focus now at the complement of the
trefoil knot.

We begin by discussing \ in some detail the fiber bundle construction for
the trefoil knot whose Seifert surface $S$ is the punctured torus. The
homeomorphisms of $S$ are generated by the sequence of Dehn twists. The
torus \ can be visualized in a variety of ways. For example, as some cell of
the square-type lattice whose opposite sides are being identified. The
punctured torus would require us in addition to get rid of the vertices and
of neighborhoods of these vertices. \ The sequence of Dehn twists acting on
the punctured torus relates different square-type lattices to each other.
Should the puncture be absent, the elementary cells of these different
lattices upon identification of opposite sides would be converted to
different tori. Closed geodesics on these tori would correspond to different
torus knots. Presence of the puncture complicates matters considerably, e.g.
read [$83$].

Let $G=\pi _{1}(S)$ be the fundamental group of a surface and $P\subseteq G$
be the set of \textit{peripheral} \textit{elements}, that is those elements
of $G$ \ which correspond to loops freely homotopic to the boundary
components. For the punctured torus, $G$ is just a free group of two
generators $a$ and $b.$ That is $G=<a,b\mid >.$ The subset $P$ is determined
by the commutator $aba^{-1}b^{-1}\equiv \lbrack a,b].$ If generators $a$ and 
$b$ are represented by the matrices, then it can be shown that the presence
of the puncture is reflected in the fact that $tr[a,b]=-2.$ The question
arises: is it possible \ to find $a$ and $b$ explicitly based on this
information? \ Notice, because of the puncture Euler characteristic $\chi $
of the torus becomes -1. This means that the covering space is \textbf{H}$%
^{2}$ that is it is the Poincar$e^{\prime }$ upper half plane or, which is
equivalent, the Poincare disc $\mathcal{D}$. The group of isometries of $%
\mathcal{D}$ is $PSL(2,\mathbf{R}).$ Its closest relative is $SL(2,\mathbf{R}%
)$. It is possible to do all calculations using $SL(2,\mathbf{R})$ and only
in the end to use the projective representation. For matrices of $SL(2,%
\mathbf{R})$ the following identities are known%
\begin{eqnarray}
2+tr[a,b] &=&\left( tra\right) ^{2}+\left( trb\right) ^{2}+\left(
trab\right) ^{2}-\left[ tra\right] \text{ }\cdot \left[ trb\right] \cdot 
\text{ }\left[ trc\right] ,  \notag \\
\left[ tra\right] \text{ }\cdot \left[ trb\right] &=&trab+trab^{-1}. 
\TCItag{7.7}
\end{eqnarray}

Since we know already that $tr[a,b]=-2$ the above identities can be
rewritten as 
\begin{equation}
x^{2}+y^{2}+z^{2}=xyz;\text{ }xy=z+w.  \tag{7.8}
\end{equation}%
Clearly, $x=tra,y=trb,z=trab$ and $w=trab^{-1}.$ It happens that already the
first identity is sufficient for restoration of matrices $a$ and $b$. These
are given by 
\begin{equation}
a=\frac{1}{z}\left( 
\begin{array}{cc}
xz-y & x \\ 
x & y%
\end{array}%
\right) ,\text{ }b=\frac{1}{z}\left( 
\begin{array}{cc}
yz-x & -y \\ 
-y & x%
\end{array}%
\right) .  \tag{7.9}
\end{equation}%
For the integer values of x, y and z the first of identities in eq.(7.8) is
known as equation for the Markov triples. It was discovered in the number
theory by Markov\footnote{%
E.g. read all needed references in [$81$].}. By introducing the following
redefinitions 
\begin{equation}
x=3m_{1},y=3m_{2},z=3m_{3}\text{ }  \tag{7.10}
\end{equation}%
the first identity in eq.(7.8) acquires standard form used in number theory%
\begin{equation}
m_{1}^{2}+m_{2}^{2}+m_{3}^{2}=3m_{1}m_{2}m_{3}.  \tag{7.11}
\end{equation}%
The simplest solution of this equation is $m_{1}^{2}=m_{2}^{2}=m_{3}^{2}=1.$
To generate additional solutions it is convenient to introduce the notion of
trace maps. In the present case \ we have to investigate the map $\mathcal{F}
$ defined as 
\begin{equation}
\mathcal{F}:\left( 
\begin{array}{c}
x \\ 
y \\ 
z%
\end{array}%
\right) \rightarrow \left( 
\begin{array}{c}
3yz-x \\ 
y \\ 
z%
\end{array}%
\right)  \tag{7.12}
\end{equation}%
which possess an "integral of motion" $I(x,y,z)=x^{2}+y^{2}+z^{2}-3xyz$. By
definition, it remains unchanged \ under the action of $\mathcal{F}$. From
the theory of Teichm\"{u}ller spaces [$84$] it is known that the length $%
l(\gamma )$ of closed geodesics associated with $\gamma \in G$ is given by 
\begin{equation}
tr^{2}\gamma =4\cosh ^{2}(\frac{1}{2}l(\gamma )).  \tag{7.13}
\end{equation}%
In the present case, we have 
\begin{equation}
x^{2}=4\cosh ^{2}(\frac{1}{2}l(\gamma ))  \tag{7.14}
\end{equation}%
with analogous results for $y$ and $z$. In view of eq.(7.10) for the Markov
triple (1,1,1) we obtain: 
\begin{equation}
l(\gamma )=2\cosh ^{-1}(\frac{3}{2})=2\ln (\frac{1}{2}(3+\sqrt{5}))\equiv
2\ln \lambda .  \tag{7.15}
\end{equation}%
Here $\lambda $ is the eigenvalue of the monodromy matrix to be determined
momentarily. For this purpose we notice that the larger root of the equation%
\begin{equation}
\Delta _{K_{8}}(t)=t^{2}-3t+1=0  \tag{7.16}
\end{equation}%
for the Alexander polynomial of figure 8 knot is equal to $\lambda .$ Taking
into account that eq.(7.4) can be equivalently presented as 
\begin{equation}
\Delta _{K}(t)=t^{2}-\left( trM\right) t+\det M  \tag{7.17a}
\end{equation}%
and, in view of the fact that $\Delta _{K}(0)=a_{0}=a_{2g}=\pm 1,$ eq.(7.17
a) can be rewritten as

\begin{equation}
\Delta _{K}(t)=t^{2}-\left( trM\right) t\pm 1.  \tag{7.17b}
\end{equation}%
From the knot theory [$55$] it is known that for any knot $\Delta
_{K}(1)=\pm 1.$ Therefore, we are left with the condition 
\begin{equation}
\Delta _{K}(t)=1-\left( trM\right) t\pm 1=\pm 1.  \tag{7.18}
\end{equation}%
This equation leaves us with two options: $trM=3$ or $1$. In view of
eq.(7.17), the first option leads us to the Alexander polynomial for the
figure eight knot while the 2nd- to the Alexander polynomial for the trefoil
knot. From here we recover the monodromy matrices $M$ for the figure 8 and
trefoil knots. These are given by 
\begin{equation}
M_{8}=\left( 
\begin{array}{cc}
2 & 1 \\ 
1 & 1%
\end{array}%
\right) \text{ and }M_{\text{T}}=\left( 
\begin{array}{cc}
1 & 1 \\ 
-1 & 0%
\end{array}%
\right) \text{ or }\left( 
\begin{array}{cc}
1 & -1 \\ 
1 & 0%
\end{array}%
\right) .  \tag{7.19}
\end{equation}%
Obtained results are in accord with BWa) and BWb) where they were given
without derivation. In addition, notice that the obtained result for $M_{8}$
linked with the motion along the geodesics is not applicable for $M_{\text{T}%
}.$ From here several conclusions can be drawn:

a) \ Closed geodesics in hyperbolic space are related to hyperbolic knots;

b) \ Not all (even hyperbolic) knots can be associated with closed geodesics
[$81,85$]. In this work we are not going to discuss these exceptional cases.

Notice that the same results can be obtained differently. There is a good
reason for doing so as we would like to demonstrate \ now. For instance, use
of the solution $x=y=z=3$ in eq.s (7.9) results in the matrices%
\begin{equation}
a=\left( 
\begin{array}{cc}
2 & 1 \\ 
1 & 1%
\end{array}%
\right) \text{ and }b=\left( 
\begin{array}{cc}
2 & -1 \\ 
-1 & 1%
\end{array}%
\right)  \tag{7.20}
\end{equation}%
yielding the trace of commutator $[a,b]$ being equal to $-2$ as required. To
understand the physical meaning of these results, we introduce two new
matrices 
\begin{equation}
L=\left( 
\begin{array}{cc}
1 & 0 \\ 
1 & 1%
\end{array}%
\right) \text{ and }R=\left( 
\begin{array}{cc}
1 & 1 \\ 
0 & 1%
\end{array}%
\right) .  \tag{7.21}
\end{equation}%
These are easily recognizable as \ Logitudinal and Meridional elementary
Dehn twists [$55$], page 24. Remarkably, now we are having the following
chain of identities%
\begin{equation}
a=RL=M_{8}\text{ .}  \tag{7.22}
\end{equation}%
Also, 
\begin{equation}
b=L^{-1}aR^{-1},  \tag{7.23}
\end{equation}%
where 
\begin{equation*}
L^{-1}=\left( 
\begin{array}{cc}
1 & 0 \\ 
-1 & 1%
\end{array}%
\right) \text{ and \ }R^{-1}=\left( 
\begin{array}{cc}
1 & -1 \\ 
0 & 1%
\end{array}%
\right) .
\end{equation*}%
Because of this, it is convenient to introduce yet another matrix 
\begin{equation}
\hat{I}=\pm \left( 
\begin{array}{cc}
0 & -1 \\ 
1 & 0%
\end{array}%
\right) ,\hat{I}^{2}=\mp \left( 
\begin{array}{cc}
1 & 0 \\ 
0 & 1%
\end{array}%
\right) =\mp I  \tag{7.24}
\end{equation}%
giving us a chance to represent $L^{-1}$ and $R^{-1}$ as%
\begin{equation}
L^{-1}=\hat{I}R\hat{I}\text{ and }R^{-1}=\hat{I}L\hat{I}  \tag{7.25a}
\end{equation}%
and, in addition, to obtain%
\begin{equation}
L\hat{I}L=R,\text{ }R\hat{I}R=L,\text{ }R\hat{I}L=\hat{I}\text{ and }L\hat{I}%
R=\hat{I}.  \tag{7.25b}
\end{equation}%
Consider now the "word" of the type 
\begin{equation}
W_{1}=L^{\alpha _{1}}R^{\beta _{1}}\cdot \cdot \cdot L^{\alpha _{r}}R^{\beta
_{r}}  \tag{7.26}
\end{equation}%
in which $\alpha _{i}$ and $\beta _{i}$ are some integers. In addition, we
could consider words with insertions of $\hat{I}$. The analysis done in [$81$%
] spares us from the necessity of doing so thanks to eq.s(7.25). In the end,
only words of the type $W_{1}$ and $\hat{I}W_{1}$ should be
considered.\bigskip

\textbf{Remark 7.5.} The obtained results are useful to compare against
those in the paper by Ghys [$25$]. In it, Ghys takes $U=-\hat{I}$ and $V=\pm
R^{-1}$ as two letters used in \ creation of the word $W$ of the type%
\begin{equation}
W=UV^{\varepsilon _{1}}UV^{\varepsilon _{2}}\cdot \cdot \cdot
UV^{\varepsilon _{n}}  \tag{7.27}
\end{equation}%
with each $\varepsilon _{i}=\pm 1.$ It is clear, however, that our $W_{1}$
is equivalent to $W$ (e.g. see Birman [$86$], page 35). \ To relate $W$ (or $%
W_{1}$) to Lorenz knots/links is a simple matter at this point. This is \
done by realizing that encoding of\ dynamical semiflows on Lorenz template $%
\mathcal{T}$ is isomorphic with the sequence of letters LR in eq.(7.26).
Details are given in Ghys [$25$] and Birman[$86$].

Instead of copying results of these authors, it is of interest to arrive at
these results differently. This \ will help us \ to shorten our description
of semiflows \ connected with the figure 8 knot by adding few details to the
paper by Miller [$85$] in which results of BWb) reobtained and simplified

\ To describe the detour, we do not need to do more than we did already. We
only need to present results in the appropriate order. The key observations
are.

1. Words of the type W$_{1}$ are represented as matrices%
\begin{equation*}
M=\left( 
\begin{array}{cc}
\alpha & \beta \\ 
\gamma & \delta%
\end{array}%
\right) ,\text{ \ }\alpha \delta -\gamma \beta =1.
\end{equation*}

2. Matrix entries are determined by the Markov triples.

3. Traces of these matrices are related to the lengths $l(\gamma )$ of
closed geodesics via eq.(7.14).

4. At the same time, the remarkable relation $l(\gamma )=2\ln \lambda $
connects these lengths with

\ \ \ \ the largest eigenvalue of the monodromy matrix $M$ \ entering the
definition of

\ \ \ \ the Alexander polynomial which itself is a topological invariant.

4. The representation of the fundamental group of the complement of the
trefoil knot

\ \ \ \ in $S^{3}$ is \ given either as 
\begin{equation}
<x,y\mid x^{3}=y^{2}>  \tag{7.28a}
\end{equation}

\ \ \ \ or as 
\begin{equation}
<a,b\mid aba=bab>=<a,b,c\mid ca=bc,ab=c>.  \tag{7.28b}
\end{equation}%
In view of eq.(6.3) the 2nd form of representation is easily recognizable as
B$_{3.}.$ The centralizer $Z$ of B$_{3}$ is the combination ($aba$)$^{4}$.
However, already the combination $\Delta _{3}=$($aba$)$^{2}$ is of
importance since, on one hand, it can be proven [$87$] that $\sigma
_{i}\Delta _{3}=\Delta _{3}\sigma _{3-i}$ while, on another, this operation
physically means the following. While \ keeping the top of the braid fixed,
the bottom is turned by an angle $\pi .$ Accordingly, for Z we have to
perform a twist by $2\pi .$ If we consider a quotient B$_{3}/\mathbf{Z}_{3}$
it will bring us back to the braids representation associated with
transverse knots/links discussed in Section 6. In such a case Kassel and
Turaev [$87$] demonstrated that the presentation for the quotient can be
written as 
\begin{equation}
<a,b\mid a^{3}=b^{2}=1>  \tag{7.29}
\end{equation}%
This result can be realized in terms of matrices we had obtained already.
Specifically, we can identify $b$ with $\hat{I}$ $\ $\ and $a$ with \ $\hat{I%
}R$. By direct computation we obtain: $\hat{I}$ $^{2}=\left( \hat{I}R\right)
^{3}.$ The presentation, eq.(7.29), \ is graphically depicted in Fig. 10a).\ 

\begin{figure}[tbp]
\begin{center}
\includegraphics[scale=1.2]{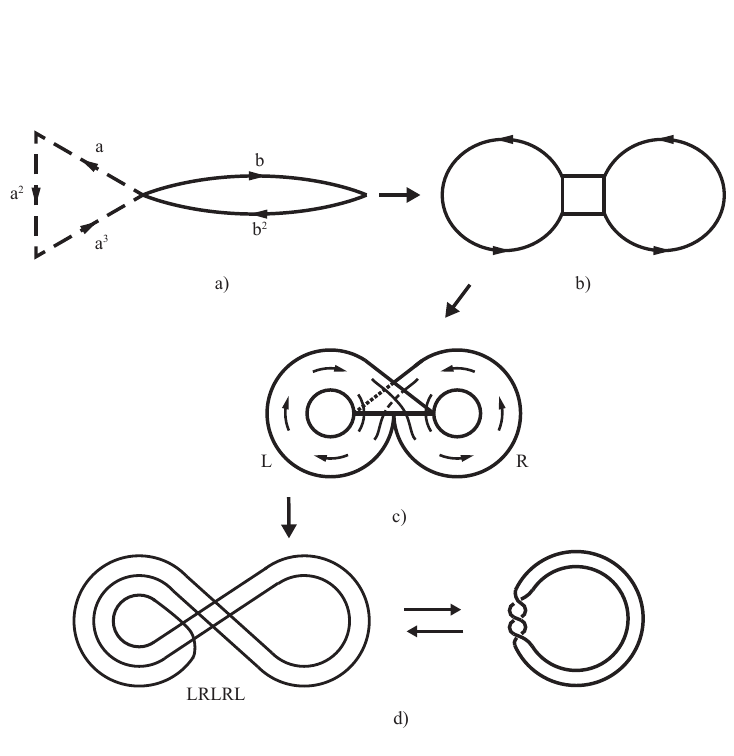}
\end{center}
\caption{From random walks on the graph for the figure eight knot to
dynamics on the Lorenz template $\mathcal{T}$ }
\end{figure}

Following our previous work [$88$] we also can consider some kind of a
random walk on the figure eight (e.g. see Fig.10b). This process \ also can
be looked upon as a diagram \ of states for finite state automaton. From
here is the connection with the symbolic dynamics (e.g. subshifts) [$37$].
Alternatively, the same process can be considered as depicting a motion on
the (Lorenz) template (Fig.10c)). The thick dark horizontal line on the
Lorenz template is made out of two parts: left (L) and right (R). The
representative closed trajectory passing through such a template is encoded
by the sequence of L and R's (Fig.10d) left). It is in one-to-one
correspondence with the (Lorenz) knot (Fig.10d) right). The word W in
eq,(7.27) keeps track record of wandering of the walker (of the mover) on
the figure eight or, which is equivalent, on the Lorenz template. \bigskip

7.6. Geodesic knots in the figure eight knot complement\medskip \bigskip

By analogy with eq.s(7.28) we begin with presentation of the fundamental
group for the figure eight knot complement. The standard form can be found
on page 58 of Rolfsen [$55$] 
\begin{equation}
\pi (S^{3}-K_{8})=<a,b.c,d\mid r_{1},r_{2},r_{3},r_{4}>  \tag{7.30}
\end{equation}%
Here $r_{1}=bcb^{-1}=a,r_{2}=ada^{-1}=b,r_{3}=d^{-1}bd=c,r_{4}=c^{-1}ac=d$ .
This presentation is not convenient however for the tasks we are having in
mind. To avoid ambiguities caused by some mistakes we found in literature,
we derive needed presentations in Appendix G. Our first task is to find a
presentation which looks analogous to that given by eq.(7.28b). This is
given by eq.(G.7) which we rewrite here as 
\begin{equation}
<a,b,w\mid wa=bw,w=b^{-1}aba^{-1}>.  \tag{7.31}
\end{equation}%
By achieving the desired correspondence with eq.(7.28b) we are interested in
obtaining \ the analog of eq.(7.29) as well. This is achieved in several
steps. They would be unnecessary, should all presentations done in Miller [$%
85$] be correct. Unfortunately, they are not. For this reason her results
are reconsidered in Appendix D. The analog of presentation eq.(7.29) now is
given by generators and relations of the dihedral group D$_{2}$%
\begin{equation}
\text{D}_{2}=<t,s\mid t^{2}=s^{2}=1,ts=st>.  \tag{7.32}
\end{equation}%
As in the case of a trefoil, where the Lorenz template was recovered, the
above presentation allows us to reobtain the universal template $\mathcal{T}$
\ of Ghrist. The successive steps are depicted in Fig.11. Their description
is analogous to those depicted in Fig.10.

\begin{figure}[tbp]
\begin{center}
\includegraphics[scale=1.5]{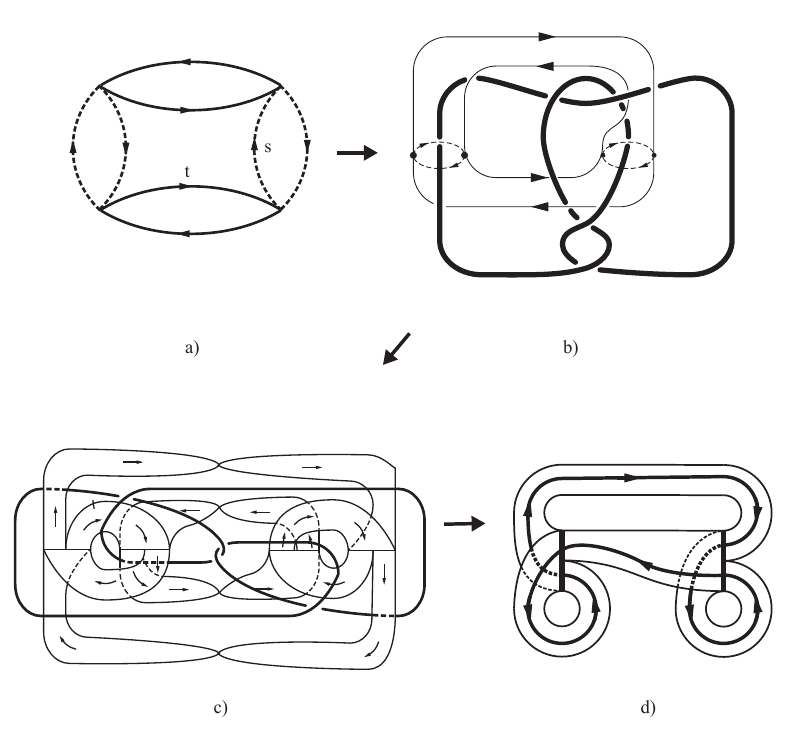}
\end{center}
\caption{From random walks on the graph for the figure eight knot to
dynamics on the universal template $\mathcal{T}$ of Ghrist }
\end{figure}
To deal with the fundamental group of the figure eight knot, Bowditch [$89$]
extended theory of Markov triples to hyperbolic space \textbf{H}$^{3}.$ To
do so requires equations for Markov triples, eq.s(7.8), to be solved in the
complex domain. In \textbf{H}$^{3}$ the isometry group is PSL(2,\textbf{C})
as compared with the isometry group of \textbf{H}$^{2}$ which is PSL(2,%
\textbf{R}). As before, we can do all calculations using SL(2,\textbf{C})
and only at the end of calculations switch to PSL(2,\textbf{C}). Because of
this, \ it is useful to know [$90$] that SL(2,\textbf{C}) can be generated
by just two elements%
\begin{equation}
SL(2,\mathbf{C})=\{V=\left( 
\begin{array}{cc}
1 & \alpha \\ 
0 & 1%
\end{array}%
\right) ,\hat{I}=\pm \left( 
\begin{array}{cc}
0 & -1 \\ 
1 & 0%
\end{array}%
\right) ,\alpha \in \mathbf{C}\}.  \tag{7.33}
\end{equation}%
It is also helpful to compare these elements with earlier derived matrices $%
\hat{I}$ and $R$. Clearly, $V\rightarrow R$ when $\alpha \rightarrow 1.$In
the case of figure eight knot it can be checked directly [$81$] that
matrices $a$ and $b$ in eq.(7.31) can be presented as follows: 
\begin{equation}
a=\left( 
\begin{array}{cc}
1 & 1 \\ 
0 & 1%
\end{array}%
\right) \text{ and }b=\left( 
\begin{array}{cc}
1 & 0 \\ 
-\omega & 1%
\end{array}%
\right) ,  \tag{7.34}
\end{equation}%
where $\omega $ is one of the solutions of the equation $\omega ^{2}+\omega
+1=0.$ Typically, one chooses $\omega =\frac{1}{2}(-1+3i).$ This result has
number-theoretic significance explained in the Appendix $B$ [$81$]. More
deeply, theory of arithmetic hyperbolic 3-manifolds is explained in the book
by [$91$]. Some physical (cosmological) implications of this arithmeticity
are discussed in our work [$81$]. Interested readers are encouraged to read
these references. By noticing that any matrix 
\begin{equation}
SL(2,\mathbf{C})=\left( 
\begin{array}{cc}
\alpha & \beta \\ 
\gamma & \delta%
\end{array}%
\right) ,\text{ \ }\alpha \delta -\gamma \beta =1\text{ }  \tag{7.35}
\end{equation}%
can be represented in terms of $V$ and $\hat{I}$ \ as demonstrated in [$90$%
], it is clear that \ now it is possible to represent all words in the form
analogous to that given in eq.(7.26) or (7.27) (if the dihedral
presentation, eq.(7.32), is used). Furthermore, the fundamental relation,
eq.(7.13), for closed geodesics can be used for \textbf{H}$^{3}$ also so
that now we are dealing with the \textsl{geodesic hyperbolic knots}. The
difference between the complements of the trefoil and that for figure eight
knots lies in the fact that now the length $l$($\gamma )$ of geodesics
becomes a complex number. It can be demonstrated [$89$], page 717, that Re$l$%
($\gamma )>0.\bigskip $

\textbf{Remark 7.6.} Although the dihedral presentation is given in [$85$]
incorrectly, it is possible to prove [$90$] that the finite subgroups of
PSL(2,\textbf{C}) contain all group of orientation-preserving isometries of
Euclidean regular solids and their subgroups. Evidently, the presence of
actual symmetries depends upon the values of complex parameters $\alpha
,\beta ,\gamma ,\delta $. The choice of D$_{2}$ leads to the universal
template of Ghrist.\bigskip

\textbf{Remark 7.7.} Instead of the real hyperbolic space \textbf{H}$^{3}$
in which the complement of the figure eight knot lives, it is possible to
use the complex hyperbolic space \textbf{H}$_{\mathbf{C}}^{3}$ advocated by
Goldman [$92$]. It is possible to consider a complement of the figure eight
in \textbf{H}$_{\mathbf{C}}^{3}$ as well [$93$]. It happens that isometry
group of the boundary of \textbf{H}$_{\mathbf{C}}^{3}$ is the Heisenberg
group, e.g. see eq.s(6.14). Thus, in dealing with problems related to the
complement of figure eight knot the contact geometry and topology again can
be successfully used in accord with result obtained by Etnyre[$75$].\bigskip

7.7. Cosmetic knots and gravity\bigskip

Using results of part I, the famous Gordon and Luecke theorem [$72$] stating
that in $S^{3}$ knots are determined by their complements can be restated
now in physical terms. For this we need to recall some facts from general
relativity. The original formulation by Einstein makes heavy emphasis on
study of the Schwarzschild solution of Einstein's equations originating from
the Einstein field equations without sources with prescribed (spherical)
symmetry in which the test mass enters as an adjustable parameter [$94$].
Using these equations the curvature tensor is obtained. If masses are
associated with knots, then knot complements producing curved spaces around
knots can be used for parametrizing the masses. In which case \textsl{the
Gordon and Luecke theorem is telling us that the mass} \textsl{spectrum is
discrete}. \ In studying knotted geodesics in the complement of the trefoil
and figure eight knot we followed Einsteinian methodology without
recognizing Einsteinian influence. Indeed, we completely ignored the fact
that each nontrivial knot in the complement of the trefoil/figure eight knot
creates its own complement. Thus, the space around, say, the trefoil knot is
not just a complement of the trefoil knot but it is the space made out of
complements of all knots existing in the trefoil knot complement! This fact
is ignored \ in calculations presented above. Therefore, these
"complementary" knots are, in fact, closed geodesics on the punctured torus.
Very much like in Einsteinian theory of gravity the motion of the point-like
objects whose mass is ignored is taking place along the geodesics around the
massive body whose field of gravity is described by the Schwarzschild
solution. To account for extended sizes of particles in both general
relativity and Yang-Mills field theory \ is always a great challenge, e.g.
read page 97 of \ [$6$]. In the case of knotted geodesics the effect of
extended size (that is of finite thickness of a geodesic) is also a
difficult problem which is not solved systematically as can be seen from \ [$%
85$].

To be specific, we would like to talk about the complement of the figure
eight knot $K_{8}$. We begin with $S^{3}\smallsetminus K_{8}.$ This is a
hyperbolic manifold. Next, we drill in it a closed curve $c$ (a tube of
finite thickness or, which is the same, a tubular neighborhood, e.g. see
Fig.3a)) which is freely homotopic to some geodesic $\gamma $. Such closed
curve forms a link with $K_{8}.$ When we remove the content of
knotted/linked tube from $S^{3}\smallsetminus K_{8}$ we obtain a manifold
which is complementary to the link $K_{8}\cup c$ and, if we are interested
in computation of the hyperbolic volume of such hyperbolic link, the result
is going to depend very sensitively on the tube thickness. Thus, indeed, we
end up with the situation encountered in gravity. Furthermore, additional
complications might arise because of the following.

The Gordon and Luecke theorem [$72$] was proven for $S^{3}$ only. In the
present case, say, if we are dealing with $M=S^{3}\smallsetminus K_{8}$ ,
the knots $K_{i}$ $,i=1,2....,$ living \ in $S^{3}\smallsetminus K_{8}$ also
will have their complements, that is $M\diagdown K_{i}$ and, because of
this, we arrive at the following

\textbf{Question}: Suppose in some 3-manifold $M$ (other than $S^{3})$ there
are two knots $K_{1}$ and $K_{2}$ so that the associated complements are $%
M\smallsetminus K_{1}$ and $M\smallsetminus K_{2}$. By analogy with Fig.12
suppose that there is \ a homeomorphism $h$: \ $h$($M\smallsetminus
K_{1})=(M\smallsetminus K_{2}).$ Will such a homeomorphism imply that $%
K_{1}=K_{2}?$

If the answer to this question is negative, this then would imply that
different knots would have the same complement in $M$. If we are interested
in attaching some physics to these statements then, we \ should only look
for situations analogous to those in $S^{3}.$ This means that all physical
processes should be subject to selection rules (just like in quantum
mechanics) making such degenerate cases physically forbidden. \ Evidently,
all these considerations about the selection rules presuppose that the
degeneracy just described can be realized in Nature. And indeed, it can!
This leads us to the concept of \textit{cosmetic knots}.\bigskip

\textbf{Definition 7.8. }Let $M$ be some 3-manifold (other than $S^{3}$ and $%
S^{2}\times S^{1})$ and $K$ be some non-trivial knot "living" in $M.$ If
performing Dehn surgery on $K$ results in the same manifold $M$, then, the
pair $(M,K)$ is called \textit{cosmetic pair}.\bigskip

\textbf{Remark 7.9.} To perform Dehn surgery physically is not possible.
However, one can imagine a situation when \ there are two different knots $%
K_{1}\neq $ $K_{2}$ such that their complements $M-K_{1}$ and $M-K_{2}$ are
homeomorphic while knots themselves are not. If such situation can be
realized then, if we want to keep the knot-mass correspondence alive, one
should impose selection rules which will forbid such \ cosmetic
states.\bigskip

Following Mathieu [$95$], consider now several details. To begin, we \ have
to notice similarities between the operations of cabling as depicted in
Fig.7 and Dehn surgery. \ After reading the description related to Fig.7 in
the main text and the Definition F.1., \ it is appropriate to consider some
knot $K$ in $S^{3}$ so that its complement $S^{3}\smallsetminus K=M$ is some
3-manifold with boundary. Let $k$ and $k^{\prime }$ be the cores of the
respective surgeries for this knot in $M$. \ There should be a homeomorphism
from $S^{3}\smallsetminus K$ to $M\smallsetminus k$ sending the curve $J$
(for $K$) onto meridian $\mathfrak{m}$ of $k.$ At the same time, there
should be a homeomorphism from $S^{3}\smallsetminus K$ to $M\smallsetminus
k^{\prime }$ sending another curve $\tilde{J}$ (for $K$) onto meridian $%
\mathfrak{m}^{\prime }$ of $k^{\prime }$. Should $k$ and $k^{\prime }$ be
equivalent, then by using composition of mappings it would be possible to
find a homeomorphism in $M=S^{3}\smallsetminus K$ -from $J$ to $\tilde{J}$ -%
\textit{without} preserving the meridian of $K.$ Notice that, by design, $J$ 
$\neq $ $\pm \tilde{J}$ making such homeomorphism impossible. Therefore, $k$
and $k^{\prime }$ are not equivalent while the complements $M\smallsetminus
k $ and $M\smallsetminus k^{\prime }$ are since they are
homeomorphic.\bigskip

\textbf{Remark 7.10. }Let $K$ be a trefoil knot. Then in its complement
there will be cosmetic knots. In fact, all Lorenz knots are cosmetic. The
same is true for\ all knots living in the complement of the figure eight
knot. Evidently, such cosmetic knots cannot be associated with physical
masses.

\bigskip

\bigskip \textbf{Appendix A.} \ \textbf{The Beltrami equation and contact
geometry}

Consider 3-dimensional Euclidean space. Let \textbf{A} be some vector in
this space so that $\mathbf{A}=\dsum\nolimits_{i=1}^{3}A_{i}\mathbf{e}_{i}$
. Introduce next the $\flat -$operator via%
\begin{equation}
\left( \mathbf{A}\right) ^{\flat }=\dsum\nolimits_{i=1}^{3}A_{i}dx_{i}. 
\tag{A.1}
\end{equation}%
Use of $\flat $ allows us now to write 
\begin{equation}
d\left( \mathbf{A}\right) ^{\flat }=(\frac{\partial A_{2}}{\partial x_{1}}-%
\frac{\partial A_{1}}{\partial x_{2}})dx_{1}\wedge dx_{2}+(\frac{\partial
A_{1}}{\partial x_{3}}-\frac{\partial A_{3}}{\partial x_{1}})dx_{3}\wedge
dx_{1}+(\frac{\partial A_{3}}{\partial x_{2}}-\frac{\partial A_{2}}{\partial
x_{3}})dx_{2}\wedge dx_{3}  \tag{A.2}
\end{equation}%
Next, we obtain, 
\begin{equation}
\ast d\left( \mathbf{A}\right) ^{\flat }=(\frac{\partial A_{2}}{\partial
x_{1}}-\frac{\partial A_{1}}{\partial x_{2}})dx_{3}+(\frac{\partial A_{1}}{%
\partial x_{3}}-\frac{\partial A_{3}}{\partial x_{1}})dx_{2}+(\frac{\partial
A_{3}}{\partial x_{2}}-\frac{\partial A_{2}}{\partial x_{3}})dx_{1}. 
\tag{A.3}
\end{equation}%
Finally, by using an\ inverse operation to $\flat $ we obtain, 
\begin{equation}
\left[ \ast d\left( \mathbf{A}\right) ^{\flat }\right] ^{\natural }=(\frac{%
\partial A_{2}}{\partial x_{1}}-\frac{\partial A_{1}}{\partial x_{2}})%
\mathbf{e}_{3}+(\frac{\partial A_{1}}{\partial x_{3}}-\frac{\partial A_{3}}{%
\partial x_{1}})\mathbf{e}_{2}+(\frac{\partial A_{3}}{\partial x_{2}}-\frac{%
\partial A_{2}}{\partial x_{3}})\mathbf{e}_{1}=curl\mathbf{A.}  \tag{A.4}
\end{equation}%
Using these results we obtain as well%
\begin{equation}
\ast d\left( \mathbf{A}\right) ^{\flat }=\left( \mathbf{\nabla }\times 
\mathbf{A}\right) ^{\flat }.  \tag{A.5}
\end{equation}%
If $\left( \mathbf{A}\right) ^{\flat }=\alpha ,$ where $\alpha $ is contact
1-form, \ then the Beltrami equation reads as 
\begin{equation}
\ast d\alpha =\kappa \alpha .  \tag{A.6}
\end{equation}

\textbf{Appendix B.} \ \textbf{Derivation of the force-free equation from
the source-free Maxwell's equations\bigskip }

\bigskip In Appendix 4 of part I we mentioned about works by Chu and Ohkawa [%
$96$] and Brownstein [$97$] in which the force-free equation is derived from
the source-free Maxwell's equations. In this appendix we use some results
from the paper by Uehara, Kawai and Shimoda [$98$] in which the same
force-free equation was obtained in the most transparent way. In the system
of units in which $\varepsilon _{0}=\mu _{0}=1$ the source-free Maxwell's
equations acquire the following form 
\begin{equation}
\frac{\partial }{\partial t}\mathbf{E}=\mathbf{\nabla }\times \mathbf{B}%
\text{ \ and \ }-\frac{\partial }{\partial t}\mathbf{B}=\mathbf{\nabla }%
\times \mathbf{E,}  \tag{B.1}
\end{equation}%
provided that div$\mathbf{B}=$div$\mathbf{E}=0.$ Let $\mathbf{E}=\mathbf{v}(%
\mathbf{r},t)\cos f(\mathbf{r},t)$ and $\mathbf{B}=\mathbf{v}(\mathbf{r}%
,t)\sin f(\mathbf{r},t)$. In view of the fact that $\mathbf{\nabla }\times (a%
\mathbf{v})=a\mathbf{\nabla }\times \mathbf{v}-\mathbf{v\times \nabla }a$ \
use of \textbf{E} and \textbf{B }in Maxwell's equations leads to the
following results%
\begin{equation}
-\dot{f}\mathbf{v}\sin f+\mathbf{\dot{v}}\cos f=\sin f(\nabla \times \mathbf{%
v})-\left( \mathbf{v\cdot \nabla }f\right) \cos f  \tag{B.2a}
\end{equation}%
and%
\begin{equation}
\dot{f}\mathbf{v}\cos f+\mathbf{\dot{v}}\sin f=-\cos f(\nabla \times \mathbf{%
v})-\left( \mathbf{v\cdot \nabla }f\right) \sin f  \tag{B.2b}
\end{equation}%
where, as usual, the dot $\cdot $ over $f$ denotes time differentiation.
Analysis shows that eq.(B.2a) and (B.2b) equivalent to the following set of
equations%
\begin{equation}
\dot{f}\mathbf{v=-\nabla \times v}  \tag{B.3a}
\end{equation}%
and 
\begin{equation}
\mathbf{\dot{v}}=-\mathbf{v\times \nabla }f  \tag{B.3b}
\end{equation}%
while the incompressibility equations div$\mathbf{B}=$div$\mathbf{E}=0$ are
converted into the requirements 
\begin{subequations}
\begin{equation}
\text{div}\mathbf{v}=0  \tag{B.4a}
\end{equation}%
and 
\end{subequations}
\begin{equation}
\mathbf{v}\cdot \nabla f=0.  \tag{B.4b}
\end{equation}%
If now we identify $\mathbf{v}$ with the fluid velocity, then both
eq.s(B.4a) and (B.4b) can be converted into one equation div$(f\mathbf{v})=0$
which we had encountered already in section 4 of part I. E.g. read the text
below the eq.(4.3). This observation allows us to reinterpret eq.s(B.3a) and
(B.3b). Specifically, if we let $\dot{f}=-\kappa (\mathbf{r})$ in eq.(B.3a),
then the force-free equation 
\begin{equation}
\mathbf{\nabla \times v=}\kappa \mathbf{v}  \tag{B.5}
\end{equation}%
is obtained. By \ applying the curl operator to both sides of this equation
and taking into account eq.(B.4a) we obtain 
\begin{equation}
-\nabla ^{2}\mathbf{v=}\kappa ^{2}\mathbf{v}+\left( \mathbf{\nabla }\kappa
\right) \times \mathbf{v.}  \tag{B.5}
\end{equation}%
Now eq.(B.3b) can be rewritten as $\mathbf{\dot{v}}=t(\mathbf{v\times \nabla 
}\kappa ).$This result can be formally used in eq.(B.5). The Hemholtz
equation 
\begin{equation}
\nabla ^{2}\mathbf{v+}\kappa ^{2}\mathbf{v=}0  \tag{B.6}
\end{equation}%
is obtained if $\left( \mathbf{\nabla }\kappa \right) \times \mathbf{v=}0$.
Since in addition $\kappa $ must also satisfy eq.(B.4b), we conclude that $%
\kappa =const$ is the only admissible solution. As discussed in [$6$], the
Helmholtz eq.(B.6) is related to the force-free eq.(B.5) \ as the
Klein-Gordon equation is related to the Dirac equation. This means that
every solution of eq.(B.5) is also a solution of eq.(B.6) but not another
way around.

Using just obtained results, the energy density, eq.(2.11) of part I,
manifestly time-independent. In the non-Abelian case of Y-M fields this
would correspond to the case of monopoles in accord with statements made in
part I.

Consider now briefly the case when \ \textbf{v} is time-depemdent. Then
eq.(B.5) can be equivalently rewritten as 
\begin{equation}
\mathbf{\dot{v}=-}t\mathbf{(}\nabla ^{2}\mathbf{v+}\kappa ^{2}\mathbf{v).} 
\tag{B.7}
\end{equation}%
Let now \textbf{v}(\textbf{r},t) =\textbf{V}(\textbf{r})T(t). Using this
result in eq. (B.7) we obtain 
\begin{equation}
\frac{-1}{tT(T)}\dot{T}(t)=\frac{1}{\mathbf{V}}(\nabla ^{2}\mathbf{V+}\kappa
^{2}\mathbf{V).}  \tag{B.8}
\end{equation}%
If $\ \nabla ^{2}\mathbf{V+}\kappa ^{2}\mathbf{V=-}E\mathbf{V},$ then the
Helmholtz eq.(B.6) should be replaced by the same type of equation in which
we have to make a substitution $\kappa ^{2}$ $\rightarrow \kappa ^{2}+E=%
\mathcal{K}^{2}$ and then to treat $\mathcal{K}^{2}$ as an eigenvalue \ of
thus modified eq.(B.6) with the same boundary conditions, e.g. \ see
eq.(2.1b). The situation in the present case is completely analogous that
for \ the Navier-Stokes equation, e.g. read Majda and Bertozzi [$22$],Chr.2.
Therefore,\ the solution $\mathbf{v}(\mathbf{r},t)=\mathbf{V}(\mathbf{r}%
)T(t) $ of the Navier-Stokes equation is represented as the Fourier
expansion over the egenfunctions of the force-free (or Beltrami) eq.(B.5) $%
\mathbf{V}(\mathbf{r})\rightarrow \mathbf{V}_{m}(\mathbf{r})$. Using
eq.(B.8) $T(t)$ is straightforwardly obtainable. By applying the appropriate
reparametrization $T(t)$ can be brought to the form (e.g. see eq.(3.8) of
part I and comments next to it) suggested by Donaldson [$99$]. Thus, even in
the case of time-dependent solutions \ the obtained results can be made
consistent with \ those suggested by Floer and Donaldson which were
discussed in part I. Because of this, it is sufficient to treat only the
time-independent case. In electrodynamics this corresponds to use of the
constant electric and magnetic transversal fields.

\bigskip

\bigskip \textbf{Appendix C \ Some facts from symplectic and contact geometry%
}

\textbf{\bigskip }

Consider a symplectic 2-form $\omega =\dsum\nolimits_{j=1}^{m}dp_{j}\wedge
dq_{j}.$ Introduce a differentiable function $f^{(\mathbf{v})}$ associated
with the vector field $\mathbf{v}$ \ via skew-gradient 
\begin{equation}
\mathbf{v}=(sgrad)f^{(\mathbf{v})}=\dsum\nolimits_{i}[\frac{\partial f^{(%
\mathbf{v})}}{\partial p_{i}}\frac{\partial }{\partial q_{i}}-\frac{\partial
f^{(\mathbf{v})}}{\partial q_{i}}\frac{\partial }{\partial p_{i}}] 
\tag{C.1a}
\end{equation}%
and such that $-df^{(\mathbf{v})}=i_{\mathbf{v}}\omega $. This \ result can
be checked by direct computation. By definition, a vector field\ $\mathbf{v}$
is symplectic if $\mathcal{L}_{\mathbf{v}}\omega =0.$ Since $\mathcal{L}_{%
\mathbf{v}}=d\circ i_{\mathbf{v}}+i_{\mathbf{v}}\circ d$ , we obtain: $%
\mathcal{L}_{\mathbf{v}}\omega =-ddf^{(\mathbf{v})}=0,$ as required.
Introduce now another \ vector field $\mathbf{w}$ via%
\begin{equation}
\mathbf{w}=(sgrad)f^{(\mathbf{w})}=\dsum\nolimits_{i}[\frac{\partial f^{(%
\mathbf{w})}}{\partial p_{i}}\frac{\partial }{\partial q_{i}}-\frac{\partial
f^{(\mathbf{w})}}{\partial q_{i}}\frac{\partial }{\partial p_{i}}] 
\tag{C.1b}
\end{equation}%
and calculate $i_{\mathbf{w}}\circ i_{\mathbf{v}}\omega .$ We obtain:%
\begin{eqnarray}
i_{\mathbf{w}}\circ i_{\mathbf{v}}\omega &=&i_{\mathbf{w}}(-df^{(\mathbf{v}%
)})=i_{\mathbf{w}}\dsum\nolimits_{i}(-\frac{\partial f^{(\mathbf{v})}}{%
\partial p_{i}}dp_{i}-\frac{\partial f^{(\mathbf{v})}}{\partial q_{i}}%
dq_{i})=\dsum\nolimits_{i}[\frac{\partial f^{(\mathbf{v})}}{\partial p_{i}}%
\frac{\partial f^{(\mathbf{w})}}{\partial q_{i}}-\frac{\partial f^{(\mathbf{v%
})}}{\partial q_{i}}\frac{\partial f^{(\mathbf{w})}}{\partial p_{i}}]  \notag
\\
&\equiv &\{f^{(\mathbf{v})},f^{(\mathbf{w})}\}=\omega (\mathbf{v},\mathbf{w}%
)=\mathbf{v}f^{(\mathbf{w})}=-\mathbf{w}f^{(\mathbf{v})}.  \TCItag{C.2}
\end{eqnarray}%
The combination $\{f^{(\mathbf{v})},f^{(\mathbf{w})}\}$ is instantly
recognizable as the Poisson bracket. Let now $f^{(\mathbf{v})}=H,$ where $H$
is the Hamiltonian, then $-dH=i_{\mathbf{v}}\omega $ $=-\dsum\nolimits_{i}(%
\dfrac{\partial H}{\partial q_{i}}dq_{i}+\dfrac{\partial H}{\partial p_{i}}%
dp_{i})$ as required. At the same time, since $\dfrac{\partial H}{\partial
q_{i}}=-\dot{p}_{i}$ and $\dfrac{\partial H}{\partial p_{i}}=\dot{q}_{i}$ we
can rewrite eq.(C.1a) as 
\begin{equation}
\mathbf{v}_{H}=(sgrad)H=\dsum\nolimits_{i}[\dot{q}_{i}\frac{\partial }{%
\partial q_{i}}+\dot{p}_{i}\frac{\partial }{\partial p_{i}}].  \tag{C.1c}
\end{equation}%
By design, we obtain then $dH(\mathbf{v}_{H})=0$. \ \bigskip

\bigskip \textbf{Appendix D \ Some facts about the Morse-Smale flows\bigskip 
}

Let $\mathcal{M}$ be some smooth $n-$dimensional manifold. To describe the
motion on $\mathcal{M}$ \ we associate with each $t\in \mathbf{R}$ \ (or 
\textbf{R}$^{+})$ a mapping $g^{t}:\mathcal{M}\rightarrow \mathcal{M}$ -a
group (semigroup) of diffeomorphisms with the property $%
g^{t+s}=g^{t}g^{s},g^{0}=\mathbf{1}_{M}.$ For some $\mathbf{x}\in \mathcal{M}
$ suppose that some point-like particle was initially located at $\mathbf{x}$%
. Then after time $t$ the \textit{(phase)flow (trajectory)} will carry this
particle to $g^{t}\mathbf{x}$ so that the velocity $\mathbf{v}(\mathbf{x})=%
\frac{d}{dt}\mid _{t=0}g^{t}\mathbf{x}.$ The function $t\longrightarrow g^{t}%
\mathbf{x}$ is called the \textit{motion}. In physics this is written as $%
\mathbf{\dot{x}}=\mathbf{v}(\mathbf{x}).$ This is a system of ordinary
differential equations. It is supplemented by the initial condition $\mathbf{%
x}(0)=\mathbf{x}_{0}.$ The vector $\mathbf{v}(\mathbf{x})$ belongs to the
tangent space of $\mathcal{M}$ at $\mathbf{x}$, that is $\mathbf{v}(\mathbf{x%
})\in T_{x}\mathcal{M}.$ Sometimes it happens that $g^{t}\mathbf{\tilde{x}=%
\tilde{x}}$ $\forall t.$ Such point $\mathbf{\tilde{x}}$ is called the 
\textit{equilibrium point. }Otherwise, there could be some (period) $T$ such
that $g^{t+T}\mathbf{\tilde{x}=\tilde{x}.}$ In this case the point $\mathbf{%
\tilde{x}}$ is called \textit{periodic}. It belongs to the periodic (closed)
trajectory. Suppose now that we are having another manifold $\mathcal{N}$ \
so that there is a map (a homomorphism) $h:\mathcal{M}\rightarrow \mathcal{N}%
.$ Dynamical systems (flows) $g^{t}$on $\mathcal{M}$ and $\check{g}^{t}$on $%
\mathcal{N}$ are \textit{topologically} \textit{conjugate} if $h\circ g^{t}=%
\check{g}^{t}\circ h.$ With the topological conjugacy associated structural
stability. Omitting some fine details which can be found, for example, in
Arnol'd [$100$], we shall not be making distinctions between topological
conjugacy and structural stability.

The analysis of the flow described by $\mathbf{\dot{x}}=\mathbf{v}(\mathbf{x}%
)$ \ begins with linearization. For this purpose we select some point $%
\mathbf{\tilde{x}}$ along the trajectory. Typically such a point is the
equilibrium point. Let $\mathbf{x}=\mathbf{\tilde{x}+x-\tilde{x}\equiv 
\tilde{x}+\delta x.}$ Using this result in $\mathbf{\dot{x}}=\mathbf{v}(%
\mathbf{x})$ we obtain%
\begin{equation}
\mathbf{\delta }\dot{x}_{i}\text{\textbf{=}}\dsum\limits_{j}\frac{\partial
v_{i}}{\partial x_{j}}\mid _{\mathbf{x}=\mathbf{\tilde{x}}}\delta x_{j}. 
\tag{D.1}
\end{equation}%
The equilibrium point $\mathbf{\tilde{x}}$ is called \textit{hyperbolic }if
all the eigenvalues of the matrix $A_{ij}=\frac{\partial v_{i}}{\partial
x_{j}}\mid _{\mathbf{x}=\mathbf{\tilde{x}}}$ are real (that is not complex).
It is called \textit{nondegenerate} if all eigenvalues are nonzero. Since
eq.(D.1) is in $T\mathcal{M}$ space it is clear that replacing \textbf{R}$%
^{n}$ by $\mathcal{M}$ and vice versa introduces nothing new into \emph{%
local }study of the system of equations $\mathbf{\dot{x}}=\mathbf{v}(\mathbf{%
x}).$ The question of central importance is the following. How can solution
of eq.(D.1) help us in solving equation $\mathbf{\dot{x}}=\mathbf{v}(\mathbf{%
x})?$ The answer is given by the following\bigskip\ [$101$]

\textbf{Theorem D.1. }\textit{(Grobman -Hartman) If the point }$\tilde{x}$%
\textit{\ is hyperbolic, then there is a homeomorphism defined on some
neighborhood U of }$\tilde{x}$\textit{\ in R}$^{n}$\textit{\ locally taking
orbits of the nonlinear flow }$g^{t}$\textit{\ to those of linear flow
defined by }$exp(tA_{ij}).$ \bigskip

The above theorem is being strengthened by the following\bigskip

\textbf{Theorem D.2. }\textit{Suppose that equation }$\mathbf{\dot{x}}=%
\mathbf{v}(\mathbf{x})$\textit{\ has a hyperbolic fixed point }$\mathbf{%
\tilde{x}}$\textit{. Then there exist local stable and unstable manifolds }$%
W^{s}(\mathbf{\tilde{x}}),W^{u}(\mathbf{\tilde{x}})$\textit{\ of the same
dimensions }$d_{s}=d_{u}$\textit{\ as those of eigenspaces }$E^{s}$\textit{\
(}$\lambda _{l}\func{negative})$\textit{\ and }$E^{u}(\lambda _{l}$\textit{\
positive), }$\lambda _{l}$\textit{\ are eigenvalues of the matrix }$A_{ij}.$%
\textit{\ }$W^{s}(\mathbf{\tilde{x}})$\textit{\ and }$W^{u}(\mathbf{\tilde{x}%
})$\textit{\ are tangent to }$E^{s}$\textit{\ and }$E^{u}$\textit{\
respectively.\bigskip }

\textbf{Remark D.3. }Eigenspaces\textbf{\ }$E^{s}$ and $E^{u}$ are made of
respective eigenvector spaces. As in quantum mechanics, it is possible to
make egenvectors mutually orthogonal. If this is possible, the associated
stable and unstable manifolds are \textit{transversal} to each other. These
manifolds are considered to be transversal to each other even when the
associated eigenvectors are not orthogonal (since they can be made
orthogonal). The existence and uniqueness of the solution of $\mathbf{\dot{x}%
}=\mathbf{v}(\mathbf{x})$ ensures that two stable/unstable manifolds of
distinct fixed points, say $\mathbf{\tilde{x}}_{1}$ and $\mathbf{\tilde{x}}%
_{2},$ cannot intersect. For the same reason $W^{s}(\mathbf{\tilde{x}})$ and 
$W^{u}(\mathbf{\tilde{x}})$ cannot self-intersect. However, intersections of
stable \ and unstable manifolds of distinct fixed points or even of the same
fixed point are allowed. To define the Morse-Smale flows we need to
introduce the following$.$ \bigskip

\textbf{Definition D.4.} A point $p$ is called \textit{nonwandering} for the
flow $g^{t}$ if \ for any

neighborhood $U$ of $p$ \ there exists arbitrary large $t$ such that $%
g^{t}(U)$ $\cap U\neq \varnothing $ .

From here it follows, in particular, that fixed points and periodic orbits
are non wandering.\bigskip

\textbf{Definition D.5.} A Morse-Smale dynamical system is the one for which

the following holds.

1. The number of fixed points and periodic orbits is finite and each is
hyperbolic.

2. All stable and unstable manifolds intersect transversely.

3. The non wandering set consists of fixed points and periodic orbits
alone.\bigskip

\textbf{Corollary D.6.} The Morse- Smale systems are structurally stable.
Furthermore,

compact manifolds can possess only a finite number of periodic orbits and
fixed points.\bigskip

\textbf{Definition D.7.} The nonsingular Morse-Smale flow (NMS) does not
contain fixed points.\bigskip

\textbf{Corollary D.8.} The Poincar$e^{\prime }$-Hopf index theorem permits
NMS flow only on manifolds

whose Euler characteristic is zero. Therefore, such flows cannot

be realized, say, on $S^{2}$ but can be realized on $T^{2}.$ On $S^{3}$ such
flows also can exist.

Details are discussed in the main text.\medskip \bigskip

\textbf{Appendix E \ \ \ \ Some basic facts about Seifert fibered
spaces\medskip \bigskip }

A Seifert fibered manifold $M$ is a fiber bundle $(M,N,\pi ,S^{1})$ with $M$
being the total space, $N$ being the base, called the \textit{orbit}
manifold/orbit space. Typically, it is an orbifold, that is the manifold
with corners/cones). $\pi $ is the projection, $\pi :M\rightarrow N,$ and $%
S^{1}$ is the fiber. \ Let $\mathcal{F}$ be one of such fibers. A Seifert
fibered manifold is characterized by the property that each $\mathcal{F}$ \
has a \textit{fibered} \textit{neighborhood} \ which can be mapped into a 
\textit{fibered solid torus}. \ It is made out of a fibered cylinder $%
D^{2}\times I$ in which the fibers are $x\times I,$ $x\in D^{2}.$ By
rotating $D^{2}\times I$ through the angle $2\pi $ ($\nu /\mu )$ while
keeping $D^{2}\times 0$ fixed\ and identifying $D^{2}\times 0$ with $%
D^{2}\times I$ a \textit{fibered solid torus }is obtained. Upon such an
identification \ the fibers of $D^{2}\times I$ are being decomposed into
classes in such a way that each class contains exactly $\mu $ lines which
match together to give one fiber of the solid torus, e.g. see Fig.4a),
except that which is belonging to the class containing the axis of $%
D^{2}\times I.$ This fiber remains unchanged after identification. It
belongs to the class called \textit{singular} (exceptional, middle) \ [$56$]
. When $\mu =1$ the fiber bundle (still \ to be constructed) is trivial. The
number $\mu $ is called \textit{multiplicity} of the singular fiber. There
always exist a fiber-preserving homeomorphism into fibered solid torus such
that $\mathcal{F}$ can be mapped into a singular fiber. Reciprocally, from
this property\ it follows that each fibered neighborhood of $\mathcal{F}$
should have a boundary. Naturally, it should be $T^{2}.$ This means that
there should be a homeomorphism placing $\mathcal{F}$ onto $T^{2}$ in the
form of some closed curve $J$. Such a curve is being defined by its \textit{%
meridian} $\mathfrak{M}$ and \textit{longitude} $\mathfrak{L.}$ The
longitude $\mathfrak{L}$ of the solid torus is a simple closed curve on $%
T^{2}$ which intersects $\mathfrak{M}$ \ in exactly one point. \ Thus
formally, we can write 
\begin{equation}
J\sim \nu \mathfrak{M}+\mu \mathfrak{L.}  \tag{E.1a}
\end{equation}%
The numbers $\nu $ and $\mu $ are the same for all fibered neighborhoods of $%
\mathcal{F}$. Furthermore, from the way the solid torus was constructed, it
easily follows that the number $\nu $ is defined not uniquely: $\nu _{1}=\nu
_{2}$ $\left( \func{mod}\mu \right) .$ Therefore fibrations $2\pi $ ($\nu
/\mu )$ and $2\pi $ ($\nu +n\mu /\mu )$ are indistinguishable. In view of
previous results, \ we shall call the nonsingular $\ $fiber $\mathcal{F}$
for $\mu >1$\textit{\ as regular}$.$ \ The following theorem is of central
importance\medskip

\textbf{Theorem E.1. }\textit{A closed (compact) Seifert fibered space
contains at most finitely many singular fibers.\medskip \medskip }

From here it follows that all singular fibers are isolated. It is possible
to design Seifert manifolds without singular fibers though. In view of their
importance, we briefly sketch their design. Let $N$ \ be an \textit{orbit
space}. Suppose that it is a compact orientable surface with non-empty
boundary \ (made of $\partial D_{i}^{2}$ , $i=1,...,n).$ Then $M=N\times
S^{1}$ is a Seifert fibered manifold (a trivial fiber bundle) with fibers $%
x\times I,$ $x\in N.$ An orientable saddle, Fig.3a), defined in section 5 is
an example of such manifold.

Using thus designed (nonsingular) Seifert manifold, it is possible now to
design a manifold with singular fibers. For this purpose we select $n$ pairs
of coprime integers $(\alpha _{i},\beta _{i}),$ then we select a base $N$
which \ we had already described. It is a compact orientable surface from
which $n$ open discs $\mathring{D}_{i}^{2}=D_{i}^{2}$ $\smallsetminus
\partial D_{i}^{2}$ are removed so that the obtained surface $\tilde{N}=N-(%
\mathring{D}_{1}^{2}\cup \cdot \cdot \cdot \cup \mathring{D}_{n}^{2}).$ By
design, such Seifert manifold $\tilde{M}=\tilde{N}\times S^{1}$ is \textit{%
Seifert manifold without singular fibers}. Because of Theorem E.1. it is
appropriate to associate the locations of singular fibers with the locations
of the discs $D_{i}^{2}.$ \ By doing so we are going to construct a Seifert
manifold with singular fibers. For this purpose we need to notice that with
the boundary $\partial D_{i}^{2}$ of each disc $i$ the map $\pi
^{-1}(\partial D_{i}^{2})\rightarrow T_{i}^{2}$ can be associated. Choose on 
$T_{i}^{2}$ the meridian $\mathfrak{M}_{i}=\partial D_{i}^{2}\times \{\ast
\} $ and the longitude $\mathfrak{L}_{i}=\{\ast \}_{i}\times S^{1}$
consistent with the orientation of $\tilde{M}.$ At the boundary $\partial
D_{i}^{2}\times S^{1}=$ $T_{i}^{2}$ we can construct a curve $J_{i}$ of the
same type as given by eq.(E.1a)%
\begin{equation}
J_{i}\sim \alpha _{i}\mathfrak{M}_{i}+\beta _{i}\mathfrak{L}_{i}  \tag{E.1b}
\end{equation}%
(we use the pair $(\alpha _{i},\beta _{i})$ for its description). Take now
trivially fibered solid torus $D^{2}\times S^{1}$and glue it to $T_{i}^{2}$
in such a way that its meridian $\mathfrak{\tilde{M}}_{i}$ is glued to $%
J_{i}\in T_{i}^{2}.$ For each $i$ the image under gluing of the curve $%
\{0\}\times S^{1}\in D^{2}\times S^{1}$ is going to become the i-th singular
fiber. This gluing homeomorphism is the last step in designing the Seifert
fibered manifold with $n$ singular fibers. A particular Seifert fibered
manifold can be now described via the following specification: $%
M=M(N;(\alpha _{1},\beta _{1}),...,(\alpha _{n},\beta _{n})).$ In section 5
an example of such a manifold is given by the non-orientable saddle, e.g.
see Fig.4a). It is the Seifert fibered manifold of the type $%
M=M(D^{2};(2,1)) $ with one singular fiber. Incidentally, if the orbit
manifold is $S^{2}$, then the obtained Seifert manifold \ is already familiar%
\textsl{\ Hopf fibration. It is a Seifert fibered manifold without singular
fibers }[$71$], page 87. Other Seifert-fibered manifolds without singular
fibers having $S^{2}$ as their orbit space are $S^{2}\times S^{1}$ and the
Lens spaces of the type $L(n,1)$. For $n=1$ we get back the Hopf fibration,
for $n=2$ we get the projective space \textbf{P}$^{3}$, while the $n=0$ case
is identified with $S^{2}\times S^{1}.$ Finally, a complement of any torus
knot in $S^{3}$ is Seifert fibered space too [$71$], page 87.\bigskip

\textbf{Appendix F Some facts about Kirby moves\medskip /calculus\bigskip }

In the famous theorem by Gordon and Luecke [$72$] it is proven that knots
are determined by their complements. That is to say, a non-trivial Dehn
surgery on a non-trivial knot in $S^{3}$ does not yield back $S^{3}$. This
theorem does not cover the case of links as Fig.12 illustrates.

\begin{figure}[tbp]
\begin{center}
\includegraphics[scale=1.2]{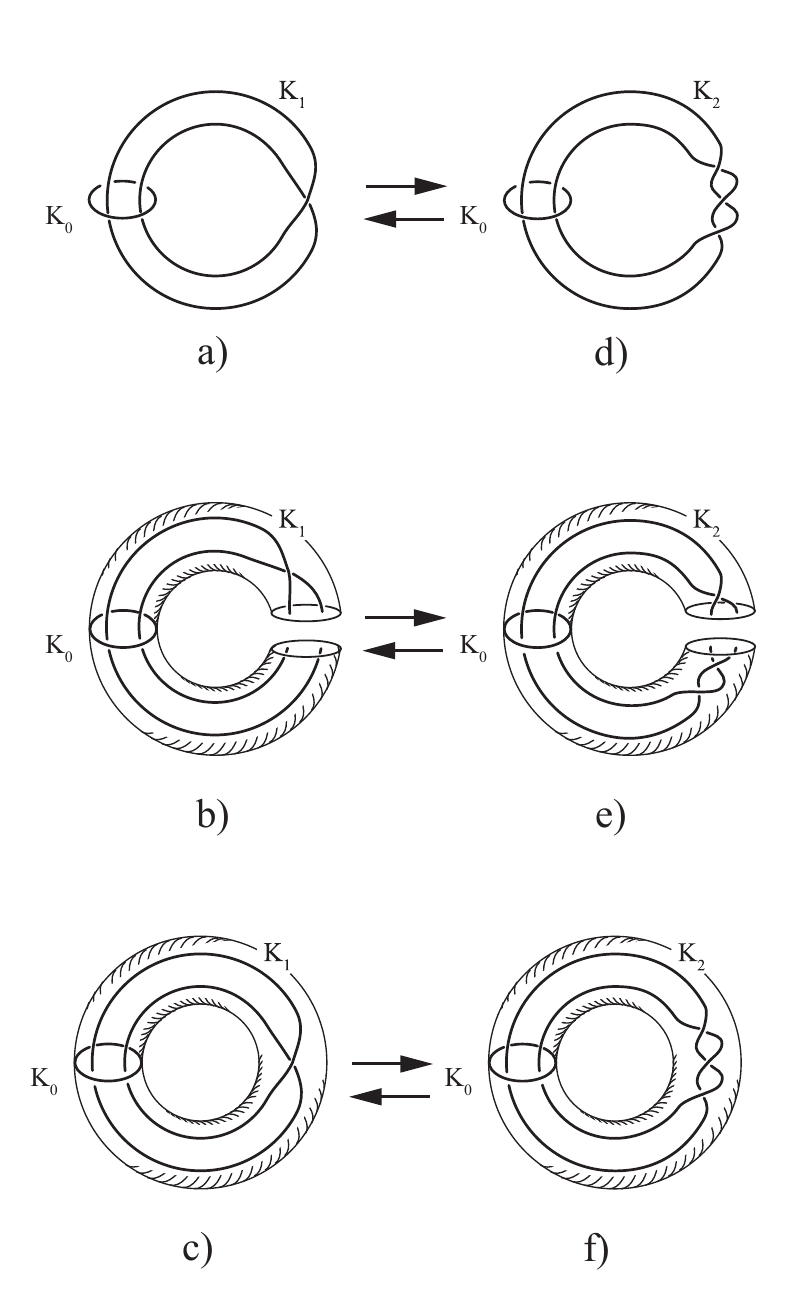}
\end{center}
\caption{The generic case (Rolfsen1976), page 49, of generating
topologically different links having the same complement in $S^{3} $}
\end{figure}

In section 7.7 we argued that the theorem by Gordon and Luecke [$72$] could
be also violated for the case of \ some knots. In Fig.12 the initial state
is a). It is made out of two unknots (e.g. think also about the Hopf link).
The final state is d). To reach d) from a) we have to perform a reversible
homeomorphism- from b) to c). The trick is achieved by enclosing the
original link located in $S^{3}$ into a solid torus, just like it is done in
the case of Fig.3a), and then performing the Dehn twist (or the sequence of
Dehn twists). This is further illustrated in going from c) to f). \ 

This generic example can be extended in several directions. For instance,
one of these directions was developed by Robion Kirby [$102$]. He invented
moves, now known as Kirby moves/calculus, broadly generalizing situation
depicted in Fig.12. To explain what he did we have to introduce several
definitions. We begin with\bigskip

\textbf{Definition F.1.(}\textit{Dehn surgery}) Let $K$ be some knot in some
3-manifold $M$. Let $N(K)$ be a tubular neighborhood of $K$ (that is
trivially fibered (that is without a twist) solid torus). Remove now this
solid torus $N(K)$ from $M.$ Then, what is left is a 3-manifold with
boundary $\mathring{M}=M\smallsetminus intN(K)$ where \textit{int} means
"interior" and the solid torus $N(K)=$ $D^{2}\times S^{1}.$ $\mathring{M}$
is the manifold whose boundary is $T^{2}$ that is $\partial \mathring{M}%
=T^{2}.$ Apparently, $M=\mathring{M}\cup (D^{2}\times S^{1}).$ To complicate
matters we can glue the solid torus back into $\mathring{M}$ via some
homeomorphism $h:\partial D^{2}\times S^{1}\rightarrow T^{2}.$ In such a
case, instead of $M=\mathring{M}\cup (D^{2}\times S^{1}),$ we obtain $\tilde{%
M}=\mathring{M}\cup _{h}(D^{2}\times S^{1}).$ Thus, a 3-manifold $\tilde{M}$
\ is obtained from $M$ via \textit{Dehn surgery along} $K.\bigskip $

The manifold $\tilde{M}$ depends upon the specification of $h$. It can be
demonstrated that it is sufficient to use the following type of (gluing)
homeomorphism\footnote{%
It is essentially the same as was used above for description of Seifert
fibered manifolds with singular fibers}. Choose $J\sim \alpha \mathfrak{M}%
+\beta \mathfrak{L}$ at the boundary $T^{2}$ of $\mathring{M}.$ Let $%
\mathfrak{M}$ be the meridian $\partial D^{2}\times \{\ast \}$ of the solid
torus $N(K).$ Then the $h$ is defined by the map: $J=h(\partial D^{2}\times
\{\ast \}).\func{Si}$nce the pairs $(\alpha ,\beta )$ and $(-\alpha ,-\beta
) $ define the same curve $J$, it is clear that the ratio $r=$ $\alpha
/\beta $ determines $J$. \ In particular, since $1/0=\infty $ determines the
meridian, the result of any $1/0$ surgery leaves $M$ unchanged.
Incidentally, the surgery determined by $r=0$ along the trivial knot
switchers meridian with parallel (the torus switch). This type of surgery
leads to the 3-manifold $S^{2}\times S^{1}$. The types of surgeries
determined by the rational $r$ are called \textit{rational}. The surgery is 
\textit{integral} if $\beta =\pm 1.$ As demonstrated by Lickorish and
Wallace [$54$] any closed orientable 3-manifold can be obtained by an
integral surgery along a link $\mathcal{L}\subset S^{3}.\bigskip $

\textbf{Remark F.2}. Operation of Dehn surgery is technically closely
related to the cabling operation, e.g. see Fig.7.\bigskip

\textbf{Definition F.3.} The procedure of \ assigning of $r_{i}$ for each
component $\mathcal{L}_{i}$ of $\mathcal{L}$ is called \textit{%
framing.\bigskip }

It is very helpful to think about the integral surgery/framing in terms of
linking numbers. The linking number \ for oriented knots/links can be easily
defined via knot/link projection. Using the convention as depicted in
Fig.13\ 

\begin{figure}[tbp]
\begin{center}
\includegraphics[scale=1.5]{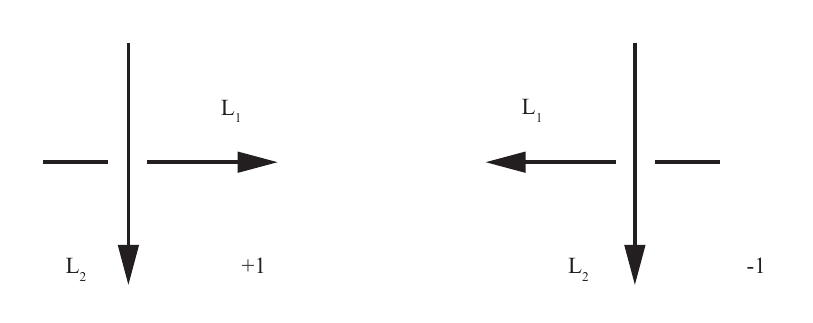}
\end{center}
\caption{The sign convention used for calculation of a) linking and b)
sel-linking (writhe) numbers for oriented links (for a) and knots (for b)
respectively}
\end{figure}

the linking number lk($L_{1}$,$L_{2})$ for oriented curves(links) $L_{1}$and 
$L_{2}$ is defined as 
\begin{equation}
lk(L_{1},L_{2})=\dsum\limits_{i}\varepsilon _{i}\text{, \ }\varepsilon =\pm
1.  \tag{F.1}
\end{equation}%
Evidently, $lk(L_{1},L_{2})=lk(L_{2},L_{1})$ and $%
lk(-L_{1},L_{2})=-lk(L_{2},L_{1})$. Here $-L_{1}$ means link $L_{1}$ with
orientation reversed.

Notice now that in the case of integral surgery the meridian $\mathfrak{M}$
is being mapped into the curve $J=\alpha \mathfrak{M}+\mathfrak{L.}$ From
here it follows that $J$ is making exactly one revolution in the direction
parallel to $\mathfrak{L}$. Locally this situation can be illustrated with
help of a ribbon depicted in Fig.14a).

\begin{figure}[tbp]
\begin{center}
\includegraphics[scale=1.5]{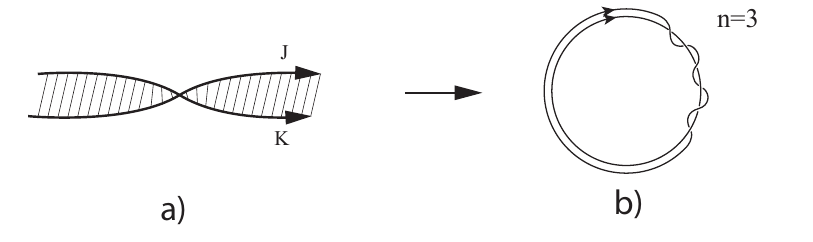}
\end{center}
\caption{Example of framing}
\end{figure}

From here we obtain the following\bigskip

\textbf{Definition F.4. \ }The integral framing of a knot $K$ corresponds to
the choice $lk(K,\mathfrak{L})=\alpha =n,$ where $n$ is an integer. An
example of integral framing is depicted in Fig.14b). In appendix E it was
stated that both Hopf links and complements of any torus knots in $S^{3}$
are Seifert fibered spaces. This does not mean \ that \ these spaces \ must
be the same. \ Following Lickorish and Wallace [$51,54$], we can construct
these Seifert fibered 3-manifolds using some sequence of \ integral
surgeries. \ This brings us to\bigskip

\textbf{The main question}: (\textit{Kirby}) How to determine when two
differently framed \ links \ produce the same 3-manifold?\bigskip

The answer to this question is given in terms of two (Kirby) moves. They are
designed as equivalence relations between links with different framings
which produce the same 3-manifolds. It is possible to inject some physics
into these equivalence moves using results of part I. Specifically,
following conventions, the value of integral framing is depicted next to the
respective projection of knot/link. For instance, if we supply the framing $%
\pm 1$ to the unknot, it is becoming the Hopf link. It was identified in
part I with the magnetic or electric charge. Accordingly, the 1st Kirby move
depicted in Fig.15 represents the charge conservation (e.g. see Fig.8).

\begin{figure}[tbp]
\begin{center}
\includegraphics[scale=1.5]{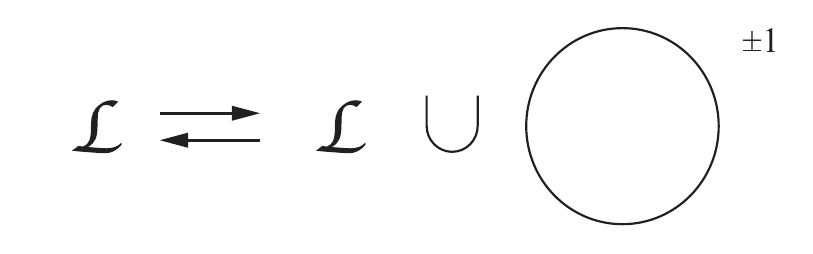}
\end{center}
\caption{The fist Kirby move}
\end{figure}

This move can be interpreted as follows: It is permissible to add or delete
an unknot with framing $\pm 1$ which does not intersect the other components
of $\mathit{L}_{i}$ to a given link $\mathcal{L}$. The 2nd Kirby move is
depicted in Fig.16.\bigskip\ 

\begin{figure}[tbp]
\begin{center}
\includegraphics[scale=1.5]{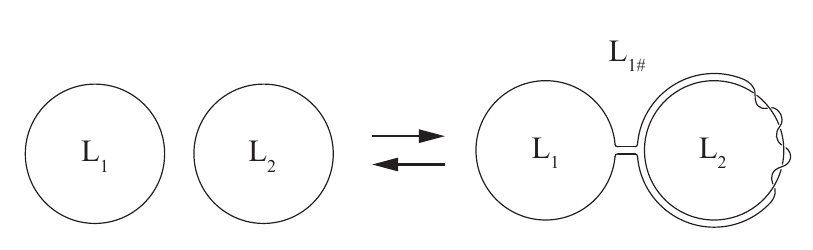}
\end{center}
\caption{The second Kirby move}
\end{figure}

Physically, it can be interpreted \ in terms of interaction between charges.
\ Mathematically this move can be interpreted as follows. Let $L_{1}$ and $%
L_{2}$ be two link components framed by the integers $n_{1}$ and $n_{2}$
respectively and $L_{2}^{\prime }$ a longitude defining the framing of $%
L_{2} $ that is $lk(L_{2},L_{2}^{\prime })=n_{2}.$ Replace now the pair $%
L_{1}$ $\cup $ $L_{2}$ by another pair $L_{1\#}$ $\cup $ $L_{2}$ in which $%
L_{1\#}=L_{1}$ $\#_{b}$ $L_{2}^{\prime }$ and $b$ is 2-sided band connecting 
$L_{1}$ with $L_{2}^{\prime }$ and disjoint from another link components.
While doing so, the rest of the link $\mathcal{L}$ remains unchanged. The
framings of all components, except $L_{1},$ are preserved while the framing
of $L_{1}$ is being changed into that for $L_{1\#}$ and is given by\ $n_{1}$ 
$+$ $n_{2}+2lk(L_{1},L_{2}).$ The computation of $lk(L_{1},L_{2})$ proceeds
in the standard way as described above (e.g. see Fig.13), provided that both 
$L_{1}$ and $L_{2}$ are oriented links.\bigskip

\textbf{Remark F.5. }Using the 1st and the 2nd\textbf{\ }Kirby moves it is
possible now to extend/improve results depicted in Fig.12. These
improvements are summarized and depicted in Fig.17.

\begin{figure}[tbp]
\begin{center}
\includegraphics[scale=1.4]{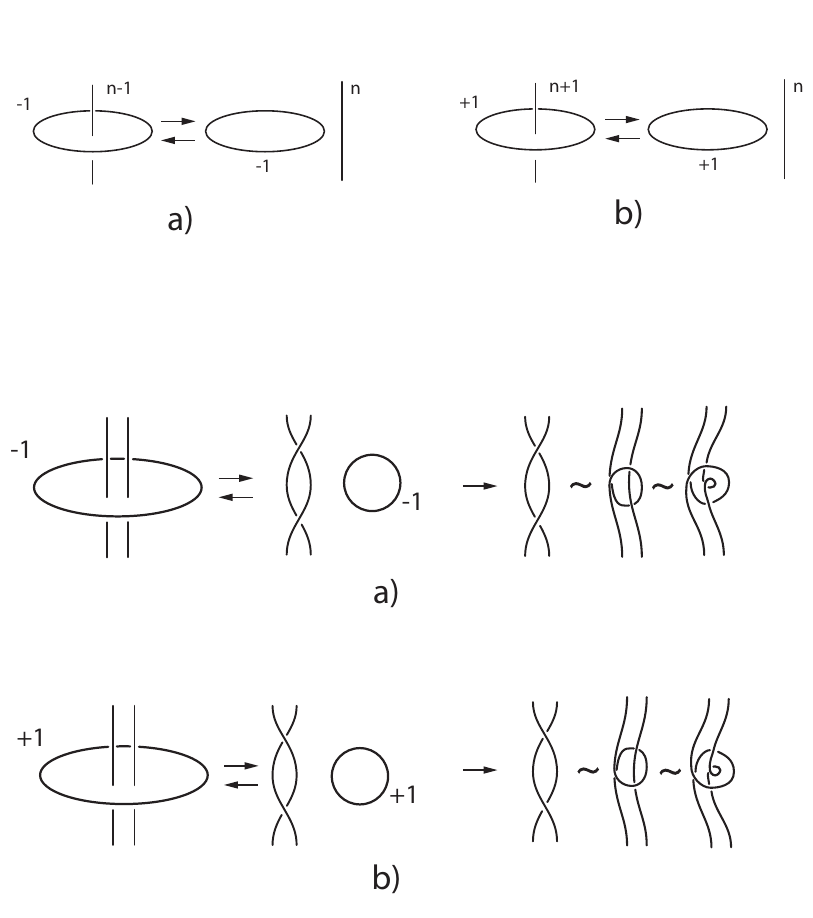}
\end{center}
\caption{ The Kirby moves, operations a) and b), are called "blow down" for
arrows pointing to the right, and "blow up" for arrows pointing to the left.
Twistings on the r.h.s. are known as the Fenn-Rourke moves.These can be used
instead of Kirby moves}
\end{figure}

The Fenn-Rourke moves are equivalent to blow ups /downs and are equivalent
to the combined first and second Kirby-type moves [$51$].\bigskip

\textbf{Remark F.6. }By comparing Kirby moves just described against those
suggested by Wada [$50$](e.g. see Fig.s 2-7 of paper by Campos et al [$61$])
it is clear that the moves suggested by Wada are just specific adaptations
of Kirby (or Fenn-Rourke) moves\medskip .\bigskip

\textbf{Appendix G Calculations of the various presentations for the
fundamental}

\textbf{group of the figure eight knot}

\textbf{\bigskip }

a) We begin with the set of relations%
\begin{equation}
r_{1}:bcb^{-1}=a,\rightarrow c=b^{-1}ab  \tag{G.1}
\end{equation}%
\begin{equation}
r_{2}:ada^{-1}=b,\rightarrow d=a^{-1}ba  \tag{G.2}
\end{equation}%
\begin{equation}
r_{3}:d^{-1}bd=c,\rightarrow \left( a^{-1}ba\right) ^{-1}b\left(
a^{-1}ba\right) =c=b^{-1}ab  \tag{G.3}
\end{equation}%
\begin{equation}
r_{4}:c^{-1}ac=d\rightarrow \left( b^{-1}ab\right)
^{-1}a(b^{-1}ab)=d=a^{-1}ba  \tag{G.4}
\end{equation}%
Noticing now that $\left( a^{-1}ba\right) ^{-1}=a^{-1}b^{-1}a$ and $\left(
b^{-1}ab\right) ^{-1}=b^{-1}a^{-1}b$ we can rewrite eq.(G.3) as 
\begin{equation}
a^{-1}b^{-1}aba^{-1}ba=b^{-1}ab\rightarrow a=ba^{-1}b^{-1}aba^{-1}bab^{-1} 
\tag{G.5}
\end{equation}%
and eq.(G.4) as 
\begin{equation}
b^{-1}a^{-1}bab^{-1}ab=a^{-1}ba\rightarrow a=ba^{-1}b^{-1}aba^{-1}bab^{-1} 
\tag{G.6}
\end{equation}%
Notice that the r.h.s. of eq.s(G.5) and (G.6) coincide. This circumstance
allows us to get rid of generators $c$ and $d$ in view of eq.s(G.3) and
(G.4). This means that we now can rewrite\footnote{%
Afrer still one line of algebra using eq.(G.5).} the presentation for $\pi
_{1}(S^{3}-K_{8})$ as follows%
\begin{equation}
<a,b,w\mid wa=bw,w=b^{-1}aba^{-1}>  \tag{G.7}
\end{equation}

b) In Miller [85] and in her PhD thesis [103] the following presentation of
the first fundamental group of the figure eight knot complement was given 
\begin{equation}
\pi _{1}(S^{3}-K_{8})=<x,y,z\mid zx^{-1}yz^{-1}x=1,xy^{-1}z^{-1}y=1> 
\tag{G.8}
\end{equation}%
without derivation. \ To avoid ambiguities, we checked this result and found
an error. The correct result is 
\begin{equation}
\pi _{1}(S^{3}-K_{8})=<x,y,z\mid zx^{-1}yxz^{-1}x=1,xy^{-1}z^{-1}y=1> 
\tag{G.9}
\end{equation}

Proof:

From eq.(G.8b) we obtain 
\begin{equation}
x^{-1}yxz^{-1}x=z^{-1}  \tag{G.9a}
\end{equation}%
and%
\begin{equation}
y^{-1}z^{-1}y=x^{-1}.  \tag{G.9b}
\end{equation}%
By comparing eq.(G.9b) with eq.(G.3) we make the following identifications%
\begin{equation}
x^{-1}=c,y=d,z^{-1}=b  \tag{G.10}
\end{equation}%
Now we can rewrite all relations in eq.s(G1-G.4) accordingly. Thus, we get: 
\begin{eqnarray}
r_{1} &:&z^{-1}x^{-1}z=a,  \notag \\
r_{2} &:&aya^{-1}=z^{-1},  \notag \\
r_{3} &:&y^{-1}z^{-1}y=x^{-1},  \notag \\
r_{4} &:&xax^{-1}=y.  \TCItag{G.11}
\end{eqnarray}%
Using eq.s(G.11), $r_{4},$ we obtain: \ $a=x^{-1}yx.$ Using this result in $%
r_{2}$ we obtain : $x^{-1}yxyx^{-1}y^{-1}x=z^{-1}.$ Using $r_{3}$ in this
relation, we obtain: $x^{-1}yxy(y^{-1}z^{-1}y)y^{-1}x=z^{-1}$ or, $%
x^{-1}yxz^{-1}x=$ $z^{-1}$ . But this result coincides with eq.(G.9a)! Next,
going to $r_{1}$ we obtain $z^{-1}x^{-1}z=x^{-1}yx$ or $x^{-1}yxz^{-1}xz=1$
in view of eq.(G.9a).

QED\bigskip

c) \ Use of symmetry enables us to embed the standard presentation eq(7.28)
for the trefoil into much more manageable representation, eq.(7.29). The
same philosophy was used by Miller (2001,2005). Unfortunately, its actual
implementation is plagued by mistakes. Specifically, Miller tried to embed
her presentation, eq.(G.8), into the dihedral group D$_{2}.$ For any
dihedral group D$_{n}$ , $n=1,2,...$ the presentation is well known [$104$]
and is given by 
\begin{equation}
\text{D}_{n}=<s,t\mid t^{n}=1,s^{2}=1,(ts)^{2}=1>.  \tag{G.12}
\end{equation}%
Miller choose $n=4$ for which she wrote the following presentation 
\begin{equation}
\text{D}_{4}=<s,t\mid t^{4}=1,s^{2}=1,s^{-1}ts=t^{-1}>  \tag{G.13}
\end{equation}%
Not only is the last relation incorrect, but, more importantly, the
graphical representation for this group given as Fig.14 of Miller [$85$] is
also incorrect.The depicted graph corresponds to the graph for D$_{2}!$ \
Since this graph matches well with BWb) graph for the template for the
figure eight knot (e.g. see Fig.1.2 of BW b)), we would like to demonstrate
that, indeed D$_{2}$ is the correct group. Thus, we need to find if we can
embed $\pi _{1}(S^{3}-K_{8})$ into D$_{2.}$. For D$_{2.}$ we have the
following presentation%
\begin{equation}
\text{D}_{2}=<s,t\mid t^{2}=1,s^{2}=1,tst=s^{-1}>  \tag{G.14}
\end{equation}%
Before proceeding with calculations, it is important to notice that
eq.(G.14) can be rewritten differently in view of the fact that for D$_{2}$
we have $ts=st.$ Indeed,%
\begin{equation}
1=(ts)^{2}=tsts=tsst=t^{2}\text{ since }s^{2}=1.  \tag{G.15}
\end{equation}%
Clearly, the relation $ts=st$ is compatible with rewriting of eq.(G.15) as $%
ts=1$ and $st=1.$ In view of these results, using eq.(G.9a), we obtain: 
\begin{equation}
xz^{-1}x=y^{-1}xz^{-1}.  \tag{G.16}
\end{equation}%
By looking at eq.(G.14) the following identification can be made: $%
x=t,z^{-1}=s$ and $y^{-1}xz^{-1}=s^{-1}$. The last result is equivalent to $%
y=ts^{2}.$ Using eq.(G.16) rewritten in just defined notations we obtain: $%
tst=ts^{2}$ or $st=s^{2}.$ If $s^{2}=1,$ then $st=1.$

At the same time, using eq$.($G$.9b)$ we obtain as well: $\left(
ts^{2}\right) ^{-1}s\left( ts^{2}\right) =t^{-1}.$ Since $s^{2}=1,$ we can
convert this result into $tst=t^{-1}$ and, if $t^{2}=1,$ then we obtain: $%
ts=1$. But we already obtained $st=1.$ Therefore, we get finally: $ts=st.$

QED

\bigskip

\bigskip

\bigskip \textbf{References\bigskip \bigskip }

[1] \ \ \ A. Enciso, D. Peralta-Salas, Ann.Math. 175 (2012) 345.

[2] \ \ \ J. Etnyre, R. Ghrist, Transactions AMS 352 (2000) 5781.

[3] \ \ \ H. Moffatt, J.Fluid Mech\textit{.} 159 (1985\textbf{)} 359.

[4] \ \ \ E. Witten, Comm.Math.Phys\textit{.}121(1989) 351.

[5] \ \ \ M. Atiyah, The Geometry and Physics of Knots,

\ \ \ \ \ \ \ Cambridge University Press, Cambridge,UK, 1990.

[6] \ \ \ A.Kholodenko, Applications of Contact Geometry and Topology

\ \ \ \ \ \ \ \ in Physics,World Scientific, Singapore, 2013.

[7] \ \ \ M. Arrayas, J.Trueba, 2012. arXiv:1106.1122.

[8] \ \ \ M. Kobayashi, M. Nitta, Phys.Lett.B 728 (2014) 314.

[9] \ \ \ A. Thomson, J. Swearngin, D. Bouwmeester, 2014. arXiv:1402.3806.

[10] \ \ A.Kholodenko, Analysis\&Math.Phys. 5 (2015)

\ \ \ \ \ \ \ \ http://link.springer.com/article/10.1007\%2Fs13324-015-0112-6

[11] \ \ C.Misner, J.Wheeler, Ann Phys. 2 (1957) 525.

[12] \ \ M.Atiyah, N. Manton, B.Schroers, Proc.Roy.Soc.A 2141 (2012) 1252.

[13] \ \ A.Kholodenko, IJMPA 30 (2015) 1550189.

[14] \ \ H. Geiges,\ \ An Introduction to Contact Topology,

\ \ \ \ \ \ \ \ \ Cambridge University Press, Cambridge, UK, 2008.

[15] \ \ N. Zung, A.Fomenko, Russian Math.Surveys 45 (1990) 109.

[16] \ \ J. Birman,R.Williams, Topology 22 (1983) 47.\ 

[17] \ \ J. Birman, R.Williams, Contemp. Math. 20 (1983) 1.

[18] \ \ R. Ghrist, \ Topology 36 (1997) 423.

[19] \ \ A.Ranada, \ Lett.Math.Phys. 18 (1989) 97.

[20] \ \ A.Ranada, \ J.Phys.A 25 (1992)1621.

[21] \ \ H. Kedia, I. Bialynicki-Birula, D. Peralta-Salas, W.Irvine,W 2013

\ \ \ \ \ \ \ \ Phys.Rev.Lett 111(2013)150404.

[22] \ \ A. Majda, A. Bertozzi, Votricity and Incompressible Flow,

\ \ \ \ \ \ \ \ Cambridge University Press, Cambridge, UK, 2003.

[23] \ \ A. Enciso, D. Peralta-Salas,\ Procedia IUTAM \textbf{7} (2013)13.

[24] \ \ A. Fomenko, in The Geometry of Hamiltonian Systems,

\ \ \ \ \ \ \ T. Ratiu Editor, pp.131-340, Springer-Verlag, Berlin, 1991.

[25] \ \ E.Ghys, \ ICM Proceedings, pp 247-277, Madrid, Spain, 2006.

[26] \ \ D. Kleckner,W. Irvine, Nature Physics 9 (2013) 253.

[27] \ \ J. Etnyre,R. Ghrist, Nonlinerarity 13 (2000) 441.

[28] \ \ S.Chern, R. Hamilton, LNM 1111 (1985) 279.

[29] \ \ A. Fischer, V. Moncrief, \ Class.Quantum Grav.18 (2001) 4493.

[30] \ \ A.Kholodenko, J.Geom.Phys. 58 (2008) 259.

[31] \ \ V.Arnol'd, Mathematical Methods of Classical Mechanics,

\ \ \ \ \ \ Springer-Verlag, Berlin, 1989.

[32] \ V.Arnol'd, B.Khesin, Topological Methods in Hydrodynamics,

\ \ \ \ \ \ \ Springer-Verlag, Berlin, 1998.

[33] \ A. Fomenko, A.Bolsinov, Integrable Hamiltonian Systems:

\ \ \ \ \ \ \ Geometry,Topology and Classification, CRC Press LLC, \ \ \ \ \
\ 

\ \ \ \ \ \ \ Boca Raton, Florida, 2004.

[34] \ B. Jovanovi\v{c}, The Teaching of Mathematics 13 (2011\textbf{)} 1.

[35] \ O. Calin, Ch.Der-Ch, Sub-Riemannian Geometry,

\ \ \ \ \ \ \ Cambridge University Press, Cambridge, UK, 2009.

[36] \ \ A. Kamchatov,\ Sov.Phys. JETP 55 (1982) 69.

[37] \ R.Ghrist, P. Holmes, M. Sullivan, Knots and links in three
dimensional flows,

\ \ \ \ \ \ \ Lecture Notes in Mathematics\textit{\ }1654\textbf{, }%
Springer-Verlag, Berlin ,1997.

[38] \ R. Gilmore, M.Lefranc, The Topology of Chaos\textit{\ }

\ \ \ \ \ \ \ Wiley-Interscience Inc. New York, 2002.

[39] \ \ J.Franks, M. Sullivan M 2002 Flows with knotted closed orbits in

\ \ \ \ \ \ \ \ Handbook of Geometric Topology, pp 471-497,

\ \ \ \ \ \ \ \ North-Holland, Amsterdam, 2002.

[40] \ \ H. Hofer,\ \ Dynamics, topology and holomorphic curves,

\ \ \ \ \ \ \ \ Documenta Mathematica, Extra Volume, ICM, 1-27.

[41] \ \ M. Hutchings, \ AMS Bulletin \ 47 (2009) 73.

[42] \ \ H. Hofer,\ Inv.Math. 114 (1993) 515.

[43] \ \ V. Ginzburg,\ in The Breadth of Symplectic and Poisson\textit{\ }

\ \ \ \ \ \ \ \ Geometry, pp. 139-172, Birkh\"{a}user, Boston, 2005.

[44] \ \ H. Hofer, K.Wysocki, E. Zehnder, Ann.Math. 148 (1998) 197.

[45] \ \ J. Morgan, Topology 18 (1978$)$ 41.

[46] \ \ M. Hirch,S. Smale, Differential Equations, Dynamical Systems\textit{%
\ }

\ \ \ \ \ \ \ \ and Linear Algebra\textit{,} Academic Press, New York, 1974.

[47] \ \ C.Nash, S.Sen, \ Topology and Geometry for Physicists\textit{,}

\ \ \ \ \ \ \ \ Academic Press Inc., New York, 1983.

[48] \ \ T. Frankel,\ The Geometry of Physics,

\ \ \ \ \ \ \ \ Cambridge University Press, Cambridge, UK,1997.

[49] \ \ \ M. Guest,\ Morse theory in the 1990's,

\ \ \ \ \ \ \ \ in Invitation to Geometry and Topology\textit{,}pp. 146-207,

\ \ \ \ \ \ \ \ Oxford U. Press, Oxford,UK, 2001.

[50] \ \ M.Wada,\ \ J.Math.Soc.Japan 41(1989) 405.

[51] \ \ \ V. Prasolov, A.Sossinsky, Knots, Links, Braids and 3-Manifolds%
\textit{,}

\ \ \ \ \ \ \ \ \ AMS Publishers, Providence,RI, 1997.

[52] \ \ \ P. Scott, Bull London Math. Soc\textit{.} 15 (1983) 401.

[53] \ \ \ A. Fomenko, S. Matveev, Algorithmic and Computer Methods\textit{\ 
}

\ \ \ \ \ \ \ \ \ for Three-Manifolds\textit{, }Kluver Academic Publishers,
Boston, MA, 1997.

[54] \ \ \ N. Saveliev, Lectures on the Topology of 3-Manifolds\textit{,}

\ \ \ \ \ \ \ \ \ Walter de Gruyter Gmbh \&Co, Berlin, 2012.

[55] \ \ \ D. Rolfsen, \ Knots and Links\textit{, }

\ \ \ \ \ \ \ \ \ Publish or Perish, Inc., Houston, TX, 1976.

[56] \ \ \ H.Seifert, W. Threlfall, A Textbook on Topology\textit{,}

\ \ \ \ \ \ \ \ \ Academic Press, New York , 1980.

[57] \ \ \ S. Matveev, A. Fomenko, Russian Math.Surveys 43 (1988\textbf{)} 3.

[58] \ \ \ J.Kock, Frobenius Algebras and 2d Topological Quantum

\ \ \ \ \ \ \ \ \ Field Theories, Cambridge University Press,
Cambridge,UK,2004.

[59] \ \ \ V. Manturov, Virtual Knots, World Scientific, Singapore, 2013.

[60] \ \ \ L.Kauffman, European J. Combin. 20 (1999) 663.

[61] \ \ \ B.Campos, J. Martinez-Alfaro, P.Vindel,

\ \ \ \ \ \ \ \ \ \ \ J.of Bifurcation and Chaos 7 (1999\textbf{)} 1717.

[62] \ \ \ W. Menasco, Geometry and Topology 5 (2001) 651.

[63] \ \ \ J. Birman, N. Wrinkle, J.Diff.Geom. 55 (2000) 325.

[64] \ \ \ K.\ Murasugi, \ Knot Theory and Its Applications\textit{,}

\ \ \ \ \ \ \ \ \ Birkh\"{a}user, Boston,MA, 1996.

[65] \ \ \ W. Jaco, P. Shalen, Seifert Fibered Spaces in 3-Manifolds,

\ \ \ \ \ \ \ \ \ AMS Memoirs 21, Number 220,

\ \ \ \ \ \ \ \ \ AMS Publishers, Providence, RI, 1979.

[66] \ \ \ J. Hempel,\ AMS Proceedings 15 (1964)154.

[67] \ \ \ J. Milnor,\ Singular Points of Complex Hypersurfaces\textit{,}

\ \ \ \ \ \ \ \ \ Princeton University Press, Princeton, NJ,1968.

[68] \ \ \ D. Eisenbud,W.Neumann, Three-dimensional Link Theory and

\ \ \ \ \ \ \ \ \ Invariants of Plane Curve Singularities\textit{,}

\ \ \ \ \ \ \ \ \ Princeton University Press, Princeton, NJ,1985.

[69] \ \ \ \ M. Dennis, R. King , B. Jack, K. O'Holleran, M. Padgett,

\ \ \ \ \ \ \ \ \ \ Nature Physics 6 (2010) 118.

[70] \ \ \ \ T. Machon, G.Alexander, PNAS 110 (2013) 14174.

[71] \ \ \ \ W. Jaco, \ Lectures on Three-Manifold Topology\textit{,}

\ \ \ \ \ \ \ \ \ \ AMS Publishers, Provuidence, RI, 1980.

[72] \ \ \ \ \ C. Gordon, J. Luecke,\ J.Amer.Math.Soc. 2 (1989\textbf{)} 371.

[73] \ \ \ \ \ H. Schubert, Acta Math. 90 (1953) 131.

[74] \ \ \ \ \ V. Arnol'd, Russian Math. Surveys 41(1986) 1.

[75] \ \ \ \ \ J.Etnyre,\ \ 2004.\ \ arxiv:math/0306256 v2.

[76] \ \ \ \ \ S. Overkov,V.Shevchisin, J.Knot Theory and Ramifications 12
(2003) 905.

[77] \ \ \ \ \ D. Benequin, Asterisque 197 (1983) 87.

[78] \ \ \ \ \ A. Hurtado,C. Rosales, Math.Ann. 340 (2008)\textbf{\ } 675.

[79] \ \ \ \ \ R.Ghrist, Chaos, Solitons and Fractals 9 (1998) 583.

[80] \ \ \ \ \ H. Fr\"{o}hlich, Proc.Phys.Soc.London 87 (1966) 330.

[81] \ \ \ \ \ A. Kholodenko,\ J.Geom.Phys. 38\ (2001) 81.

[82] \ \ \ \ \ E. Ghys, J Leys, Lorenz and modular flows: A visual
introduction.

\ \ \ \ \ \ \ \ \ \ \ Monthly Essays on Mathematical Topics, AMS Feature
Column, 2011.

[83] \ \ \ \ \ Y. Minsky, Ann.Math\textit{.}149 (1999) 559.

[84] \ \ \ Y. Imayoshi, M.Taniguchi, An introduction to Teichm\"{u}ller
Spaces\textit{,}

\ \ \ \ \ \ \ \ \ Springer-Verlag, Berlin, 1992.

[85] \ \ \ S. Miller,\ Experimental Mathematics 10 (2001)\ 419.

[86] \ \ \ J. Birman, \ Lorenz knots and links\ (transparencies of the talk
given

\ \ \ \ \ \ \ \ \ on \ Feb13, 2009).

[87] \ \ \ C.Kassel,V.Turaev, Braid Groups\textit{,} Springer-Verlag,
Berlin, 2008.

[88] \ \ \ A. Kholodenko, 1999. arXiv: cond. mat/9905221.

[89] \ \ \ B. Bowditch, Proc. London Math.Soc. 77 (1998) 697.

[90] \ \ \ J. Elstrodt, F. Grunewald, J. Mennicke, Groups Acting on
Hyperbolic Space\textit{,}

\ \ \ \ \ \ \ \ \ Springer-Verlag, Berlin, 1998.

[91] \ \ \ C. Maclachlan, A. Reid,\ \textit{T}he Arithmetic Hyperbolic
3-Manifolds

\ \ \ \ \ \ \ \ \ Springer-Verlag, Berlin, 2003.

[92] \ \ \ W. Goldman, Complex Hyperbolic Geometry,

\ \ \ \ \ \ \ \ \ Clarendon Press, Oxford, 1999.

[93] \ \ \ E.Falbel, J.Diff.Geometry 79 (2008)\ 69.

[94] \ \ \ H. Stephani, \ \ General Relativity\textit{,}

\ \ \ \ \ \ \ \ \ Cambridge University Press,Cambridge, UK,1990.

[95] \ \ \ Y. Mathieu, J.of Knot Theory and Its Ramifications\ 1 \ (1992)
279.

[96] \ \ \ C. Chu, T. Ohkawa, Phys.Rev.Lett.48 (1982) 837.

[97] \ \ \ K. Brownstein, Phys.Rev. A 35 (1987) 4856.

[98] \ \ \ K.\ Uehara, T.Kawai, K. Shimoda, \ J.Phys.Soc.Japan 58 (1989)
3570.

[99] \ \ \ S. Donaldson, \ Floer Homology Groups in Yang-Mills Theory\textit{%
,}

\ \ \ \ \ \ \ \ \ Cambridge University Press, Cambridge, UK, 2002.

[100] \ \ V. Arnol'd, Geometrical Methods in the Theory of Ordinary
Differential

\ \ \ \ \ \ \ \ \ Equations, Springer-Verlag, Berlin, 1988.

[101] \ \ J. Guckenheimer, P. Holmes, \ Nonlinear Oscillations, Dynamical

\ \ \ \ \ \ \ \textit{\ \ }Systems and Bifurcations of Vector Fields\textit{,%
} Springer-Verlag, Berlin 1983.

[102] \ \ R. Kirby, Inv.Math. 45 (1978) 36.

[103] \ \ S. Miller,\ Geodesic Knots in Hyperbolic 3-Manifolds,

\ \ \ \ \ \ \ \ \ \ PhD Tesis, Department of Mathematics and Statistics,

\ \ \ \ \ \ \ \ \ \ U.of Melbourne, Australia, 2005.

[104] \ \ \ I.Grosman, W.Magnus, Groups and Their Presentations,

\ \ \ \ \ \ \ \ \ \ The Random House, New York, 1964.

\bigskip

\bigskip

\bigskip

\bigskip

\bigskip

\bigskip

\end{document}